\documentclass[12pt]{article}

  \usepackage{graphicx}
  \usepackage{epsfig}
  \usepackage{amssymb}
  \usepackage{subfigure}
  \usepackage{axodraw}
    \usepackage{feynmf}%%%{feynmp}
\evensidemargin - 1.truecm
\begin{document}
\begin{fmffile}{1Mffa}

\newcommand{\be}{\begin{equation}}
\newcommand{\ee}{\end{equation}}
\newcommand{\nn}{\nonumber}
\newcommand{\bea}{\begin{eqnarray}}
\newcommand{\eea}{\end{eqnarray}}
\newcommand{\bfig}{\begin{figure}}
\newcommand{\efig}{\end{figure}}
\newcommand{\bc}{\begin{center}}
\newcommand{\ec}{\end{center}}
\newcommand{\bd}{\begin{displaymath}}
\newcommand{\ed}{\end{displaymath}}

\begin{titlepage}
\nopagebreak
{\flushright{
        \begin{minipage}{5cm}
	Rome1-1353/03\\
        Freiburg-THEP 04/01\\
        {\tt hep-ph/0304028}\\
        \end{minipage}        }

}
\renewcommand{\thefootnote}{\fnsymbol{footnote}}
\vspace*{-1.5cm}                       
\vskip 3.5cm
\begin{center}
\boldmath
{\Large\bf Master integrals with one massive propagator\\[3mm]
for the two-loop electroweak form factor }\unboldmath
\vskip 1.cm
{\large U. Aglietti\footnote{Email: Ugo.Aglietti@roma1.infn.it}},
\vskip .2cm
{\it Dipartimento di Fisica Universit\`a di Roma 
"La Sapienza" and INFN \\ 
Sezione di Roma, 
I-00185 Roma, Italy} 
\vskip .2cm
{\large  R. Bonciani\footnote{Email:
Roberto.Bonciani@physik.uni-freiburg.de}
\footnote{This work was supported by the European Union under
contract HPRN-CT-2000-00149},}
\vskip .2cm
{\it Physikalisches Institut,
%Facult\"at f\"ur Mathematik und Physik, 
Albert-Ludwigs-Universit\"at
Freiburg, \\ 
D-79104 Freiburg, Germany} 
\end{center}
\vskip 1.2cm

\begin{abstract}

We compute the master integrals containing one massive propagator 
entering the two-loop electroweak form factor, i.e. the process 
$ f \bar{f}\rightarrow X $, where  $f\bar{f}$ is an on-shell massless 
fermion pair and $X$ is a singlet particle under 
$SU(2)_{L} \times U(1)_{Y}$, such as a virtual gluon or an hypothetical 
$Z'$. The method used is that of the differential equation 
in the evolution variable $x = -s/m^2$, where $s$ is the c.m. energy 
squared and $m$ is the mass of the $W$ or $Z$ bosons (assumed to be 
degenerate). The $1/\epsilon$ poles and the finite parts are computed 
exactly in terms of one-dimensional harmonic polylogarithms of the 
variable $x$, $H(\vec{w};x)$, with $\epsilon=2-D/2$ and $D$ the 
space-time dimension. We present large-momentum expansions of the 
master integrals, i.e. expansions for $|s| \gg m^2$, which are relevant 
for the study of infrared properties of the Standard Model. We also 
derive small-momentum expansions of the master integrals, i.e. 
expansions in the region $|s| \ll m^2$, related to the threshold 
behaviour of the form factor (soft probe). Comparison with previous 
results in the literature is performed finding complete agreement.


\end{abstract}
\vfill
\end{titlepage}    

\section{Introduction \label{Intro}}

Present day and planned high-energy experiments often require accurate 
perturbative calculations. While in quantum-choromodynamics (QCD) higher
order computations are available for various processes, in the 
electroweak (EW) case progress is more difficult because of the presence
of the large, non negligible, masses of $W$, $Z$ and Higgs bosons.
A full calculation of two-loop electroweak corrections to a 
typical process such as $e^{+} e^{-} \rightarrow \mu^{+} \mu^{-}$
is still outside present-day technology. 
It involves indeed the computation of a large number of massive 
two-loop box diagrams having thresholds at $s=0,~m^2,~4m^2$ and $s=9m^2$, 
assuming $m\sim m_W \sim m_Z \sim m_H$. To have an explicit estimate 
of the size of two-loop electroweak corrections, we consider then the 
model process: 
\be
f(p_{1}) + \bar{f}(p_{2})  \rightarrow X(q),
\label{basic}
\ee  
where $f$ ($\bar{f}$) is a massless fermion (antifermion) on its 
mass-shell, $p_{1}^{2} = 0$ ($p_{2}^{2} = 0$),
and $X$ is any particle without electromagnetic or weak charge, i.e. a 
singlet under the electroweak gauge group $SU(2)_{L} \times U(1)_{Y}$, 
such as a (time-like) gluon or an hypothetical $Z'$. Eq. (\ref{basic}) 
also represents the simplest process containing electroweak double 
logarithms, considered for the first time, as far as we know, in 
\cite{ciafcom}. Reaction (\ref{basic}) is the electroweak analog of the
QCD processes involving a color-neutral probe, such as Drell-Yan 
production of vector bosons or muon pairs: 
$q \overline{q}^{(')}  \rightarrow W, Z,\mu^{+}\mu^{-}$
or Higgs production by gluon-gluon fusion: $g g \rightarrow H$.

At two-loop level, the annihilation in Eq. (\ref{basic}) involves the 
emission of a pair of virtual bosons among $\gamma$, $W$, $Z$ and $H$'s.
In the case of double photon emission, no mass is present in the loops 
and we recover the known results for the QED or QCD amplitudes. 
The latter, for dimensional reasons, are of the form:
\begin{equation}
\! \! \! {\rm (massless~amp.)} =
G(\epsilon) \left(\frac{-s}{\mu^{2}}\right)^{-2\epsilon} s^{4-n_d} =
a^{4-n_d}\left(\frac{\mu^2}{a}\right)^{2\epsilon}G(\epsilon) 
x^{4-n_d-2\epsilon},
\end{equation}
where $s$ is the c.m. energy squared,
$G(\epsilon)$ is a function having a Laurent expansion in 
$\epsilon=2-D/2$ (with $D$ the space-time dimension), $\mu$ is the
mass-scale of Dimensional Regularization and $n_{d}$ is the number 
of scalar propagators. We have defined:
\begin{equation}
x = - \frac{s}{m^2} \, ,
\end{equation}
where $m$ is the mass of the $W$, $Z$ or Higgs bosons\footnote
{The minus sign in the definition of $x$ 
is inserted for convenience.}.
On the other hand, if $W$, $Z$ or $H$'s are also 
emitted, the dependence on $x$ is not factorized in a single 
power and the amplitude has the more general form
\begin{equation}
{\rm (massive~amp.)}=
a^{4-n_d}\left(\frac{\mu^2}{a}\right)^{2\epsilon}G(x;\epsilon). 
\end{equation}
The amplitudes involving a single massive propagator, i.e. a single $W$ 
or $Z$ emission, have thresholds only at $s = 0$ and $s = m^2$. As a 
consequence of this simple structure, these amplitudes can be 
exactly computed in terms of one-dimensional harmonic polylogarithms 
\cite{Polylog,Polylog3} (a generalization of the Nielsen's
polylogarithms \cite{Nielsen,Kolbig}) of argument $x$, $H(\vec{w},x)$, 
with weight up 
to $w = 4$ included. In this paper we present the evaluation of the 
master integrals entering the above-mentioned amplitudes, leaving to a 
future work the more complicated amplitudes with two and three massive 
propagators, involving also the threshold $s = 4 m^2$.
Since external fermions are massless and are taken on their mass-shell, 
infrared divergences (soft and collinear) occur in the amplitudes. 
When an intermediate vector boson is radiated, its mass $m$ acts as an 
infrared regulator and mass logarithms do appear in four dimensions,
i.e. terms of the form log$^k(-s/m^2)$ with $k$ up to two included. 
Infrared divergences related to photon emission instead are not 
regulated by any physical parameter. Our computation is done within 
the Dimensional Regularization scheme \cite{DimReg}, in which the 
latter manifest themselves as poles in $1/\epsilon$.
These poles are unphysical and, in the physical cross-section, 
are canceled by the corresponding ones appearing in the real photon 
emission contributions or are factorized into QED structure functions. 
Ultraviolet divergences --- related to short-distance behaviour and 
leading to additional $1/\epsilon$ poles --- are also regulated within 
Dimensional Regularization and can be subtracted with ordinary 
renormalization  prescriptions, such as, for example, the 
$\overline{MS}$ scheme.

The computation of the form factor is naturally divided in two steps.

The first one involves the reduction of the original tensor diagrams,
given in Figs.~\ref{fig1} and \ref{fig1bis}, into a minimal set of 
scalar amplitudes, the so-called master integrals. This step is 
discussed in section \ref{Master} and involves in turn three different 
operations: $a)$ the decomposition of tensor amplitudes into
scalar amplitudes by means of projectors on invariant
form factors, see section 2.1;  
$b)$ the transformation to linearly independent scalar amplitudes
by means of scalar product rotations, as 
discussed in section 2.2; 
$c)$ the reduction of all the independent 
scalar amplitudes generated to a small subset of 
them --- the so called master integrals --- by means of the 
integration-by-parts identities \cite{Chet} and other symmetry 
relations, as discussed in section 2.3.

The second step involves the explicit evaluation of the master 
integrals, which we perform in all the massive cases with the method 
of differential equations on the external kinematical invariants 
\cite{Kotikov1,Kotikov2,Kotikov3,Rem1,Rem2,Rem3,Bon1}. A 
set of special functions of the polilogarithm kind is needed to 
represent the pole and finite parts of the master integrals. We choose 
the basis of the (one-dimensional) harmonic polylogarithms 
\cite{Polylog,Polylog3}. This step is described in section 
\ref{diffeqs}.

In section \ref{Results} we list the results for the pole and finite 
parts of the master integrals containing up to five denominators 
included.
In section \ref{6Denominatori} we give the expressions of all the scalar
integrals containing six denominators. The planar topologies
are reducible while the crossed ladders --- 
two amplitudes --- are master integrals. We decided to list them 
together because all the six-denominator amplitudes are interesting by 
themselves for reference. We also compare with previous results in the 
literature.
Section \ref{GrandiP} is devoted to the large momentum expansion
$|s| \gg m^2$ of all the six-denominator amplitudes. As already 
said, this expansion is relevant for the study of asymptotic 
properties of multiple electroweak radiation.
Section \ref{PiccoliP} deals with the small momentum expansion --- i.e. 
the expansion for $|s| \ll m^2$ --- of all the six-denominator
amplitudes.
In the conclusions we summarize the results obtained and we discuss 
future perspectives.
To not bore the reader with too many technical details or formulas --- 
which are anyway worth reporting --- we added a few appendices.
Appendix \ref{app1} contains the expressions of all the denominators 
entering the two-loop amplitudes.
In appendix \ref{app2} we give the results of the one-loop amplitudes
which enter the factorized two-loop diagrams, i.e. the diagrams in 
which the two loops do not have common propagators.
Finally, in appendix \ref{app3}, we list the results for all the 
independent topologies of Figs. \ref{fig2}, \ref{fig3} and \ref{fig4} 
which are not master integrals but can be useful for general reference.
%
%
%%%%%%%%%%%%%%%%%%%% 2-loop Vertex %%%%%%%%%%%%%%%%%%%%%%%%%%%%%%%%%%%%%
\bfig
\bc
\subfigure[]{
\begin{fmfgraph*}(35,35)
\fmfleft{i1,i2}
\fmfright{o}
\fmf{photon}{i1,v1}
\fmf{photon}{i2,v2}
\fmf{fermion}{v5,o}
\fmflabel{$p_{2}$}{i1}
\fmflabel{$p_{1}$}{i2}
\fmfv{l=$p_{1} \! - \! k_{1} \! + \! k_{2}$,l.a=15,l.d=.1w}{v3}
\fmfv{l=$p_{2} \! + \! k_{1} \! - \! k_{2}$,l.a=-15,l.d=.1w}{v4}
\fmf{photon,tension=.3,label=$p_{1} \! - \! k_{1}$,
                       label.side=left}{v2,v3}
\fmf{photon,tension=.3}{v3,v5}
\fmf{photon,tension=.3,label=$p_{2} \! + \! k_{1}$,
                       label.side=right}{v1,v4}
\fmf{photon,tension=.3}{v4,v5}
\fmf{photon,tension=0,label=$k_{1}$,label.side=right}{v2,v1}
\fmf{photon,tension=0,label=$k_{2}$,label.side=left}{v4,v3}
\end{fmfgraph*} }
%
%%%%%%%%%%%%%%%%%%%%%%%
%
\hspace{4mm}
\subfigure[]{
\begin{fmfgraph*}(35,35)
\fmfleft{i1,i2}
\fmfright{o}
\fmf{photon}{i1,v1}
\fmf{photon}{i2,v2}
\fmf{fermion}{v5,o}
\fmflabel{$p_{2}$}{i1}
\fmflabel{$p_{1}$}{i2}
\fmfv{l=$p_{1} \! - \! k_{1} \! + \! k_{2}$,l.a=15,l.d=.1w}{v3}
\fmfv{l=$p_{2} \! + \! k_{1} \! - \! k_{2}$,l.a=-15,l.d=.1w}{v4}
\fmf{photon,tension=.3,label=$p_{1} \! - \! k_{1}$,
                       label.side=left}{v2,v3}
\fmf{photon,tension=.3}{v3,v5}
\fmf{photon,tension=.3,label=$p_{2} \! + \! k_{1}$,
                       label.side=right}{v1,v4}
\fmf{photon,tension=.3}{v4,v5}
\fmf{fermion,tension=0,label=$k_{1}$,label.side=right}{v2,v1}
\fmf{photon,tension=0,label=$k_{2}$,label.side=left}{v4,v3}
\end{fmfgraph*} }
%
%%%%%%%%%%%%%%%%%%%%%%%
%
\hspace{4mm}
\subfigure[]{
\begin{fmfgraph*}(35,35)
\fmfleft{i1,i2}
\fmfright{o}
\fmf{photon}{i1,v1}
\fmf{photon}{i2,v2}
\fmf{fermion}{v5,o}
\fmflabel{$p_{2}$}{i1}
\fmflabel{$p_{1}$}{i2}
\fmfv{l=$p_{1} \! - \! k_{1} \! + \! k_{2}$,l.a=15,l.d=.1w}{v3}
\fmfv{l=$p_{2} \! + \! k_{1} \! - \! k_{2}$,l.a=-15,l.d=.1w}{v4}
\fmf{photon,tension=.3,label=$p_{1} \! - \! k_{1}$,
                       label.side=left}{v2,v3}
\fmf{photon,tension=.3}{v3,v5}
\fmf{photon,tension=.3,label=$p_{2} \! + \! k_{1}$,
                       label.side=right}{v1,v4}
\fmf{photon,tension=.3}{v4,v5}
\fmf{photon,tension=0,label=$k_{1}$,label.side=right}{v2,v1}
\fmf{fermion,tension=0,label=$k_{2}$,label.side=left}{v4,v3}
\end{fmfgraph*} } \\
%
%%%%%%%%%%%%%%%%%%%%%%%
%
\subfigure[]{
\begin{fmfgraph*}(35,35)
\fmfleft{i1,i2}
\fmfright{o}
\fmf{photon}{i1,v1}
\fmf{photon}{i2,v2}
\fmf{fermion}{v5,o}
\fmflabel{$p_{2}$}{i1}
\fmflabel{$p_{1}$}{i2}
\fmfv{l=$p_{1} \! - \! k_{1} \! + \! k_{2}$,l.a=15,l.d=.1w}{v3}
\fmfv{l=$p_{2} \! + \! k_{1} \! - \! k_{2}$,l.a=-15,l.d=.1w}{v4}
\fmf{fermion,tension=.3,label=$p_{1} \! - \! k_{1}$,
                        label.side=left}{v2,v3}
\fmf{photon,tension=.3}{v3,v5}
\fmf{photon,tension=.3,label=$p_{2} \! + \! k_{1}$,
                       label.side=right}{v1,v4}
\fmf{photon,tension=.3}{v4,v5}
\fmf{photon,tension=0,label=$k_{1}$,label.side=right}{v2,v1}
\fmf{photon,tension=0,label=$k_{2}$,label.side=left}{v4,v3}
\end{fmfgraph*} }
%
%%%%%%%%%%%%%%%%%%%%%%%
%
\hspace{4mm}
\subfigure[]{
\begin{fmfgraph*}(35,35)
\fmfleft{i1,i2}
\fmfright{o}
\fmfforce{0.05w,0.7h}{v11}
\fmfforce{0.05w,0.3h}{v12}
\fmf{photon}{i1,v1}
\fmf{photon}{i2,v2}
\fmf{fermion}{v5,o}
\fmflabel{$p_{2}$}{i1}
\fmflabel{$p_{1}$}{i2}
\fmfv{l=$p_{1} \! - \! k_{1} \! + \! k_{2}$,l.a=15,l.d=.1w}{v3}
\fmfv{l=$p_{2} \! + \! k_{1} \! - \! k_{2}$,l.a=-15,l.d=.1w}{v4}
\fmfv{l=$k_{1}$,l.a=0}{v11}
\fmfv{l=$k_{2}$,l.a=0}{v12}
\fmf{photon,tension=.3,label=$p_{1} \! - \! k_{1}$,
                       label.side=left}{v2,v3}
\fmf{photon,tension=.3}{v3,v5}
\fmf{photon,tension=.3,label=$p_{2} \! - \! k_{2}$,
                       label.side=right}{v1,v4}
\fmf{photon,tension=.3}{v4,v5}
\fmf{photon,tension=0}{v2,v4}
\fmf{photon,tension=0}{v1,v3}
\end{fmfgraph*} }
%
%%%%%%%%%%%%%%%%
%
\hspace{4mm}
\subfigure[]{
\begin{fmfgraph*}(35,35)
\fmfleft{i1,i2}
\fmfright{o}
\fmfforce{0.05w,0.7h}{v11}
\fmfforce{0.05w,0.3h}{v12}
\fmf{photon}{i1,v1}
\fmf{photon}{i2,v2}
\fmf{fermion}{v5,o}
\fmflabel{$p_{2}$}{i1}
\fmflabel{$p_{1}$}{i2}
\fmfv{l=$p_{1} \! - \! k_{1} \! + \! k_{2}$,l.a=15,l.d=.1w}{v3}
\fmfv{l=$p_{2} \! + \! k_{1} \! - \! k_{2}$,l.a=-15,l.d=.1w}{v4}
\fmfv{l=$k_{1}$,l.a=0}{v11}
\fmfv{l=$k_{2}$,l.a=0}{v12}
\fmf{photon,tension=.3,label=$p_{1} \! - \! k_{1}$,
                       label.side=left}{v2,v3}
\fmf{photon,tension=.3}{v3,v5}
\fmf{photon,tension=.3,label=$p_{2} \! - \! k_{2}$,
                       label.side=right}{v1,v4}
\fmf{photon,tension=.3}{v4,v5}
\fmf{fermion,tension=0}{v2,v4}
\fmf{photon,tension=0}{v1,v3}
\end{fmfgraph*} } \\
%
%%%%%%%%%%%%%%%%
%
\subfigure[]{
\begin{fmfgraph*}(35,35)
\fmfleft{i1,i2}
\fmfright{o}
\fmfforce{0.2w,0.9h}{v2}
\fmfforce{0.2w,0.1h}{v1}
\fmfforce{0.2w,0.5h}{v3}
\fmfforce{0.8w,0.5h}{v5}
\fmf{photon}{i1,v1}
\fmf{photon}{i2,v2}
\fmf{fermion}{v5,o}
\fmflabel{$p_{2}$}{i1}
\fmflabel{$p_{1}$}{i2}
\fmfv{l=$k_{1} \! + \! k_{2}$,l.a=10,l.d=.1w}{v3}
\fmf{photon,tension=0,label=$p_{1} \! - \! k_{1}$,
                      label.side=left}{v2,v5}
\fmf{photon,tension=0}{v3,v4}
\fmf{photon,tension=.4,label=$p_{2} \! - \! k_{2}$,
                       label.side=right}{v1,v4}
\fmf{photon,tension=.4,label=$p_{2} \! + \! k_{1}$,
                       label.side=right}{v4,v5}
\fmf{photon,tension=0,label=$k_{2}$,label.side=left}{v1,v3}
\fmf{photon,tension=0,label=$k_{1}$,label.side=right}{v2,v3}
\end{fmfgraph*} }
%
%%%%%%%%%%%%%%%%%%%%%%%
%
\hspace{4mm}
\subfigure[]{
\begin{fmfgraph*}(35,35)
\fmfleft{i1,i2}
\fmfright{o}
\fmfforce{0.2w,0.9h}{v2}
\fmfforce{0.2w,0.1h}{v1}
\fmfforce{0.2w,0.5h}{v3}
\fmfforce{0.8w,0.5h}{v5}
\fmf{photon}{i1,v1}
\fmf{photon}{i2,v2}
\fmf{fermion}{v5,o}
\fmflabel{$p_{2}$}{i1}
\fmflabel{$p_{1}$}{i2}
\fmfv{l=$k_{1} \! + \! k_{2}$,l.a=10,l.d=.1w}{v3}
\fmf{photon,tension=0,label=$p_{1} \! - \! k_{1}$,
                      label.side=left}{v2,v5}
\fmf{photon,tension=0}{v3,v4}
\fmf{fermion,tension=.4,label=$p_{2} \! - \! k_{2}$,
                        label.side=right}{v1,v4}
\fmf{photon,tension=.4,label=$p_{2} \! + \! k_{1}$,
                       label.side=right}{v4,v5}
\fmf{photon,tension=0,label=$k_{2}$,label.side=left}{v1,v3}
\fmf{photon,tension=0,label=$k_{1}$,label.side=right}{v2,v3}
\end{fmfgraph*} } 
%
%%%%%%%%%%%%%%%%%%%%%%%
%
\hspace{4mm}
\subfigure[]{
\begin{fmfgraph*}(35,35)
\fmfleft{i1,i2}
\fmfright{o}
\fmfforce{0.2w,0.9h}{v2}
\fmfforce{0.2w,0.1h}{v1}
\fmfforce{0.2w,0.5h}{v3}
\fmfforce{0.8w,0.5h}{v5}
\fmf{photon}{i1,v1}
\fmf{photon}{i2,v2}
\fmf{fermion}{v5,o}
\fmflabel{$p_{2}$}{i1}
\fmflabel{$p_{1}$}{i2}
\fmfv{l=$k_{1} \! + \! k_{2}$,l.a=10,l.d=.1w}{v3}
\fmf{photon,tension=0,label=$p_{1}-k_{1}$,label.side=left}{v2,v5}
\fmf{photon,tension=0}{v3,v4}
\fmf{photon,tension=.4,label=$p_{2}-k_{2}$,label.side=right}{v1,v4}
\fmf{photon,tension=.4,label=$p_{2}+k_{1}$,label.side=right}{v4,v5}
\fmf{photon,tension=0,label=$k_{2}$,label.side=left}{v1,v3}
\fmf{fermion,tension=0,label=$k_{1}$,label.side=right}{v2,v3}
\end{fmfgraph*} }
%
%%%%%%%%%%%%%%%%%%%%
\vspace*{8mm}
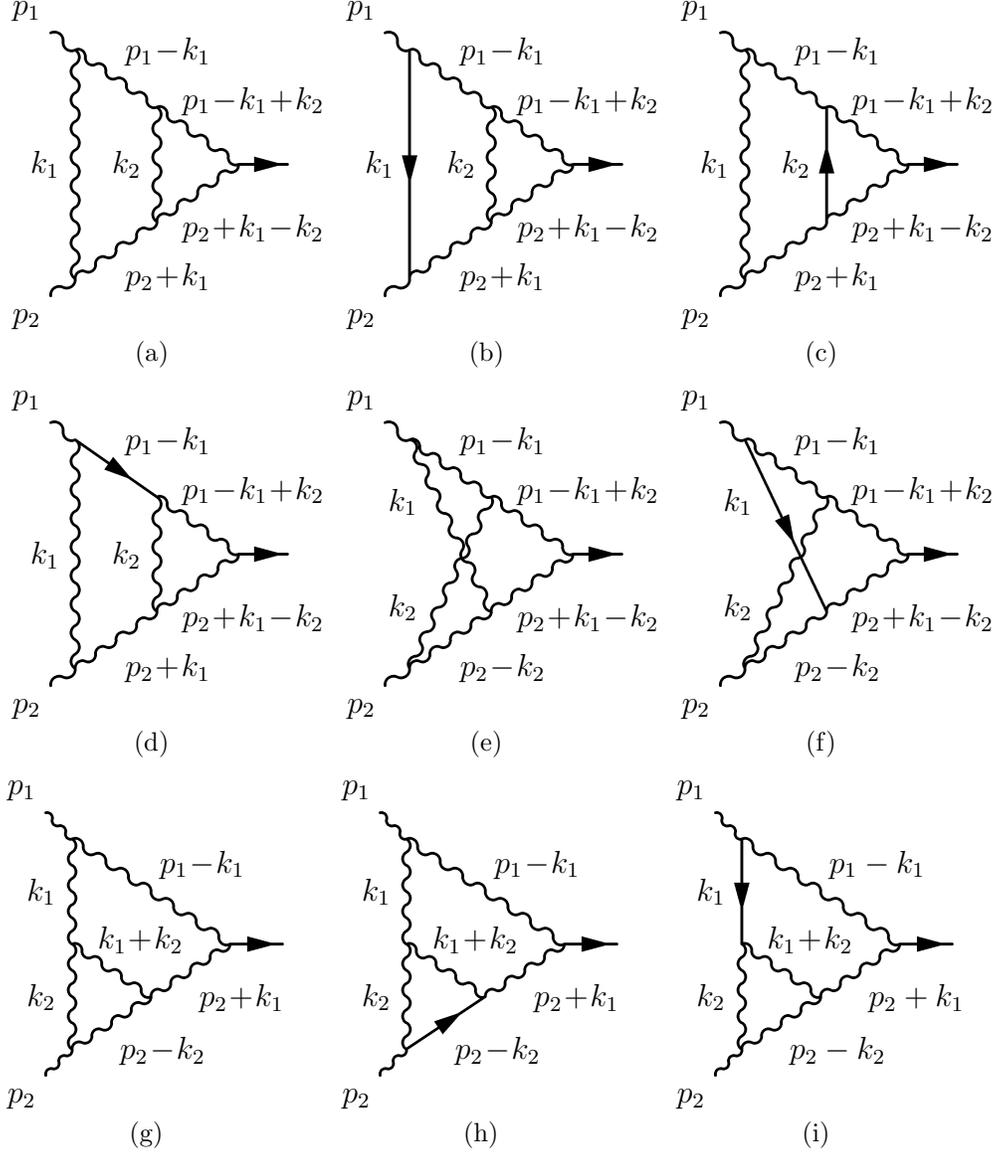
\caption{\label{fig1} two-loop vertex diagrams with up to one massive
propagator. They involve the ladder, crossed-ladder and 
vertex-insertion topologies and are {\it real} six-denominator 
topologies (see text).}
\ec
\efig
%%%%%%%%%%%%%%%%%%%%%%%%%%%%%%%%%%%%%%%%%%%%%%%%%%%%%%%%%%%%%%%%%%%%%%%%
%
%
%
%
%
%%%%%%%%%%%%%%%%%%%% 2-loop Vertex %%%%%%%%%%%%%%%%%%%%%%%%%%%%%%%%%%%%%
\bfig
\bc
\subfigure[]{
\begin{fmfgraph*}(35,35)
\fmfleft{i1,i2}
\fmfright{o}
\fmfforce{0.2w,0.93h}{v2}
\fmfforce{0.2w,0.07h}{v1}
\fmfforce{0.8w,0.5h}{v5}
\fmfforce{0.2w,0.4h}{v9}
\fmfforce{0.5w,0.45h}{v10}
\fmfforce{0.2w,0.5h}{v11}
\fmf{photon}{i1,v1}
\fmf{photon}{i2,v2}
\fmf{fermion}{v5,o}
\fmflabel{$p_{2}$}{i1}
\fmflabel{$p_{1}$}{i2}
\fmfv{l=$k_{2}$,l.a=180,l.d=0.05w}{v10}
\fmfv{l=$k_{1}$,l.a=180,l.d=.06w}{v11}
\fmf{photon,label=$p_{1} \! - \! k_{1}$,label.side=left}{v2,v3}
\fmf{photon,tension=.25,right}{v3,v4}
\fmf{photon,tension=.25,label=$p_{1} \! - \! k_{1} \! - \! k_{2}
$,label.side=left}{v3,v4}
\fmf{photon,label=$p_{1} \! - \! k_{1}$,label.side=left}{v4,v5}
\fmf{photon,label=$p_{2} \! + \! k_{1}$,label.side=right}{v1,v5}
\fmf{photon}{v1,v2}
\end{fmfgraph*} }
%
%%%%%%%%%%%%%%%%%%%%%%%
%
\hspace{6mm}
\subfigure[]{
\begin{fmfgraph*}(35,35)
\fmfleft{i1,i2}
\fmfright{o}
\fmfforce{0.2w,0.93h}{v2}
\fmfforce{0.2w,0.07h}{v1}
\fmfforce{0.8w,0.5h}{v5}
\fmfforce{0.2w,0.4h}{v9}
\fmfforce{0.5w,0.45h}{v10}
\fmfforce{0.2w,0.5h}{v11}
\fmf{photon}{i1,v1}
\fmf{photon}{i2,v2}
\fmf{fermion}{v5,o}
\fmflabel{$p_{2}$}{i1}
\fmflabel{$p_{1}$}{i2}
\fmfv{l=$k_{2}$,l.a=180,l.d=0.05w}{v10}
\fmfv{l=$k_{1}$,l.a=180,l.d=.06w}{v11}
\fmf{photon,label=$p_{1} \! - \! k_{1}$,label.side=left}{v2,v3}
\fmf{fermion,tension=.25,right}{v3,v4}
\fmf{photon,tension=.25,label=$p_{1} \! - \! k_{1} \! - \! k_{2}
$,label.side=left}{v3,v4}
\fmf{photon,label=$p_{1} \! - \! k_{1}$,label.side=left}{v4,v5}
\fmf{photon,label=$p_{2} \! + \! k_{1}$,label.side=right}{v1,v5}
\fmf{photon}{v1,v2}
\end{fmfgraph*} }
%
%%%%%%%%%%%%%%%%%%%%%%%
%
\hspace{6mm}
\subfigure[]{
\begin{fmfgraph*}(35,35)
\fmfleft{i1,i2}
\fmfright{o}
\fmfforce{0.2w,0.93h}{v2}
\fmfforce{0.2w,0.07h}{v1}
\fmfforce{0.8w,0.5h}{v5}
\fmfforce{0.2w,0.4h}{v9}
\fmfforce{0.5w,0.45h}{v10}
\fmfforce{0.2w,0.5h}{v11}
\fmf{photon}{i1,v1}
\fmf{photon}{i2,v2}
\fmf{fermion}{v5,o}
\fmflabel{$p_{2}$}{i1}
\fmflabel{$p_{1}$}{i2}
\fmfv{l=$k_{2}$,l.a=180,l.d=0.05w}{v10}
\fmfv{l=$k_{1}$,l.a=180,l.d=.06w}{v11}
\fmf{photon,label=$p_{1} \! - \! k_{1}$,label.side=left}{v2,v3}
\fmf{photon,tension=.25,right}{v3,v4}
\fmf{photon,tension=.25,label=$p_{1} \! - \! k_{1} \! - \! k_{2}
$,label.side=left,l.d=0.05w}{v3,v4}
\fmf{photon,label=$p_{1} \! - \! k_{1}$,label.side=left}{v4,v5}
\fmf{photon,label=$p_{2} \! + \! k_{1}$,label.side=right}{v1,v5}
\fmf{fermion}{v2,v1}
\end{fmfgraph*} } \\
%
%%%%%%%%%%%%%%%%%%%%%%%
%
\subfigure[]{
\begin{fmfgraph*}(35,35)
\fmfleft{i1,i2}
\fmfright{o}
\fmfforce{0.2w,0.93h}{v2}
\fmfforce{0.2w,0.07h}{v1}
\fmfforce{0.2w,0.3h}{v3}
\fmfforce{0.2w,0.7h}{v4}
\fmfforce{0.8w,0.5h}{v5}
\fmf{photon}{i1,v1}
\fmf{photon}{i2,v2}
\fmf{fermion}{v5,o}
\fmflabel{$p_{2}$}{i1}
\fmflabel{$p_{1}$}{i2}
\fmf{photon,label=$p_{1} \! - \! k_{1}$,label.side=left}{v2,v5}
\fmf{photon,label=$k_{1}$,label.side=left}{v1,v3}
\fmf{photon,label=$k_{1}$,label.side=right}{v2,v4}
\fmf{photon,label=$p_{2} \! + \! k_{1}$,label.side=right}{v1,v5}
\fmf{photon,right,label=$k_{1} \! \! + \! \! k_{2}$,label.side=left,
     l.d=0.03w}{v4,v3}
\fmf{photon,right,label=$k_{2}$,label.side=right}{v3,v4}
\end{fmfgraph*} }
%
%%%%%%%%%%%%%%%%%%%%%%%
%
\hspace{6mm}
\subfigure[]{
\begin{fmfgraph*}(35,35)
\fmfleft{i1,i2}
\fmfright{o}
\fmfforce{0.2w,0.93h}{v2}
\fmfforce{0.2w,0.07h}{v1}
\fmfforce{0.2w,0.3h}{v3}
\fmfforce{0.2w,0.7h}{v4}
\fmfforce{0.8w,0.5h}{v5}
\fmf{photon}{i1,v1}
\fmf{photon}{i2,v2}
\fmf{fermion}{v5,o}
\fmflabel{$p_{2}$}{i1}
\fmflabel{$p_{1}$}{i2}
\fmf{photon,label=$p_{1} \! - \! k_{1}$,label.side=left}{v2,v5}
\fmf{photon,label=$k_{1}$,label.side=left}{v1,v3}
\fmf{fermion,label=$k_{1}$,label.side=right}{v2,v4}
\fmf{photon,label=$p_{2} \! + \! k_{1}$,label.side=right}{v1,v5}
\fmf{photon,right,label=$k_{1} \! \! + \! \! k_{2}$,label.side=left,
     l.d=0.03w}{v4,v3}
\fmf{photon,right,label=$k_{2}$,label.side=right}{v3,v4}
\end{fmfgraph*} }
%
%%%%%%%%%%%%%%%%
%
\hspace{6mm}
\subfigure[]{
\begin{fmfgraph*}(35,35)
\fmfleft{i1,i2}
\fmfright{o}
\fmfforce{0.2w,0.9h}{v2}
\fmfforce{0.2w,0.1h}{v1}
\fmfforce{0.2w,0.5h}{v3}
\fmfforce{0.8w,0.5h}{v5}
\fmf{photon}{i1,v1}
\fmf{photon}{i2,v2}
\fmf{fermion}{v5,o}
\fmflabel{$p_{2}$}{i1}
\fmflabel{$p_{1}$}{i2}
\fmf{photon,tension=0,label=$p_{1}-k_{1}$,label.side=left}{v2,v5}
%\fmf{photon,tension=0,label=$k_{1}+k_{2}$,label.side=left}{v3,v4}
\fmf{fermion,left=90,label=$k_{2}$,label.side=right}{v3,v3}
\fmf{photon,tension=0,label=$p_{2}+k_{1}$,label.side=right}{v1,v5}
\fmf{photon,tension=0,label=$k_{1}$,label.side=left}{v1,v3}
\fmf{photon,tension=0,label=$k_{1}$,label.side=right}{v2,v3}
\end{fmfgraph*} }
%
%%%%%%%%%%%%%%%%%%%%
\vspace*{8mm}
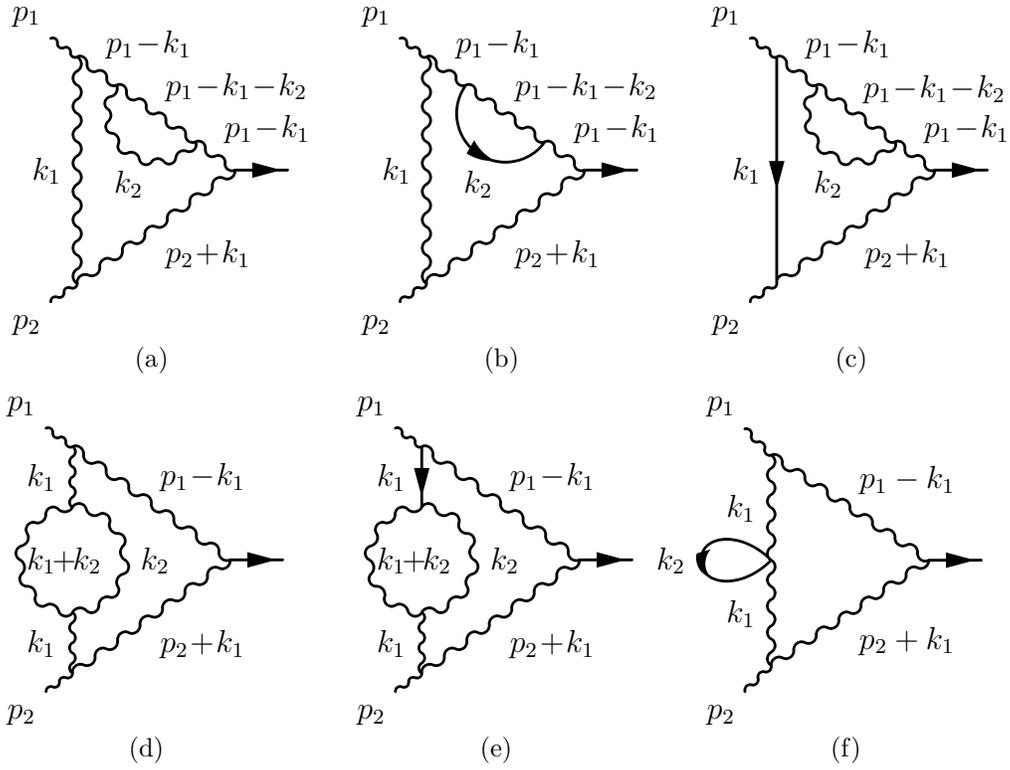
\caption{\label{fig1bis} two-loop vertex diagrams with up to one massive
propagator involving a self-energy correction to the basic one-loop 
vertices. The related topologies are five and four denominator 
topologies (see text).}
\ec
\efig
%%%%%%%%%%%%%%%%%%%%%%%%%%%%%%%%%%%%%%%%%%%%%%%%%%%%%%%%%%%%%%%%%%%%%%%%

\section{The reduction to master integrals \label{Master}}

All the two-loop vertex diagrams containing six denominators with at 
most one massive propagator are reported in Figs. \ref{fig1} and 
\ref{fig1bis}\footnote{Actually, diagram $(f)$ in Fig. \ref{fig1bis} 
contains only five denominators, but we list it in this figure as it 
belongs to the same class of the six-denominator ones.}.
The graphical conventions we use are the following. A massless 
propagator is represented by a wavy internal line, while a massive 
propagator is denoted by a straight segment. An external line carrying 
the light-cone momentum $p_1$ or $p_2$ is represented by an external 
wavy line, while the external particle $X$ carrying the general 
momentum $q$ is depicted as a straight external line. This set of 
conventions is clearly a generalization of the well-known QED ones.
In Fig. \ref{fig1} we consider the basic three topologies with the 
various mass configurations: 
\begin{itemize}
\item
(a), (b), (c) and (d) are ladder diagrams, i.e.
corrections to the hard vertex of the basic one-loop amplitude;
\item
(e) and (f) are crossed ladders --- the non-planar topology;
\item
(g), (h) and (i) finally are the vertex-insertion diagrams, 
containing a correction to the vertices with momentum $p_1$ or $p_2$.
\end{itemize}
The number of internal lines with different momenta entering a diagram 
is an indicator of the {\it topology} of the diagram. 
The diagrams in Fig. \ref{fig1} are {\it real} six-denominator 
amplitudes and the related topologies are given in Fig. \ref{fig1ter}.

In Fig. \ref{fig1bis} the diagrams containing self-energy corrections 
to the internal lines of the one-loop vertices are plotted. Note that, 
in contrast to Fig. \ref{fig1}, diagrams (a),  (b), (c) and (d)
in Fig. \ref{fig1bis} involve only five distinct denominators --- one 
of them being squared --- as two internal lines carry the same momentum.
These diagrams then, do not represent {\it real} six-denominator 
topologies and are more naturally grouped among the five-denominator 
ones. The topologies of diagrams (a), (b), (c) and (d) 
are shown in Fig.~\ref{fig2} in (h), (e), (f) and (s) 
respectively. Diagram (e) of Fig. \ref{fig1bis} involves two internal 
lines with the same momenta but with different masses.
It can be expressed as a superposition of integrals belonging to 
topologies (s) and (t) of Fig. \ref{fig2}, by means of partial 
fractioning in the loop momentum $k_{1}$:
\be
\frac{1}{k_{1}^{2} \, (k_{1}^{2}+a)} = \frac{1}{a} \left[
\frac{1}{k_{1}^{2}} - \frac{1}{k_{1}^{2}+a} \right] \, .
\ee
The scalar product of two
vectors is defined as: $a \cdot b = -a_0 b_0 + \vec{a}\cdot\vec{b}$ 
\footnote{Note a difference of sign 
with respect to the more common definition.}.
In graphical form, the decomposition reads:
\be
\parbox{15mm}{
\begin{fmfgraph*}(15,15)
\fmfleft{i1,i2}
\fmfright{o}
\fmfforce{0.2w,0.93h}{v2}
\fmfforce{0.2w,0.07h}{v1}
\fmfforce{0.2w,0.3h}{v3}
\fmfforce{0.2w,0.7h}{v4}
\fmfforce{0.8w,0.5h}{v5}
\fmf{photon}{i1,v1}
\fmf{photon}{i2,v2}
\fmf{plain}{v5,o}
\fmf{photon}{v2,v5}
\fmf{photon}{v1,v3}
\fmf{plain}{v2,v4}
\fmf{photon}{v1,v5}
\fmf{photon,right}{v4,v3}
\fmf{photon,right}{v3,v4}
\end{fmfgraph*} } 
\; = \;\frac{1}{a} \left[ \; 
\parbox{15mm}{
\begin{fmfgraph*}(15,15)
\fmfleft{i1,i2}
\fmfright{o}
\fmfforce{0.2w,0.9h}{v2}
\fmfforce{0.2w,0.1h}{v1}
\fmfforce{0.2w,0.55h}{v3}
\fmfforce{0.2w,0.15h}{v5}
\fmfforce{0.8w,0.5h}{v4}
\fmf{photon}{i1,v1}
\fmf{photon}{i2,v2}
\fmf{plain}{v4,o}
\fmf{photon}{v2,v3}
\fmf{photon,left}{v3,v5}
\fmf{photon,right}{v3,v5}
\fmf{photon}{v1,v4}
\fmf{photon}{v2,v4}
\end{fmfgraph*} }  \; -  \; 
\parbox{15mm}{
\begin{fmfgraph*}(15,15)
\fmfleft{i1,i2}
\fmfright{o}
\fmfforce{0.2w,0.9h}{v2}
\fmfforce{0.2w,0.1h}{v1}
\fmfforce{0.2w,0.55h}{v3}
\fmfforce{0.2w,0.15h}{v5}
\fmfforce{0.8w,0.5h}{v4}
\fmf{photon}{i1,v1}
\fmf{photon}{i2,v2}
\fmf{plain}{v4,o}
\fmf{plain}{v2,v3}
\fmf{photon,left}{v3,v5}
\fmf{photon,right}{v3,v5}
\fmf{photon}{v1,v4}
\fmf{photon}{v2,v4}
\end{fmfgraph*} }  \; \right] \, .
\ee
Finally, diagram (f) is actually a four-denominator topology, 
indicated in Fig. \ref{fig3} (g).

To conclude, let us note that the topologies related to diagrams of 
Fig. \ref{fig1bis} can be obtained by properly shrink one internal line 
(or two internal lines in the case of diagram (f)) among those in the 
diagrams of Fig. \ref{fig1ter}.

\subsection{Decomposition of a tensor amplitude into
            scalar amplitudes \label{toscalar}}

In this section we consider the decomposition of a general tensor 
amplitude among those shown in Figs. \ref{fig1} and \ref{fig1bis} 
into scalar amplitudes.
Even though the decomposition is 
completely general, let us pick up, for clarity's sake, a specific case:
the ladder diagram with a mass on the external
boson line (the diagram (b) in Fig. \ref{fig1}).  
Omitting an overall constant dependent on the couplings of the theory, 
this diagram involves a double integral 
over the loop momenta $k_{1}$ and $k_{2}$ and can be written, in
all generality as\footnote{
Depending on the gauge choice for vector particles, some of the
denominators $P_{i}$ may actually appear to power two instead of one: 
the reduction scheme is clearly unchanged. We do not consider
the non covariant gauges, 
leading to additional denominators and scalar products
involving a constant gauge vector $c^{\mu}$.}:
\be
{\mathcal I}_{\mu\nu i j ...}(p_1,p_2) = \int \{d^{D}k_{1}\}\{d^{D}k_{2}\}
\frac{{\mathcal N}_{\mu\nu i j ...}(p_{1},p_{2},k_{1},k_{2};a)}{%
P_{1}P_{2}P_{3}P_{4}P_{5}P_{6}},  
\label{generalamp1}
\ee
where $\{ d^D k \}$ is the loop integration measure (see
section \ref{Results}). According to our routing of the loop 
momenta, we have defined the denominators: 
\bea
P_{1} &=& k_{1}^{2}+a,               \\ 
P_{2} &=& (p_{1}-k_{1})^{2},         \\ 
P_{3} &=& (p_{2}+k_{1})^{2},         \\ 
P_{4} &=& k_{2}{}^{2},               \\ 
P_{5} &=& (p_{1}-k_{1}+k_{2})^{2},   \\ 
P_{6} &=& (p_{2}+k_{1}-k_{2})^{2}.   
\eea
The numerator ${\mathcal N}_{\mu\nu i j ...}$ depends on the spin of 
the particles entering the diagram and on the type of interaction; 
the indices $\mu\nu i j ...$ denote collectively Lorentz indices,
Dirac indices, etc. 
By using projectors on invariant form factors, we can always restrict
ourselves to consider scalar numerators (see for example 
\cite{remiddivecchio}). The latter typically involve a trace over 
spinor indices and a sum over the polarizations of the vector particles.
Therefore, from now on, let us consider:
\begin{equation}
{\mathcal I}(p_1 \cdot p_2) = \int \{d^{D}k_{1}\}\{d^{D}k_{2}\}
\frac{{\mathcal N}(S_{1},S_{2},S_{3},S_{4},S_{5},S_{6},S_{7})}{
P_{1}P_{2}P_{3}P_{4}P_{5}P_{6}},  
\label{generalamp}
\end{equation}
where we have explicitly indicated the dependence on the seven
invariants, dependent on the loop momenta $k_{1}$ and 
$k_{2}$: 
\begin{eqnarray}              
S_{1} &=& k_{1}^{2},      
\label{invariant1}  \\ 
S_{2} &=& k_{2}^{2},      
\label{invariant2}  \\ 
S_{3} &=& k_{1}\cdot p_{1}, 
\label{invariant3}  \\ 
S_{4} &=& k_{1}\cdot p_{2}, 
\label{invariant4}  \\ 
S_{5} &=& k_{2}\cdot p_{1}, 
\label{invariant5}  \\ 
S_{6} &=& k_{2}\cdot p_{2}, 
\label{invariant6}  \\ 
S_{7} &=& k_{1}\cdot k_{2}.
\label{invariant7}
\end{eqnarray}
The scalar numerator can be expanded in a power series of the 
invariants as: 
\begin{equation}
{\mathcal N}(S_{1},S_{2},S_{3},S_{4},S_{5},S_{6},S_{7}) \! = 
\! \! \! \! 
\sum_{l_{1},\cdots l_{7}\geq 0} \! \! b(l_{1},l_{2},\cdots , l_{7}) \,
S_{1} ^ {l_{1}}
S_{2} ^ {l_{2}}
S_{3} ^ {l_{3}}
S_{4} ^ {l_{4}}
S_{5} ^ {l_{5}}
S_{6} ^ {l_{6}}
S_{7} ^ {l_{7}},
\label{Nexpand}
\end{equation}
where the (known) coefficients $b(l_{1},l_{2},\cdots , l_{7})$ also 
depend on the external invariant $s$, the mass squared $a=m^2$ and 
the space-time dimension $D$. With the typical gauge interactions, we 
expect the indices $l_{i}$ to be restricted to $l_{i}\leq 3$.

\subsection{Decomposition into independent scalar amplitudes}

The scalar amplitudes in Eq. (\ref{generalamp}), with 
${\mathcal N}$ given by the expansion in Eq. (\ref{Nexpand}) are not, in
general, linearly independent on each other; we can indeed
express six of the invariants listed in 
Eqs.~(\ref{invariant1}--\ref{invariant7}) in terms of the denominators
$P_{i}$. The next step is the reduction to a set of linearly 
independent scalar amplitudes in order to
progress in the evaluation of the tensor amplitude of Eq.
(\ref{generalamp1}). 

In this work, we used two different reduction {\it schemes} of
the scalar amplitudes, using consistency of the results
as a crossed check. The first one is the scheme of the
{\it auxiliary denominator} (or {\it auxiliary diagram}), which is 
essentially 
based on the reduction of all the amplitudes to
scalar ones containing formally only denominators; the
second one is the scheme of the {\it shifts}, in which the
independent scalar amplitudes are integrals with
a subset of the scalar products at the numerator.

In the following two sections we will recall briefly both schemes,
paying more attention on the first one, not discussed often
in the literature, and just sketching the second one, for which we
refer to the bibliography. 
Our results are presented in section \ref{Results} in the framework of
this second scheme.

\subsubsection{Auxiliary-denominator scheme \label{aux}}

As already said in the previous section, we aim at expressing 
the invariants
listed in Eqs. (\ref{invariant1}--\ref{invariant7}) in terms of the 
denominators $P_{i}$ appearing in the diagram. Since there are seven
invariants and only six denominators, we introduce a seventh
auxiliary denominator, independent from the others, in such a way to 
form a basis. The choice is of course, not unique; let us take for 
example: 
\be
P_{7}=(k_{1}-k_{2})^{2} \, .
\ee
We then have the linear relations:
\begin{equation}
S_{i} ~=~ \sum_{j=1}^{7} A_{ij} (P_{j}-a_{j})~~~~~~~~(i=1,\ldots, 7) 
\, ,
\label{reexpress}
\end{equation}
where $a_{j}=\delta_{j,1}a$ and $A_{ij}$ an invertible $ 7\times 7 $ 
constant matrix.
The non-zero matrix elements are given by:
\bea
S_{1} &=& P_{1}-a, \\
S_{2} &=& P_{4}, \\
S_{3} &=& \frac{1}{2}\left( - P_{2} + P_{1} - a \right) , \\
S_{4} &=& \frac{1}{2}\left( + P_{3} - P_{1} + a \right) , \\
S_{5} &=& \frac{1}{2}\left( - P_{7} + P_{5} - P_{2} + P_{1} - a \right) 
, \\
S_{6} &=& \frac{1}{2}\left( + P_{7} - P_{6} + P_{3} - P_{1} + a \right) 
, \\
S_{7} &=& \frac{1}{2}\left( - P_{7} + P_{4} + P_{1} - a \right) \, .
\eea
Let us note that one can construct a fictitious Feynman diagram having 
all the seven denominators, called the auxiliary diagram \cite{glover}:
\begin{equation}
{\rm (aux.~diag.)} ~ = ~
\int \frac{ \{d^{D}k_{1}\}\{d^{D}k_{2}\} }
{P_{1} P_{2} P_{3} P_{4} P_{5} P_{6} P_{7} }.
\end{equation} 
%
%
%
%%%%%%%%%%%%%%%%%%%% 2-loop Box: Aux. Diagr. %%%%%%%%%%%%%%%%%%%%%%%%%%
\bfig
\bc
\begin{fmfgraph*}(80,35)
\fmfleft{i1,i2}
\fmfright{o1,o2}
\fmf{photon}{i1,v1}
\fmf{photon,tension=.5,label=$p_{1} \! - \! k_{1} \! + \! 
                k_{2}$,label.side=left}{v4,v6}
\fmf{photon,tension=.5,label=$p_{1} \! - \! 
                k_{1}$,label.side=left}{v2,v4}
\fmf{photon,tension=.5,label=$p_{2} \! + \! 
                k_{1}$,label.side=right}{v1,v3}
\fmf{photon,tension=.5,label=$p_{2} \! + \! k_{1} \! - \! 
                k_{2}$,label.side=right}{v3,v5}
\fmf{photon}{v5,o1}
\fmf{photon}{v6,o2}
\fmf{photon}{i2,v2}
\fmflabel{$p_{1}$}{i2}
\fmflabel{$p_{2}$}{i1}
\fmflabel{$p_{1}$}{o2}
\fmflabel{$p_{2}$}{o1}
\fmf{fermion,tension=0,label=$k_{1}$,label.side=right}{v2,v1}
\fmf{photon,tension=0,label=$k_{2}$,label.side=left}{v3,v4}
\fmf{scalar,tension=0,label=$k_{1} \! - \! k_{2}$,
                 label.side=right}{v5,v6}
\end{fmfgraph*}
%%%%%%%%%%%%%%%%%%%%%%%%%%%%%%%%%%%%%
\vspace*{8mm}
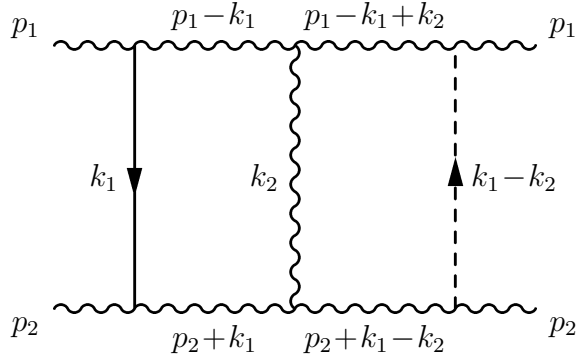
\caption{\label{fig1bisbis} Double box auxiliary diagram for the case of
the vertex diagram in Fig.~\ref{fig1} (b). The dashed line represents the
auxiliary denominator $P_{7}$.}
\ec
\efig
%%%%%%%%%%%%%%%%%%%%%%%%%%%%%%%%%%%%%%%%%%%%%%%%  %%%%%%%%%%%%%%%%%%%%
%
%
%
In our example, the auxiliary diagram is a planar double box with final 
momenta equal to the initial ones $p_{1}$ and $p_{2}$, i.e. a 
four-point forward amplitude: see Fig. \ref{fig1bisbis}.
In general, an auxiliary diagram has a number of independent 
denominators equal to the number of the invariants (seven in our case). 
The scalar numerator ${\mathcal N}$ can now be expanded in powers of 
the basis ``vectors'' $\{P_{i}\}_{i=1,7}$ as: 
\bd
{\mathcal N}=\sum_{s_{1},\cdots s_{7}\geq 0}c(s_{1},s_{2},\cdots ,
s_{7})\,P_{1}^{s_{1}}\,P_{2}^{s_{2}}\cdots P_{7}^{s_{7}} \, , 
\ed
with the new coefficients $c(s_{1},s_{2}, \cdots , s_{7})$ being linear
combinations of the old ones $b(l_{1},l_{2}\cdots , l_{7})$. Inserting 
the above expansion in the expression (\ref{generalamp}) for the 
diagram, we obtain: 
\begin{equation}
{\mathcal I}(p_1\cdot p_2) = \! \! \! \! \! \! 
\sum_{s_{1},\cdots s_{7}\geq 0}
\! \! \! c(s_{1}, \! \cdots \! , \! s_{7}) \int \! 
\frac{ \{d^{D}k_{1}\}\{d^{D}k_{2}\}  }{P_{1}^{1-s_{1}}\,P_{2}^{1-s_{2}}%
\,P_{3}^{1-s_{3}}\,P_{4}^{1-s_{4}}\,P_{5}^{1-s_{5}}\,P_{6}^{1-s_{6}}%
\,P_{7}^{-s_{7}}}.  
\label{mainres}
\end{equation}
Eq.~(\ref{mainres}) is the main result of this section. It says that the
general scalar diagram ${\mathcal I}$ is a linear combination --- with 
known coefficients --- of independent scalar amplitudes involving 
the original denominators $\{P_{i}\}_{i=1,6}$ and an additional, 
fictitious, denominator $P_{7}$.  A compact notation for the auxiliary 
scalar amplitudes is the following: 
\begin{equation}
\mathrm{Topo}\left( n_{1},n_{2},n_{3},n_{4},n_{5},n_{6};n_{7}\right) =
\int
\frac{ \{d^{D}k_{1}\}\{d^{D}k_{2}\} }{P_{1}^{n_{1}}\,P_{2}^{n_{2}}%
\,P_{3}^{n_{3}}\,P_{4}^{n_{4}}\,P_{5}^{n_{5}}\,P_{6}^{n_{6}}\,
P_{7}^{n_{7}}},
\label{Topodef}
\end{equation}
according to which Eq.~(\ref{mainres}) is written as: 
\begin{equation}
\! \! \! \! {\mathcal I}(p_1,p_2) = \! \! \! \! \! \sum_{n_{1},\cdots 
n_{6}\leq 1,\,n_{7}\leq 0} \! \! \! d(n_{1},n_{2},\cdots n_{7}) 
\mathrm{Topo}\left( n_{1},n_{2},n_{3},n_{4},n_{5},n_{6};n_{7}\right) .
\label{scalardecomp}
\end{equation}
A few remarks are in order. First, the auxiliary denominator $P_{7},$ if
it appears, it only does at the numerator: it cannot clearly be 
generated at the denominator by tensor decomposition. 
In other words, $n_{i}=1,0,-1,-2,-3,...$
for $i=1,...,6$ while $n_{7}=0,-1,-2,-3,..$.
The second remark concerns the so-called topologies. As already 
anticipated, an important quantity for a scalar amplitude is the number 
of distinct denominators with positive indices, $n_{i}>0$, which we 
call $t$: 
\be
t \equiv \sum_{i=1}^{7} \theta \left( n_{i} \right) ,
\ee
with $\theta \left( x\right) $ the unit step function defined in such a 
way that $\theta \left( 0\right) =0.$ The scalar amplitude with 
$n_{1}=2,$ $n_{2}=\cdots =n_{6}=1$ and $n_{7}=-1$ for example has 
$t=6,$ the maximal value for two-loop vertex functions.
If a scalar amplitude $\mathrm{Topo}\left(
n_{1},n_{2},n_{3},n_{4},n_{5},n_{6};n_{7}\right) $ has $n_{i}\leq 0$, 
the $i^{th}$ denominator eventually appears only at the numerator
and the $i^{th}$ internal line is shrunk to a point: the diagram then 
has less than six internal lines. Here we give an example of a 
decomposition into auxiliary amplitudes with $t=6$ and $t=5$: 
\begin{eqnarray}
\int \{d^{D}k_{1}\}\{d^{D}k_{2}\}
\frac{S_{7}}{P_{1}P_{2}P_{3}P_{4}P_{5}P_{6}} 
& = & -\frac{1}{2}\mathrm{Topo}(1,1,1,1,1,1;-1) \nn\\
& & - \frac{a}{2}\mathrm{Topo}(1,1,1,1,1,1;0) \nn\\
& & + \frac{1}{2}\mathrm{Topo}(0,1,1,1,1,1;0) \nn\\
& & + \frac{1}{2}\mathrm{Topo}(1,1,1,0,1,1;0) .
\end{eqnarray}

In general, a diagram ${\mathcal I}$ is decomposed into auxiliary 
amplitudes with decreasing $t$ starting from $t=6$ included, i.e. 
$t=6,5,4,3$\footnote{Amplitudes with $t<3$ vanish, in our case, as they 
contain a massless tadpole. }.

\subsubsection{Shift scheme \label{shifts}}

We described above the scheme of the auxiliary diagram to reduce a 
general scalar amplitude. Let us now present another scheme of reduction
which has also been used in the past (see \cite{Lap,Rem3,Bon1}).
The main results of this work are presented in sections \ref{Results} 
and \ref{6Denominatori} according to this second scheme, so let us 
briefly overview it.

In essence, it uses shifts on the loop momenta in order to
maximally simplify the structure of the sub-topologies. Let us consider a
routing with two internal lines having the loop momenta $k_1$ and $k_2$,
as for instance $P_{1}$ and $P_{4}$  in the discussion above.   
We choose for example to keep the invariant $S_7$ and we simplify, i.e. 
cancel, the factors $S_i(i=1,\ldots, 6)$ appearing at the numerator 
with the denominators by means of the formulas:
\begin{eqnarray}
S_1 &=& P_{1} - a  \\
S_2 &=& P_{4}    \\
S_3 &=& \frac{1}{2} ( - P_{2} + P_{1} + a ) \\
S_4 &=& \frac{1}{2} ( + P_{3} - P_{1} + a )  \\
S_5 &=& \frac{1}{2} ( + P_{5} - P_{4} - P_{2} ) + S_7  \\
S_6 &=& \frac{1}{2} ( - P_{6} + P_{4} + P_{3}) - S_7  \, . 
\end{eqnarray}
The simplification of a scalar product with the selected denominator
brings to the appearance of sub-topologies in the case the denominator
appears to power one.
In principle, a diagram of topology number $t$ can have $t!$
different sub-topologies. 
In our case, however, some sub-topologies actually vanish
because all the propagators, except possibly one, are massless.
Furthermore, eventual discrete symmetries of the diagram, 
such as an up-down symmetry, further reduce the number of independent 
sub-topologies.

After the cancellations between the $S_i$ and $P_j$ have been done, 
we end up with a set of subdiagrams in which the 
denominators $P_1$ and $P_4$ 
containing the loop momenta $k_1$ and $k_2$ may be absent. 
In this case,
we perform shifts on the loop momenta to reproduce denominators 
containing again $k_1$ and $k_2$. For example, if $P_1$ is absent but 
$P_2$ is still present in the given amplitude, we may do the shift 
\begin{equation}
k_{1} \rightarrow p_{1}-k_{1}.
\label{firstshift}
\end{equation}
If $P_1$ and $P_2$ are absent but $P_3$ is still 
present, we may do the shift $k_{1} \rightarrow k_{1}-p_{2}$, and so on.
In general, we construct a table of shifts on the loop momenta.
With this step, new denominators are generated. Under (\ref{firstshift})
for example,
\begin{eqnarray}
P_3 &\rightarrow& (p_1+p_2-k_1)^2,
\nonumber \\ 
P_5 &\rightarrow& (k_1+k_2)^2, 
\nonumber \\
P_6 &\rightarrow& (p_1+p_2-k_1-k_2)^2, 
\end{eqnarray}
denominators which were not initially present. We then extend the set 
of cancellation rules including those between the new denominators and 
the scalar products. The scalar product $k_1\cdot k_2$ 
cancels against the denominator $(k_1+k_2)^2$ --- no arbitrareness ---
while $(p_1+p_2-k_1)^2$
may be cancelled for example with $p_2\cdot k_1$ and 
$(p_1+p_2-k_1-k_2)^2$ with $p_2\cdot k_2$. The remaining scalar products
are then $p_1\cdot k_1$ and $p_1\cdot k_2$. Depending on the topology 
number $t$ of the diagram under consideration, we can re-express $t$ of 
the $7$ scalar products in terms of denominators and simplify such 
expressions with a power of the corresponding denominator. It follows 
that a diagram with topology number $t$ can have at most 
$\overline{t}=7-t$ 
independent scalar products in the numerator. A six-denominator 
amplitude ($t=6$), for example, has at most one scalar product, while 
a sunrise diagram ($t=3$) has at most four distinct scalar products in 
the numerator.
After all the second-step cancellations have been done,
we go for a second step of shifts of the loop momenta, and so on.
We repeat iteratively the cancellation-shift procedure until no more 
simplifications are feasible. The whole set of required denominators is 
listed in appendix \ref{app1}.
In our computation we found 9 independent scalar amplitudes with six
denominators, i.e. $t=6$, (Fig. \ref{fig1ter}), 22 independent 
amplitudes with five denominators, (Fig. \ref{fig2}), 17 independent 
amplitudes with four denominators (Fig. \ref{fig3}) and 4 
independent amplitudes with $3$ denominators (Fig. \ref{fig4}). 
This makes a total of 44 independent scalar amplitudes.

%%%%%%%%%%%%%%%%%%%% Independent diagrams 6-den %%%%%%%%%%%%%%%%%%%%%%%%
\bfig
\bc
\subfigure[]{
\begin{fmfgraph*}(20,20)
\fmfleft{i1,i2}
\fmfright{o}
\fmf{photon}{i1,v1}
\fmf{photon}{i2,v2}
\fmf{plain}{v5,o}
\fmf{photon,tension=.3}{v2,v3}
\fmf{photon,tension=.3}{v3,v5}
\fmf{photon,tension=.3}{v1,v4}
\fmf{photon,tension=.3}{v4,v5}
\fmf{photon,tension=0}{v2,v1}
\fmf{photon,tension=0}{v4,v3}
\end{fmfgraph*} }
%
%%%%%%%%%%%%%%%%%%%%%%%
%
\subfigure[]{
\begin{fmfgraph*}(20,20)
\fmfleft{i1,i2}
\fmfright{o}
\fmf{photon}{i1,v1}
\fmf{photon}{i2,v2}
\fmf{plain}{v5,o}
\fmf{photon,tension=.3}{v2,v3}
\fmf{photon,tension=.3}{v3,v5}
\fmf{photon,tension=.3}{v1,v4}
\fmf{photon,tension=.3}{v4,v5}
\fmf{plain,tension=0}{v2,v1}
\fmf{photon,tension=0}{v4,v3}
\end{fmfgraph*} }
%
%%%%%%%%%%%%%%%%%%%%%%%
%
\subfigure[]{
\begin{fmfgraph*}(20,20)
\fmfleft{i1,i2}
\fmfright{o}
\fmf{photon}{i1,v1}
\fmf{photon}{i2,v2}
\fmf{plain}{v5,o}
\fmf{photon,tension=.3}{v2,v3}
\fmf{photon,tension=.3}{v3,v5}
\fmf{photon,tension=.3}{v1,v4}
\fmf{photon,tension=.3}{v4,v5}
\fmf{photon,tension=0}{v2,v1}
\fmf{plain,tension=0}{v4,v3}
\end{fmfgraph*} }
%
%%%%%%%%%%%%%%%%%%%%%%%
%
\subfigure[]{
\begin{fmfgraph*}(20,20)
\fmfleft{i1,i2}
\fmfright{o}
\fmf{photon}{i1,v1}
\fmf{photon}{i2,v2}
\fmf{plain}{v5,o}
\fmf{plain,tension=.3}{v2,v3}
\fmf{photon,tension=.3}{v3,v5}
\fmf{photon,tension=.3}{v1,v4}
\fmf{photon,tension=.3}{v4,v5}
\fmf{photon,tension=0}{v2,v1}
\fmf{photon,tension=0}{v4,v3}
\end{fmfgraph*} }
%
%%%%%%%%%%%%%%%%%%%%%%%
%
\subfigure[]{
\begin{fmfgraph*}(20,20)
\fmfleft{i1,i2}
\fmfright{o}
\fmfforce{0.05w,0.7h}{v11}
\fmfforce{0.05w,0.3h}{v12}
\fmf{photon}{i1,v1}
\fmf{photon}{i2,v2}
\fmf{plain}{v5,o}
\fmf{photon,tension=.3}{v2,v3}
\fmf{photon,tension=.3}{v3,v5}
\fmf{photon,tension=.3}{v1,v4}
\fmf{photon,tension=.3}{v4,v5}
\fmf{photon,tension=0}{v2,v4}
\fmf{photon,tension=0}{v1,v3}
\end{fmfgraph*} } \\
%
%%%%%%%%%%%%%%%%
%
\subfigure[]{
\begin{fmfgraph*}(20,20)
\fmfleft{i1,i2}
\fmfright{o}
\fmfforce{0.05w,0.7h}{v11}
\fmfforce{0.05w,0.3h}{v12}
\fmf{photon}{i1,v1}
\fmf{photon}{i2,v2}
\fmf{plain}{v5,o}
\fmf{photon,tension=.3}{v2,v3}
\fmf{photon,tension=.3}{v3,v5}
\fmf{photon,tension=.3}{v1,v4}
\fmf{photon,tension=.3}{v4,v5}
\fmf{plain,tension=0}{v2,v4}
\fmf{photon,tension=0}{v1,v3}
\end{fmfgraph*} }
%
%%%%%%%%%%%%%%%%
%
\subfigure[]{
\begin{fmfgraph*}(20,20)
\fmfleft{i1,i2}
\fmfright{o}
\fmfforce{0.2w,0.9h}{v2}
\fmfforce{0.2w,0.1h}{v1}
\fmfforce{0.2w,0.5h}{v3}
\fmfforce{0.8w,0.5h}{v5}
\fmf{photon}{i1,v1}
\fmf{photon}{i2,v2}
\fmf{plain}{v5,o}
\fmf{photon,tension=0}{v2,v5}
\fmf{photon,tension=0}{v3,v4}
\fmf{photon,tension=.4}{v1,v4}
\fmf{photon,tension=.4}{v4,v5}
\fmf{photon,tension=0}{v1,v3}
\fmf{photon,tension=0}{v2,v3}
\end{fmfgraph*} }
%
%%%%%%%%%%%%%%%%%%%%%%%
%
\subfigure[]{
\begin{fmfgraph*}(20,20)
\fmfleft{i1,i2}
\fmfright{o}
\fmfforce{0.2w,0.9h}{v2}
\fmfforce{0.2w,0.1h}{v1}
\fmfforce{0.2w,0.5h}{v3}
\fmfforce{0.8w,0.5h}{v5}
\fmf{photon}{i1,v1}
\fmf{photon}{i2,v2}
\fmf{plain}{v5,o}
\fmf{photon,tension=0}{v2,v5}
\fmf{photon,tension=0}{v3,v4}
\fmf{plain,tension=.4}{v1,v4}
\fmf{photon,tension=.4}{v4,v5}
\fmf{photon,tension=0}{v1,v3}
\fmf{photon,tension=0}{v2,v3}
\end{fmfgraph*} } 
%
%%%%%%%%%%%%%%%%%%%%%%%
%
\subfigure[]{
\begin{fmfgraph*}(20,20)
\fmfleft{i1,i2}
\fmfright{o}
\fmfforce{0.2w,0.9h}{v2}
\fmfforce{0.2w,0.1h}{v1}
\fmfforce{0.2w,0.5h}{v3}
\fmfforce{0.8w,0.5h}{v5}
\fmf{photon}{i1,v1}
\fmf{photon}{i2,v2}
\fmf{plain}{v5,o}
\fmf{photon,tension=0}{v2,v5}
\fmf{photon,tension=0}{v3,v4}
\fmf{photon,tension=.4}{v1,v4}
\fmf{photon,tension=.4}{v4,v5}
\fmf{photon,tension=0}{v1,v3}
\fmf{plain,tension=0}{v2,v3}
\end{fmfgraph*} }
%
%%%%%%%%%%%%%%%%%%%%
\vspace*{8mm}
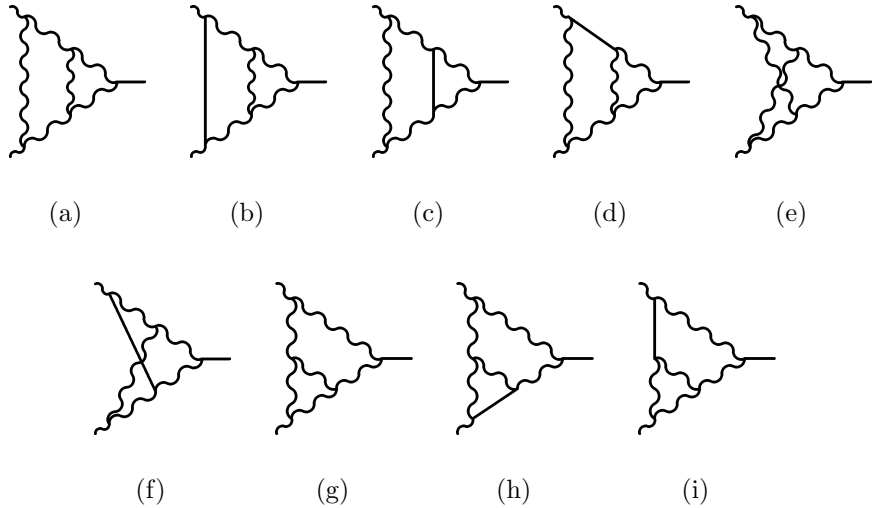
\caption{\label{fig1ter} The set of 9 independent 6-denominator 
topologies.}
\ec
\efig
%%%%%%%%%%%%%%%%%%%%%%%%%%%%%%%%%%%%%%%%%%%%%%%%%%%%%%%%%%%%%%%%%%%%%%%%

%%%%%%%%%%%%%%%%%%%%%% Independent diagrams 5-den %%%%%%%%%%%%%%%%%%%%%
\bfig
\bc
%%%%%%%%
\subfigure[]{
\begin{fmfgraph*}(20,20)
\fmfleft{i1,i2}
\fmfright{o}
\fmfforce{0.8w,0.5h}{v4}
\fmf{photon}{i1,v1}
\fmf{photon}{i2,v2}
\fmf{plain}{v4,o}
\fmf{photon,tension=.4}{v2,v3}
\fmf{photon,tension=.2}{v3,v4}
\fmf{photon,tension=.15}{v1,v4}
\fmf{photon,tension=0}{v2,v1}
\fmf{plain,tension=0}{v1,v3}
\end{fmfgraph*} }
%
%%%%%%%%%%%%%%%%%%%%%%%
%
\subfigure[]{
\begin{fmfgraph*}(20,20)
\fmfleft{i1,i2}
\fmfright{o}
\fmfforce{0.8w,0.5h}{v4}
\fmf{photon}{i1,v1}
\fmf{photon}{i2,v2}
\fmf{plain}{v4,o}
\fmf{photon,tension=.4}{v2,v3}
\fmf{photon,tension=.2}{v3,v4}
\fmf{photon,tension=.15}{v1,v4}
\fmf{plain,tension=0}{v2,v1}
\fmf{photon,tension=0}{v1,v3}
\end{fmfgraph*} }
%
%%%%%%%%%%%%%%%%%%%%%%%
%
\subfigure[]{
\begin{fmfgraph*}(20,20)
\fmfleft{i1,i2}
\fmfright{o}
\fmfforce{0.8w,0.5h}{v4}
\fmf{photon}{i1,v1}
\fmf{photon}{i2,v2}
\fmf{plain}{v4,o}
\fmf{plain,tension=.4}{v2,v3}
\fmf{photon,tension=.2}{v3,v4}
\fmf{photon,tension=.15}{v1,v4}
\fmf{photon,tension=0}{v2,v1}
\fmf{photon,tension=0}{v1,v3}
\end{fmfgraph*} }
%
%%%%%%%%%%%%%%%%%%%%%%%
%
\subfigure[]{
\begin{fmfgraph*}(20,20)
\fmfleft{i1,i2}
\fmfright{o}
\fmfforce{0.8w,0.5h}{v4}
\fmf{photon}{i1,v1}
\fmf{photon}{i2,v2}
\fmf{plain}{v4,o}
\fmf{photon,tension=.15}{v2,v4}
\fmf{photon,tension=.4}{v1,v3}
\fmf{photon,tension=.2}{v3,v4}
\fmf{photon,tension=0}{v2,v1}
\fmf{photon,tension=0}{v2,v3}
\end{fmfgraph*} }
%
%%%%%%%%%%%%%%%%%%%%%%%
%
\subfigure[]{
\begin{fmfgraph*}(20,20)
\fmfleft{i1,i2}
\fmfright{o}
\fmfforce{0.8w,0.5h}{v4}
\fmf{photon}{i1,v1}
\fmf{photon}{i2,v2}
\fmf{plain}{v4,o}
\fmf{photon,tension=.4}{v1,v3}
\fmf{photon,tension=.2}{v3,v4}
\fmf{photon,tension=.15}{v2,v4}
\fmf{photon,tension=0}{v2,v1}
\fmf{plain,tension=0,left=.5}{v3,v4}
\end{fmfgraph*} } \\
%
%%%%%%%%%%%%%%%%%%%%%%%
%
\subfigure[]{
\begin{fmfgraph*}(20,20)
\fmfleft{i1,i2}
\fmfright{o}
\fmfforce{0.8w,0.5h}{v4}
\fmf{photon}{i1,v1}
\fmf{photon}{i2,v2}
\fmf{plain}{v4,o}
\fmf{photon,tension=.4}{v1,v3}
\fmf{photon,tension=.2}{v3,v4}
\fmf{photon,tension=.15}{v2,v4}
\fmf{plain,tension=0}{v2,v1}
\fmf{photon,tension=0,left=.5}{v3,v4}
\end{fmfgraph*} }
%
%%%%%%%%%%%%%%%%%%%%%%%
%
\subfigure[]{
\begin{fmfgraph*}(20,20)
\fmfleft{i1,i2}
\fmfright{o}
\fmfforce{0.8w,0.5h}{v4}
\fmf{photon}{i1,v1}
\fmf{photon}{i2,v2}
\fmf{plain}{v4,o}
\fmf{photon,tension=.4}{v1,v3}
\fmf{photon,tension=.2}{v3,v4}
\fmf{plain,tension=.15}{v2,v4}
\fmf{photon,tension=0}{v2,v1}
\fmf{photon,tension=0,left=.5}{v3,v4}
\end{fmfgraph*} }
%
%%%%%%%%%%%%%%%%%%%%%%%
%
\subfigure[]{
\begin{fmfgraph*}(20,20)
\fmfleft{i1,i2}
\fmfright{o}
\fmfforce{0.8w,0.5h}{v4}
\fmf{photon}{i1,v1}
\fmf{photon}{i2,v2}
\fmf{plain}{v4,o}
\fmf{photon,tension=.4}{v1,v3}
\fmf{photon,tension=.2}{v3,v4}
\fmf{photon,tension=.15}{v2,v4}
\fmf{photon,tension=0}{v2,v1}
\fmf{photon,tension=0,left=.5}{v3,v4}
\end{fmfgraph*} }
%
%%%%%%%%%%%%%%%%%%%%%%%
%
\subfigure[]{
\begin{fmfgraph*}(20,20)
\fmfleft{i1,i2}
\fmfright{o}
\fmfforce{0.8w,0.5h}{v4}
\fmf{photon}{i1,v1}
\fmf{photon}{i2,v2}
\fmf{plain}{v4,o}
\fmf{plain,tension=.4}{v1,v3}
\fmf{photon,tension=.2}{v3,v4}
\fmf{photon,tension=.15}{v2,v4}
\fmf{photon,tension=0}{v2,v1}
\fmf{photon,tension=0,left=.5}{v3,v4}
\end{fmfgraph*} }
%
%%%%%%%%%%%%%%%%%%%%%%%
%
\subfigure[]{
\begin{fmfgraph*}(20,20)
\fmfleft{i1,i2}
\fmfright{o}
\fmfforce{0.2w,0.9h}{v2}
\fmfforce{0.2w,0.1h}{v1}
\fmfforce{0.2w,0.5h}{v3}
\fmfforce{0.8w,0.5h}{v4}
\fmf{photon}{i1,v1}
\fmf{photon}{i2,v2}
\fmf{plain}{v4,o}
\fmf{photon,tension=0}{v2,v3}
\fmf{photon,tension=0}{v3,v4}
\fmf{photon,tension=0}{v1,v4}
\fmf{plain,tension=0}{v2,v4}
\fmf{photon,tension=0}{v1,v3}
\end{fmfgraph*} }   \\
%
%%%%%%%%%%%%%%%%%%%%%%%
%
\subfigure[]{
\begin{fmfgraph*}(20,20)
\fmfleft{i1,i2}
\fmfright{o}
\fmfforce{0.2w,0.9h}{v2}
\fmfforce{0.2w,0.1h}{v1}
\fmfforce{0.2w,0.5h}{v3}
\fmfforce{0.8w,0.5h}{v4}
\fmf{photon}{i1,v1}
\fmf{photon}{i2,v2}
\fmf{plain}{v4,o}
\fmf{photon,tension=0}{v1,v3}
\fmf{photon,tension=0}{v3,v4}
\fmf{photon,tension=0}{v2,v4}
\fmf{plain,tension=0}{v2,v3}
\fmf{photon,tension=0}{v1,v4}
\end{fmfgraph*} }
%
%%%%%%%%%%%%%%%%%%%%%%%
%
\subfigure[]{
\begin{fmfgraph*}(20,20)
\fmfleft{i1,i2}
\fmfright{o}
\fmfforce{0.2w,0.9h}{v2}
\fmfforce{0.2w,0.1h}{v1}
\fmfforce{0.2w,0.5h}{v3}
\fmfforce{0.8w,0.5h}{v4}
\fmf{photon}{i1,v1}
\fmf{photon}{i2,v2}
\fmf{plain}{v4,o}
\fmf{photon,tension=0}{v1,v3}
\fmf{photon,tension=0}{v3,v4}
\fmf{photon,tension=0}{v2,v4}
\fmf{photon,tension=0}{v2,v3}
\fmf{photon,tension=0}{v1,v4}
\end{fmfgraph*} }
%
%%%%%%%%%%%%%%%%%%%%%%%
%
\subfigure[]{
\begin{fmfgraph*}(20,20)
\fmfleft{i1,i2}
\fmfright{o}
\fmf{photon}{i1,v1}
\fmf{photon}{i2,v2}
\fmf{plain}{v4,o}
\fmf{photon,tension=.3}{v2,v3}
\fmf{photon,tension=.3}{v1,v3}
\fmf{photon,tension=0}{v2,v1}
\fmf{photon,tension=.2,left}{v3,v4}
\fmf{photon,tension=.2,right}{v3,v4}
\end{fmfgraph*} }
%
%%%%%%%%%%%%%%%%%%%%%%%
%
\subfigure[]{
\begin{fmfgraph*}(20,20)
\fmfleft{i1,i2}
\fmfright{o}
\fmf{photon}{i1,v1}
\fmf{photon}{i2,v2}
\fmf{plain}{v4,o}
\fmf{photon,tension=.3}{v2,v3}
\fmf{photon,tension=.3}{v1,v3}
\fmf{plain,tension=0}{v2,v1}
\fmf{photon,tension=.2,left}{v3,v4}
\fmf{photon,tension=.2,right}{v3,v4}
\end{fmfgraph*} }
%
%%%%%%%%%%%%%%%%%%%%%%%
%
\subfigure[]{
\begin{fmfgraph*}(20,20)
\fmfleft{i1,i2}
\fmfright{o}
\fmf{photon}{i1,v1}
\fmf{photon}{i2,v2}
\fmf{plain}{v4,o}
\fmf{plain,tension=.3}{v2,v3}
\fmf{photon,tension=.3}{v1,v3}
\fmf{photon,tension=0}{v2,v1}
\fmf{photon,tension=.2,left}{v3,v4}
\fmf{photon,tension=.2,right}{v3,v4}
\end{fmfgraph*} }  \\
%
%%%%%%%%%%%%%%%%%%%%%%%
%
\subfigure[]{
\begin{fmfgraph*}(20,20)
\fmfforce{0.2w,0.5h}{v1}
\fmfforce{0.5w,0.8h}{v2}
\fmfforce{0.5w,0.2h}{v3}
\fmfforce{0.8w,0.5h}{v4}
\fmfleft{i}
\fmfright{o}
\fmf{plain}{i,v1}
\fmf{plain}{v4,o}
\fmf{photon,tension=.2,left=.4}{v1,v2}
\fmf{photon,tension=.2,right=.4}{v1,v3}
\fmf{photon,tension=.2,left=.4}{v2,v4}
\fmf{photon,tension=.2,right=.4}{v3,v4}
\fmf{photon,tension=0}{v2,v3}
\end{fmfgraph*} }
%
%%%%%%%%%%%%%%%%%%%%%%% 
%
\subfigure[]{
\begin{fmfgraph*}(20,20)
\fmfforce{0.2w,0.5h}{v1}
\fmfforce{0.5w,0.8h}{v2}
\fmfforce{0.5w,0.2h}{v3}
\fmfforce{0.8w,0.5h}{v4}
\fmfleft{i}
\fmfright{o}
\fmf{plain}{i,v1}
\fmf{plain}{v4,o}
\fmf{photon,tension=.2,left=.4}{v1,v2}
\fmf{photon,tension=.2,right=.4}{v1,v3}
\fmf{photon,tension=.2,left=.4}{v2,v4}
\fmf{photon,tension=.2,right=.4}{v3,v4}
\fmf{plain,tension=0}{v2,v3}
\end{fmfgraph*} } 
%
%%%%%%%%%%%%%%%%%%%%%%%%
%
\subfigure[]{
\begin{fmfgraph*}(20,20)
\fmfforce{0.2w,0.5h}{v1}
\fmfforce{0.5w,0.8h}{v2}
\fmfforce{0.5w,0.2h}{v3}
\fmfforce{0.8w,0.5h}{v4}
\fmfleft{i}
\fmfright{o}
\fmf{plain}{i,v1}
\fmf{plain}{v4,o}
\fmf{plain,tension=.2,left=.4}{v1,v2}
\fmf{photon,tension=.2,right=.4}{v1,v3}
\fmf{photon,tension=.2,left=.4}{v2,v4}
\fmf{photon,tension=.2,right=.4}{v3,v4}
\fmf{photon,tension=0}{v2,v3}
\end{fmfgraph*} }
%
%%%%%%%%%%%%%%%%%%%%%%% 
%
\subfigure[]{
\begin{fmfgraph*}(20,20)
\fmfleft{i1,i2}
\fmfright{o}
\fmfforce{0.2w,0.9h}{v2}
\fmfforce{0.2w,0.1h}{v1}
\fmfforce{0.2w,0.55h}{v3}
\fmfforce{0.2w,0.15h}{v5}
\fmfforce{0.8w,0.5h}{v4}
\fmf{photon}{i1,v1}
\fmf{photon}{i2,v2}
\fmf{plain}{v4,o}
\fmf{photon}{v2,v3}
\fmf{photon,left}{v3,v5}
\fmf{photon,right}{v3,v5}
\fmf{photon}{v1,v4}
\fmf{photon}{v2,v4}
\end{fmfgraph*} }
%
%%%%%%%%%%%%%%%%%%%%%%%
%
\subfigure[]{
\begin{fmfgraph*}(20,20)
\fmfleft{i1,i2}
\fmfright{o}
\fmfforce{0.2w,0.9h}{v2}
\fmfforce{0.2w,0.1h}{v1}
\fmfforce{0.2w,0.55h}{v3}
\fmfforce{0.2w,0.15h}{v5}
\fmfforce{0.8w,0.5h}{v4}
\fmf{photon}{i1,v1}
\fmf{photon}{i2,v2}
\fmf{plain}{v4,o}
\fmf{plain}{v2,v3}
\fmf{photon,left}{v3,v5}
\fmf{photon,right}{v3,v5}
\fmf{photon}{v1,v4}
\fmf{photon}{v2,v4}
\end{fmfgraph*} } \\
%
%%%%%%%%%%%%%%%%%%%%%%%
%
\subfigure[]{
\begin{fmfgraph*}(20,20)
\fmfleft{i1,i2}
\fmfright{o}
\fmfforce{0.1w,0.1h}{v1}
\fmfforce{0.4w,0.3h}{v2}
\fmfforce{0.1w,0.9h}{v3}
\fmfforce{0.4w,0.7h}{v4}
\fmfforce{0.9w,0.5h}{v5}
\fmf{photon}{i1,v1}
\fmf{phantom}{i2,v3}
\fmf{plain}{v5,o}
\fmf{photon,left}{v1,v2}
\fmf{plain,right}{v1,v2}
\fmf{photon}{v3,v4}
\fmf{photon}{v2,v4}
\fmf{photon}{v4,v5}
\fmf{photon}{v2,v5}
\end{fmfgraph*} }
%
%%%%%%%%%%%%%%%%%%%%%%%
%
\subfigure[]{
\begin{fmfgraph*}(20,20)
\fmfforce{0.2w,0.5h}{v1}
\fmfforce{0.5w,0.8h}{v2}
\fmfforce{0.5w,0.2h}{v3}
\fmfforce{0.8w,0.5h}{v4}
\fmfleft{i}
\fmfright{o}
\fmf{photon}{i,v1}
\fmf{photon}{v4,o}
\fmf{photon,tension=.2,left=.4}{v1,v2}
\fmf{plain,tension=.2,right=.4}{v1,v3}
\fmf{photon,tension=.2,left=.4}{v2,v4}
\fmf{photon,tension=.2,right=.4}{v3,v4}
\fmf{photon,tension=0}{v2,v3}
\end{fmfgraph*} }
%
%%%%%%%%%%%%%%%%%%%%%%% 
%
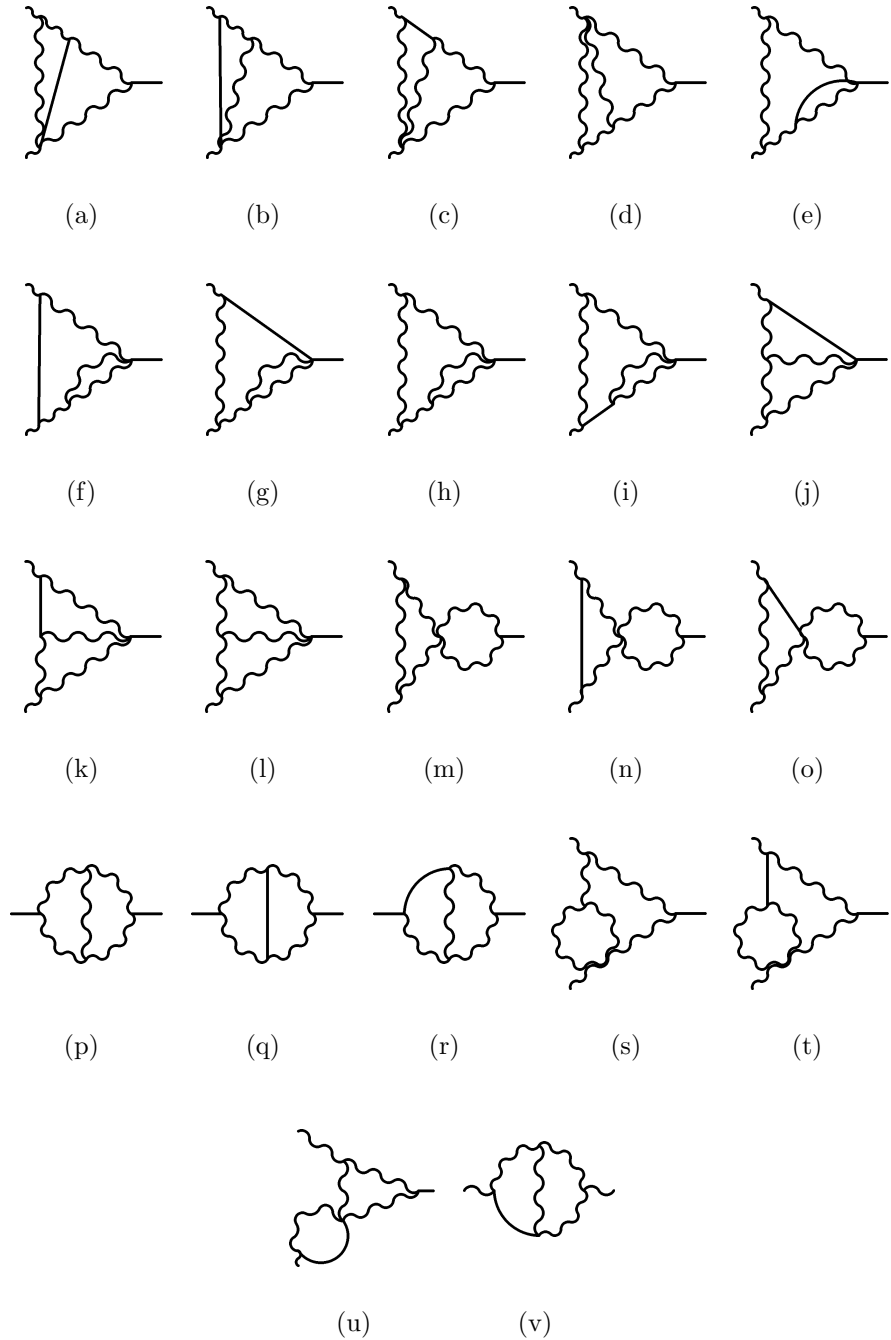
\caption{\label{fig2} The set of 22 independent 5-denominator 
topologies.}
\ec
\efig
%%%%%%%%%%%%%%%%%%%%%%%%%%%%%%%%%%%%%%%%%%%%%%%%%%%%%%%%%%%%%%%%%%%%%

%%%%%%%%% Independent diagrams 4-denominators %%%%%%%%%%%%%%%%%%%%%
\bfig
\bc
%%%%%%%%
%
\subfigure[]{
\begin{fmfgraph*}(20,20)
\fmfleft{i1,i2}
\fmfright{o}
\fmf{photon}{i1,v1}
\fmf{photon}{i2,v2}
\fmf{plain}{v3,o}
\fmf{photon,tension=.3}{v2,v3}
\fmf{photon,tension=.3}{v1,v3}
\fmf{plain,tension=0,right=.5}{v2,v1}
\fmf{photon,tension=0,right=.5}{v1,v2}
\end{fmfgraph*} }
%
%%%%%%%%%%%%%%%%%%%%%%%
%
\subfigure[]{
\begin{fmfgraph*}(20,20)
\fmfleft{i1,i2}
\fmfright{o}
\fmf{photon}{i1,v1}
\fmf{photon}{i2,v2}
\fmf{plain}{v3,o}
\fmf{photon,tension=.3}{v2,v3}
\fmf{photon,tension=.3}{v1,v3}
\fmf{photon,tension=0,right=.5}{v2,v1}
\fmf{photon,tension=0,right=.5}{v1,v2}
\end{fmfgraph*} }
%
%%%%%%%%%%%%%%%%%%%%%%%
%
\subfigure[]{
\begin{fmfgraph*}(20,20)
\fmfleft{i1,i2}
\fmfright{o}
\fmf{photon}{i1,v1}
\fmf{photon}{i2,v2}
\fmf{plain}{v3,o}
\fmf{photon,tension=.3}{v2,v3}
\fmf{photon,tension=.3}{v1,v3}
\fmf{photon,tension=0}{v2,v1}
\fmf{plain,tension=0,left=.5}{v1,v3}
\end{fmfgraph*} } 
%
%%%%%%%%%%%%%%%%%%%%%%%
%
\subfigure[]{
\begin{fmfgraph*}(20,20)
\fmfleft{i1,i2}
\fmfright{o}
\fmf{photon}{i1,v1}
\fmf{photon}{i2,v2}
\fmf{plain}{v3,o}
\fmf{photon,tension=.3}{v2,v3}
\fmf{photon,tension=.3}{v1,v3}
\fmf{plain,tension=0}{v2,v1}
\fmf{photon,tension=0,left=.5}{v1,v3}
\end{fmfgraph*} } 
%
%%%%%%%%%%%%%%%%%%%%%%%
%
\subfigure[]{
\begin{fmfgraph*}(20,20)
\fmfleft{i1,i2}
\fmfright{o}
\fmf{photon}{i1,v1}
\fmf{photon}{i2,v2}
\fmf{plain}{v3,o}
\fmf{plain,tension=.3}{v2,v3}
\fmf{photon,tension=.3}{v1,v3}
\fmf{photon,tension=0}{v2,v1}
\fmf{photon,tension=0,left=.5}{v1,v3}
\end{fmfgraph*} } \\
%
%%%%%%%%%%%%%%%%%%%%%%%
%
\subfigure[]{
\begin{fmfgraph*}(20,20)
\fmfleft{i1,i2}
\fmfright{o}
\fmf{photon}{i1,v1}
\fmf{photon}{i2,v2}
\fmf{plain}{v3,o}
\fmf{photon,tension=.3}{v2,v3}
\fmf{photon,tension=.3}{v1,v3}
\fmf{photon,tension=0}{v2,v1}
\fmf{photon,tension=0,right=.5}{v2,v3}
\end{fmfgraph*} } 
%
%%%%%%%%%%%%%%%%%%%%%%%
%
\subfigure[]{
\begin{fmfgraph*}(20,20)
\fmfleft{i1,i2}
\fmfright{o}
\fmf{photon}{i1,v1}
\fmf{photon}{i2,v2}
\fmf{plain}{v3,o}
\fmf{photon,tension=.3}{v2,v3}
\fmf{photon,tension=.3}{v1,v3}
\fmf{photon,tension=0}{v2,v1}
\fmf{plain,right=45}{v3,v3}
\end{fmfgraph*} } 
%
%%%%%%%%%%%%%%%%%%%%%%%
%
\subfigure[]{
\begin{fmfgraph*}(20,20)
\fmfleft{i}
\fmfright{o}
\fmf{plain}{i,v1}
\fmf{plain}{v3,o}
\fmf{photon,tension=.2,left}{v1,v2}
\fmf{photon,tension=.2,right}{v1,v2}
\fmf{photon,tension=.2,left}{v2,v3}
\fmf{photon,tension=.2,right}{v2,v3}
\end{fmfgraph*} }
%
%%%%%%%%%%%%%%%%%%%%%%%
%
\subfigure[]{
\begin{fmfgraph*}(20,20)
\fmfleft{i}
\fmfright{o}
\fmf{plain}{i,v1}
\fmf{plain}{v3,o}
\fmf{plain,tension=.2,left}{v1,v2}
\fmf{photon,tension=.2,right}{v1,v2}
\fmf{photon,tension=.2,left}{v2,v3}
\fmf{photon,tension=.2,right}{v2,v3}
\end{fmfgraph*} } 
%
%%%%%%%%%%%%%%%%%%%%%%%
%
\subfigure[]{
\begin{fmfgraph*}(20,20)
\fmfleft{i}
\fmfright{o}
\fmfforce{0.2w,0.5h}{v1}
\fmfforce{0.5w,0.2h}{v2}
\fmfforce{0.8w,0.5h}{v3}
\fmf{plain}{i,v1}
\fmf{plain}{v3,o}
\fmf{photon,left}{v1,v3}
\fmf{photon,right=.4}{v1,v2}
\fmf{photon,right=.4}{v2,v3}
\fmf{photon,left=.6}{v2,v3}
\end{fmfgraph*} } \\
%
%%%%%%%%%%%%%%%%%%%%%%%
%
\subfigure[]{
\begin{fmfgraph*}(20,20)
\fmfleft{i}
\fmfright{o}
\fmfforce{0.2w,0.5h}{v1}
\fmfforce{0.5w,0.2h}{v2}
\fmfforce{0.8w,0.5h}{v3}
\fmf{plain}{i,v1}
\fmf{plain}{v3,o}
\fmf{photon,left}{v1,v3}
\fmf{photon,right=.4}{v1,v2}
\fmf{photon,right=.4}{v2,v3}
\fmf{plain,left=.6}{v2,v3}
\end{fmfgraph*} } 
%
%%%%%%%%%%%%%%%%%%%%%%%
%
\subfigure[]{
\begin{fmfgraph*}(20,20)
\fmfleft{i}
\fmfright{o}
\fmfforce{0.2w,0.5h}{v1}
\fmfforce{0.5w,0.2h}{v2}
\fmfforce{0.8w,0.5h}{v3}
\fmf{plain}{i,v1}
\fmf{plain}{v3,o}
\fmf{plain,left}{v1,v3}
\fmf{photon,right=.4}{v1,v2}
\fmf{photon,right=.4}{v2,v3}
\fmf{photon,left=.6}{v2,v3}
\end{fmfgraph*} } 
%
%%%%%%%%%%%%%%%%%%%%%%%
%
\subfigure[]{
\begin{fmfgraph*}(20,20)
\fmfleft{i}
\fmfright{o}
\fmfforce{0.2w,0.5h}{v1}
\fmfforce{0.5w,0.8h}{v2}
\fmfforce{0.8w,0.5h}{v3}
\fmf{plain}{i,v1}
\fmf{plain}{v3,o}
\fmf{photon,right}{v1,v3}
\fmf{plain,left=.4}{v1,v2}
\fmf{photon,left=.4}{v2,v3}
\fmf{photon,right=.6}{v2,v3}
\end{fmfgraph*} } 
%
%%%%%%%%%%%%%%%%%%%%%%%
%
\subfigure[]{
\begin{fmfgraph*}(20,20)
\fmfleft{i}
\fmfright{o}
\fmfforce{0.2w,0.5h}{v1}
\fmfforce{0.5w,0.2h}{v2}
\fmfforce{0.8w,0.5h}{v3}
\fmf{photon}{i,v1}
\fmf{photon}{v3,o}
\fmf{plain,left}{v1,v3}
\fmf{photon,right=.4}{v1,v2}
\fmf{photon,right=.4}{v2,v3}
\fmf{photon,left=.6}{v2,v3}
\end{fmfgraph*} } 
%
%%%%%%%%%%%%%%%%%%%%%%%
%
\subfigure[]{
\begin{fmfgraph*}(20,20)
\fmfleft{i}
\fmfright{o}
\fmfforce{0.2w,0.5h}{v1}
\fmfforce{0.5w,0.2h}{v2}
\fmfforce{0.8w,0.5h}{v3}
\fmf{photon}{i,v1}
\fmf{photon}{v3,o}
\fmf{photon,left}{v1,v3}
\fmf{photon,right=.4}{v1,v2}
\fmf{photon,right=.4}{v2,v3}
\fmf{plain,left=.6}{v2,v3}
\end{fmfgraph*} } \\
%
%%%%%%%%%%%%%%%%%%%%%%%
%
\subfigure[]{
\begin{fmfgraph*}(20,20)
\fmfleft{i}
\fmfright{o}
\fmfforce{0.2w,0.5h}{v1}
\fmfforce{0.5w,0.8h}{v2}
\fmfforce{0.8w,0.5h}{v3}
\fmf{photon}{i,v1}
\fmf{photon}{v3,o}
\fmf{photon,right}{v1,v3}
\fmf{plain,left=.4}{v1,v2}
\fmf{photon,left=.4}{v2,v3}
\fmf{photon,right=.6}{v2,v3}
\end{fmfgraph*} } 
%
%%%%%%%%%%%%%%%%%%%%%%%
%
\subfigure[]{
\begin{fmfgraph*}(20,20)
\fmfbottom{v5}
\fmftop{v4}
\fmfleft{i}
\fmfright{o}
\fmf{photon}{i,v1}
\fmf{plain}{v3,o}
\fmf{photon}{v5,v2} 
\fmf{phantom}{v2,v4} 
\fmf{photon,tension=.2,left}{v1,v2}
\fmf{plain,tension=.2,right}{v1,v2}
\fmf{photon,tension=.2,left}{v2,v3}
\fmf{photon,tension=.2,right}{v2,v3}
\end{fmfgraph*} } 
%
%%%%%%%%%%%%%%%%%%%%%%%
%
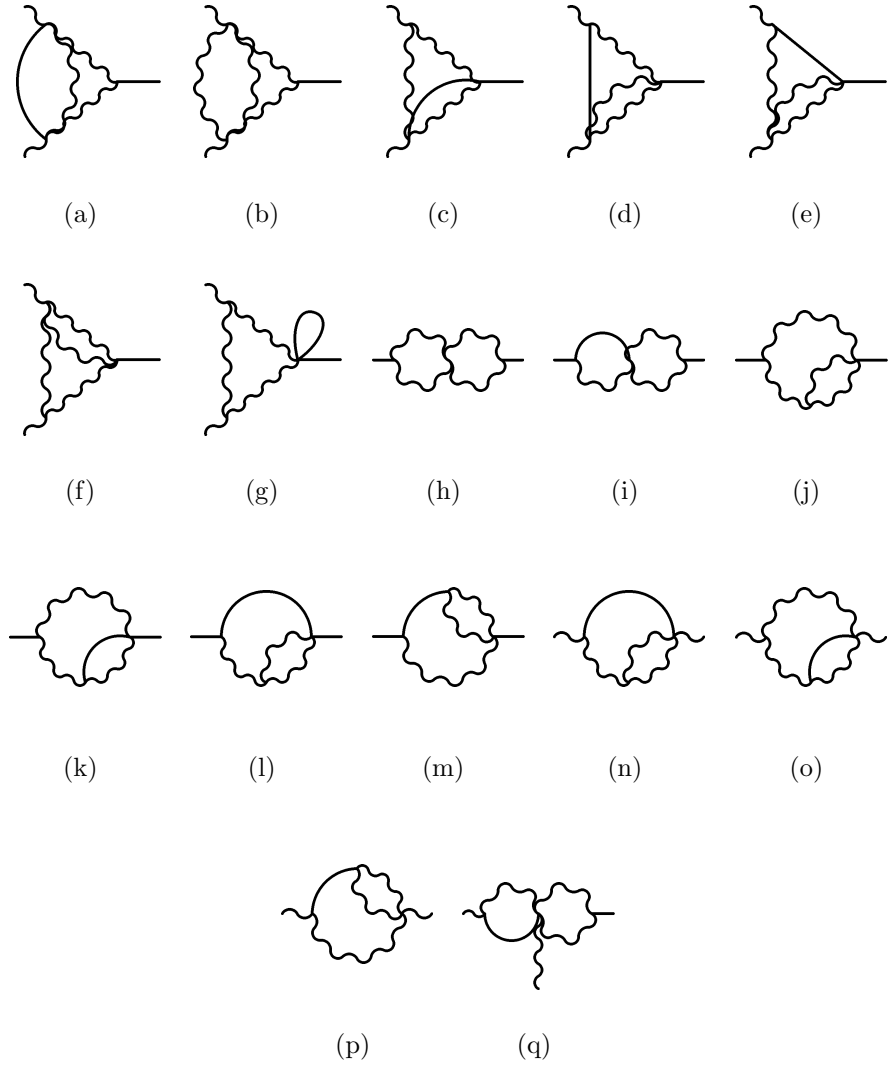
\caption{\label{fig3} The set of 17 independent 4-denominator 
topologies.}
\ec
\efig
%%%%%%%%%%%%%%%%%%%%%%%%%%%%%%%%%%%%%%%%%%%%%%%%%%%%%%%%%%%%%%%%%%%%%

%%%%%%%%% Independent diagrams 3-denominators %%%%%%%%%%%%%%%%%%%%%
\bfig
\bc
%%%%%%%%
\subfigure[]{
\begin{fmfgraph*}(20,20)
\fmfleft{i}
\fmfright{o}
\fmf{plain}{i,v1}
\fmf{plain}{v2,o}
\fmf{photon,tension=.15,left}{v1,v2}
\fmf{photon,tension=.15}{v1,v2}
\fmf{photon,tension=.15,right}{v1,v2}
\end{fmfgraph*} } 
%
%%%%%%%%%%%%%%%%%%%%%%%
%
\subfigure[]{
\begin{fmfgraph*}(20,20)
\fmfleft{i}
\fmfright{o}
\fmf{plain}{i,v1}
\fmf{plain}{v2,o}
\fmf{plain,tension=.15,left}{v1,v2}
\fmf{photon,tension=.15}{v1,v2}
\fmf{photon,tension=.15,right}{v1,v2}
\end{fmfgraph*} } 
%
%%%%%%%%%%%%%%%%%%%%%%%
%
\subfigure[]{
\begin{fmfgraph*}(20,20)
\fmfleft{i}
\fmfright{o}
\fmf{photon}{i,v1}
\fmf{photon}{v2,o}
\fmf{plain,tension=.15,left}{v1,v2}
\fmf{photon,tension=.15}{v1,v2}
\fmf{photon,tension=.15,right}{v1,v2}
\end{fmfgraph*} } 
%
%%%%%%%%%%%%%%%%%%%%%%%
%
\subfigure[]{
\begin{fmfgraph*}(20,20)
\fmfleft{i}
\fmfright{o}
\fmf{plain}{i,v1}
\fmf{plain}{v2,o}
\fmf{photon,tension=.22,left}{v1,v2}
\fmf{photon,tension=.22,right}{v1,v2}
\fmf{plain,right=45}{v2,v2}
\end{fmfgraph*} } 
%
%%%%%%%%%%%%%%%%%%%%%%%
%
%\subfigure[]{
%\begin{fmfgraph*}(20,20)
%\fmfleft{i}
%\fmfright{o}
%\fmfforce{0.5w,0.1h}{v1}
%\fmfforce{0.25w,0.62h}{v3}
%\fmfforce{0.5w,0.9h}{v7}
%\fmfforce{0.74w,0.62h}{v11}
%\fmf{plain,left=.1}{v1,v3}
%\fmf{plain,left=.5}{v3,v7}
%\fmf{photon,left=.5}{v7,v11}
%\fmf{photon,left=.1}{v11,v1}
%\fmf{photon}{v1,v7}
%\end{fmfgraph*}}
%
%%%%%%%%%%%%%%%%%%%%%%%%
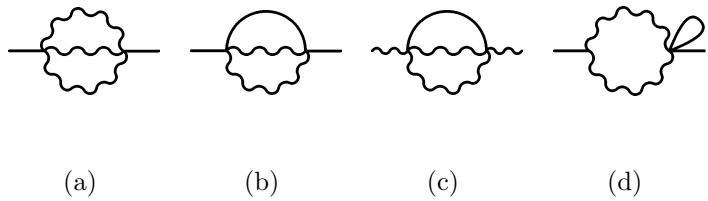
\caption{\label{fig4} The set of 4 independent 3-denominator 
topologies.}
\ec
\efig
%%%%%%%%%%%%%%%%%%%%%%%%%%%%%%%%%%%%%%%%%%%%%%%%%%%%%%%%%%%%%%%%%%%%%

Within this method, for a given topology, i.e. for a given set of 
denominators  
${\mathcal D}_{i_{1}} {\mathcal D}_{i_{2}}...{\mathcal D}_{i_{t}}$ (see 
appendix \ref{app1}), the independent scalar integrals are of the form:
\be
\mathrm{Topo}(n_1,\cdots,n_t; s_1,\cdots,s_{\overline{t}}) 
= \int \{ d^D k_1 \} \{ d^D k_2 \} \frac{S_{i_{1}}^{s_1}  \cdots   
S_{i_{\overline{t}}}^{s_{\overline{t}}}}{{\mathcal D}_{i_{1}}^{n_1} 
\cdots {\mathcal D}_{i_{t}}^{n_t}} \, , 
\label{b1} 
\ee 
where $\overline{t}=7-t$ denotes the number of independent scalar 
products $S_{i_{1}} \cdots S_{i_{\overline{t}}}$. Note that 
$s_1 , \cdots , s_{ \overline{t} } \geq 0$ while 
$n_1, \cdots , n_t \geq 1$.

The advantages of the auxiliary diagram method are the simplicity and a 
shorter code for its implementation. The main advantage of the shift 
method is that a given topology appears in a unique form, so the CPU 
time needed is less.

Let us conclude this section with a general remark. A straightforward 
strategy to evaluate ${\mathcal I}$ is that of computing all the
independent scalar amplitudes entering 
its decomposition. In simple calculations, this is probably the simplest
thing to do. In more complicated cases, however, there may be a large 
number of such scalar amplitudes and the individual computation of all 
of them may become rather laborious. In these cases, it is more
convenient to use a set of relations connecting different scalar 
amplitudes, and to compute directly only a subset of them. The latter is
our strategy that is described in the next section.

\subsection{Relation between independent scalar amplitudes: 
            the integration by parts identities \label{IBPs}}

As we have shown in the previous section, the task of evaluating a 
two-loop vertex diagram with an arbitrary scalar numerator 
${\mathcal N}$ is shifted to that of computing a set of independent 
scalar amplitudes either containing an auxiliary denominator:
$
\mathrm{Topo} ( n_{1},n_{2},n_{3},n_{4},n_{5},n_{6}; n_{7} ), 
$
or containing a selected basis of denominators and scalar products:
$
\mathrm{Topo}( n_1,\cdots,n_t;~s_1,\cdots,s_{\overline{t}} ) . 
$
Let us discuss the simpler case of the auxiliary diagram first. The 
allowed 
range of the indices are: $n_{1},\cdots ,n_{6}\leq 1$ and $n_{7}\leq 0$ 
(see Eq. (\ref{Topodef})), but in practice, we expect the relevant 
$n_{i}$'s to be restricted to $n_{i}\geq -3$, which however gives still 
a rather large set of scalar amplitudes to compute. Relations among 
different scalar amplitudes are obtained by considering the so-called 
integration-by-parts (ibp) identities \cite{Chet}:
\begin{equation}
0=\int \{d^{D}k_{1}\}\{d^{D}k_{2}\} 
\frac{\partial }{\partial k_{i}^{\mu }}\,%
\left\{ \frac{v^{\mu }}{P_{1}^{n_{1}} P_{2}^{n_{2}} P_{3}^{n_{3}} 
P_{4}^{n_{4}} P_{5}^{n_{5}} P_{6}^{n_{6}} P_{7}^{n_{7}}} \right\} ,  
\label{ibpstart}
\end{equation}
with $i=1,2$ and $v=k_{1},k_{2},p_{1},p_{2}.$ The integral above 
vanishes because of the following argument. It is the space integral in 
$D$ dimensions of the total divergence of the vector in curly brackets. 
It can then be transformed into the flux integral of this vector 
over a spherical
surface with infinite radius, $r=\infty$. For small enough dimension 
$D$, the surface integral vanishes for $r\rightarrow \infty$. It can 
then be defined to identically vanish for any $D$. By explicitly 
performing the derivations in Eq. (\ref{ibpstart}) and expressing
the scalar products in terms of the assumed basis $\{P_{i}\}_{i=1,7}$ 
according to Eqs. (\ref{reexpress}), we obtain identities of the form: 
\bea
\! \! \! \! \! \! \! \! 0 & = & c \mathrm{Topo} ( n_{1}, \! n_{2}, 
\! n_{3}, \! n_{4}, \! n_{5}, \! n_{6}; \! n_{7} ) + \! 
\sum_{i=1}^{7} n_{i} d_{i} \mathrm{Topo} ( n_{1}, \! \cdots \! ,
n_{i} \! + \! 1, \! \cdots \! , n_{7} ) \nn\\   
\! \! \! \! \! \! \! \! & & + \sum_{i\neq j}^{1,7} n_{i} e_{ij} 
\mathrm{Topo} (n_{1}, \cdots ,n_{i} \! + \! 1,\cdots ,n_{j} \! - \! 1,
\cdots ; n_{7} ) .
\label{ibpidentities}
\eea
For a given set of the indices $\{n_{i}\}_{i=1,7}$ there are eight of 
such identities; in general, however, they are not all linearly 
independent. Each identity contains, in general, three kinds of 
amplitudes:
\begin{enumerate}
\item the amplitude itself: $\mathrm{Topo}\left(
n_{1},n_{2},n_{3},n_{4},n_{5},n_{6};n_{7}\right)$;

\item amplitudes with one of the indices increased by one, 
$n_{i}\rightarrow n_{i}+1$ $(i=1,7)$: $\mathrm{Topo}\left(n_{1}, 
\cdots ,n_{i} \! + \! 1,\cdots ; n_{7} \right)$.
These terms clearly originate from the derivation of the factor $
P_{i}^{-n_{i}}\rightarrow -n_{i}P_{i}^{-n_{i}-1}$. We pulled a factor 
$n_{i}$ out of the coefficient $d_{i}$ to point out that these terms 
are absent for $n_{i}=0.$ In other words, it is not possible to 
generate an absent denominator by differentiation. This implies that 
the ibp identities for a given amplitude with topological number $t$ 
involve only amplitudes with smaller or equal topological number, 
$t^{\prime }\leq t$;

\item amplitudes with one index increased by one, $n_{i}\rightarrow n_{i}+1$
and another index decreased by one, $n_{j}\rightarrow n_{j}-1$:
$\mathrm{Topo} (n_{1}, \cdots ,n_{i} \! + \! 1,\cdots ,n_{j} \! - \! 1,
\cdots ; n_{7} )$. These terms originate from the derivation of the 
factor $P_{i}^{-n_{i}}$ and the cancellation of a power of 
$P_{j}^{-n_{j}}$ by the invariants generated at the numerator. We 
pulled out a factor $n_{i}$ also out of the coefficients $e_{ij}$ to 
point out similar properties as those discussed in $2)$.
\end{enumerate}

Other identities are obtained by considering discrete symmetries 
of the auxiliary  diagram. The ladder topologies in (a), (b) and (c) 
of Fig. \ref{fig1ter} for example have an up-down symmetry, resulting in 
the exchange of the fermion lines expressed by the relation:
\begin{equation}
\mathrm{Topo}\left( n_{1},n_{2},n_{3},n_{4},n_{5},n_{6};n_{7}\right)
=
\mathrm{Topo}\left( n_{1},n_{3},n_{2},n_{4},n_{6},n_{5};n_{7}\right)
\, .
\end{equation}
The massless crossed ladder also has an up-down symmetry, which
interchanges also the boson lines.
By solving the above identities, it is possible to express a general
scalar amplitude as a linear combination of a finite set of
$N_{F}$ basic integrals, called master integrals (MI's): 
\be
\mathrm{Topo} ( n_{1},n_{2},n_{3},n_{4},n_{5},n_{6};n_{7} )
=\sum_{i=1}^{N_{F}}c_{i} (n_{1},n_{2},n_{3},n_{4},n_{5},n_{6};n_{7} ) 
\, F_{i} .  
\label{expmi}
\ee
In more formal terms, we may say that the ibp identities induce such a
strong linear dependence between all the infinite scalar amplitudes to
project them into a finite dimensional linear space. The $F_{i}$'s 
have the same role as the basis vectors in a linear space. The choice 
of the MI's is arbitrary, as it is the choice of the basis vectors; 
only their number $N_{F}$ is fixed and equals the 
dimension of the space spanned by any choice of the 
$F_{i}^{\prime }$'s. Changing the set of MI's in the 
decomposition (\ref{expmi}) is analogous to a change of basis. 
To give an explicit example, let us report the MI decomposition 
of the basic six-denominator amplitude of the ladder with a massive 
external boson in Fig. \ref{fig1} (b), the example in the previous 
section:
\begin{eqnarray}
\! \! \! \! \! \! & & \mathrm{Topo}(1,1,1,1,1,1;0) = 2\left( 
\frac{2}{\epsilon ^{2}} - 
\frac{7}{\epsilon }+6\right) \left( \frac{1}{x^{2}}+\frac{1}{x} + 
\frac{1}{1-x}\right) \times \nn\\
\! \! \! \! \! \! & & \qquad \qquad \times \mathrm{Topo}(0,1,0,1,0,1;0) 
+ \Biggl[ \left( \frac{1}{
\epsilon ^{2}} - \frac{4}{\epsilon }+3 \right) \frac{1}{\left( 1-x 
\right) ^{2}} \nn\\
\! \! \! \! \! \! & & \qquad \qquad - 2 \left( \frac{1}{\epsilon }-1
\right) \left( \frac{1}{x}+
\frac{1}{1-x} \right) \Biggr] \bigl[ \mathrm{Topo}(1,0,0,1,1,0;0) \nn\\
\! \! \! \! \! \! & & \qquad \qquad +\mathrm{Topo}(1,0,0,0,1,1;0) \bigr] 
- \! 2 \! \left( \! \frac{1}{\epsilon ^{2}} \! - \! \frac{4}{\epsilon } 
\! + \! 4 \! \right) \mathrm{Topo}(1,1,0,1,0,1;0) \nn\\
\! \! \! \! \! \! & & \qquad \qquad - \! \left( \! \frac{1}{\epsilon } 
\! - \! 2 \right) \frac{1}{x} \mathrm{Topo}(1,0,0,1,1,1;0) - \! \Biggl[ 
\left( \! \frac{2}{\epsilon ^{2}} \! - \! \frac{9}{\epsilon } \! + \! 9 
\right) \frac{1}{\left( 1 \! - \! x\right) ^{2}} \nn\\
\! \! \! \! \! \! & & \qquad \qquad + \left( \frac{2}{\epsilon ^{2}} - 
\frac{11}{\epsilon } + 12\right) \left( \frac{1}{x}+\frac{1}{1-x} 
\right) \Biggr] \mathrm{Topo}(1,0,0,1,1,1;-1) \nn\\
\! \! \! \! \! \! & & \qquad \qquad + \left( \frac{1}{\epsilon } \! - 
\! 2\right) \frac{1}{x} \mathrm{Topo}(1,1,1,0,1,1;0) \! +\frac{2}{x} 
\mathrm{Topo}(1,1,0,1,1,1;0),
\end{eqnarray}
where we have taken for simplicity $a=1$. In this case $N_F=8$.
We have chosen MI's with a subset of the denominators with unitary 
indices: $n_{i}=1$, the remaining denominators having vanishing indices:
$n_{j}=0$. For the topology represented by the amplitude 
$\mathrm{Topo}(1,0,0,1,1,1;0)$ this was ot sufficient as two master
integrals are involved. We added 
the amplitude with the auxiliary denominator brought to the 
numerator, i.e. with index $n_{7}=-1$: $\mathrm{Topo}(1,0,0,1,1,1;-1)$;
we could have taken 
$n_2=-1,~n_3=0,~n_7=0$ or 
$n_2=0,~n_3=-1,~n_7=0$ as well.
Still another possibility would have 
been to consider amplitudes with one of the denominators squared, 
such as for example $\mathrm{Topo}(2,0,0,1,1,1;0)$. 
In general, the choice of the MI's is 
dictated by practical considerations. In some cases, for example, it may
be useful to require the absence of ultraviolet or infrared 
singularities from the MI's \cite{timovanritbergen}; in other cases, the
set of MI's can be chosen in such a way that the corresponding system of
differential equations become easier to solve (see next section).

Let us now discuss the methods to solve the ibp identities to arrive to 
the results (\ref{expmi}). The ibp identities constitute a system 
of linear equations in which the unknowns are the scalar integrals 
themselves. 
The oldest approach, developed in the original article on the ibp's, 
involves a symbolic solution of the identities, treated as recurrence
equations on the indices \cite{Chet,timovanritbergen}.
One introduces operators raising or lowering one of the indices 
\begin{equation}
{\mathbf i}^{\pm} \, \mathrm{Topo} (n_{1}, \cdots ,n_{i},\cdots ; n_{7})
= \mathrm{Topo} (n_{1}, \cdots ,n_{i}\pm 1,\cdots ; n_{7} ).
\end{equation}
Each ibp equation involves the identity operator ${\mathbf 1}$ and the 
raising operators ${\mathbf i}^{+}$ and ${\mathbf i}^+ {\mathbf j}^-$.  
If one can combine the equations in such a way that an amplitude
is written in terms of amplitudes all containing lowering operators, 
then reduction to simpler topologies has been achieved.
The advantages of the symbolic approach are the elegance and that
its implementation does not require in general a lot of CPU time. 
The main disadvantage is that a careful, case-by-case,
inspection of the equations is required.
A more recent approach is that of replacing numerical 
values for the indices and solve the resulting linear system \cite{Rem3,Bon1}.
Let us describe this method in practical terms.
An initial linear system can be obtained by assigning to the indices the values 
$n_{i}=0,1$ for $i=1,...,6$ and $n_7=0$. This is a set 
of $ 8\times 2^6 = 512 $ equations\footnote{As already said, not all the equations
in the system are linearly independent on each other and some of them 
are actually trivial  (i.e. $0=0$), such as for example
those with five indices equal to zero.}.
If these equations are not sufficient for a complete MI reduction, 
one may solve a larger system in which 
one of the indices can also take the value $n_i=-1$ for $i=1,\cdots,7$; 
this is a set of $(3 \times 2^5 \times 6 + 2^7) \times 8 = 5632 $ equations.
Let us stress that {\it all} the equations with one index $=-1$ have to be 
considered simultaneously, because the amplitudes involved are coupled 
by the ${\mathbf i}^{+} {\mathbf j}^{-}$  operators; for example:
\begin{equation}
{\mathbf 7}^{+} {\mathbf 6}^{-} : 
\mathrm{Topo}(1,1,1,1,1,0;-1) 
\rightarrow  \mathrm{Topo}(1,1,1,1,1,-1;0).
\end{equation}
One may also consider a system in which one of the indices 
takes the value $n_i=2$ for $i=1,6$;
this amounts to $3 \times 2^5 \times 6 \times 8 = 4608 $ equations.
If these systems are not sufficiently large, one may consider equations 
with indices taking both the values $2$ and $-1$ and so on. The general 
idea is that, by going to larger systems, the number of equations grows 
faster than the number of amplitudes, till a ``critical mass'' of 
equations is reached, allowing the full reduction to MI's\footnote{One of us (U.A.) 
wishes to thank S.~Laporta for a discussion on this point.} \cite{Lap}.
The linear system can be solved with the method of elimination
of the variables. In general, one has to devise a rule about which 
amplitudes in the system to solve first. Important quantities for the 
scalar integrals are the sum of the positive indices,
\begin{equation}
d = \sum_{i=1}^{7} n_{i} ~\theta(n_{i}) 
\end{equation}
and minus the sum of the negative indices,
\begin{equation}
s = \sum_{i=1}^{7} - n_{i} ~\theta(-n_{i}). 
\end{equation}
A convenient criterion is to solve for the amplitudes with the 
greatest $t$, $d$ and $s$ first. The advantage of the numerical indices 
method is that it does not require a hand inspection of the equations 
--- requiring patience and often a good mathematical intuition ---
and is therefore very general.
The main disadvantage is that it may require a large CPU time and 
very long intermediate expressions may be generated because of the 
division by the coefficients of the unknowns amplitudes.

Let us now discuss the ibp identities in the framework of the second 
reduction scheme discussed in the previous section, the one involving
independent scalar products at the numerator 
and loop momentum shifts. As with the auxiliary diagrams, the scalar 
integrals of Eq. (\ref{b1}) can be related by the ibp identities, which 
now read:
\be
\int \{ d^{D}k_{1} \} \{ d^{D}k_{2} \} \, \frac{\partial}{
\partial k_{i}^{\mu}} \left\{ v^{\mu} \, \frac{S_{i_1}^{s_{1}} \cdots 
S_{i_{\overline{t}}}^{s_{\overline{t}}}  }  
{{\mathcal D}_{i_1}^{n_{1}} \cdots {\mathcal D}_{i_t}^{n_{t}}}
\right\} = 0  \, ,
\label{b2} 
\ee
where, as before, $i=1,2$ and 
$v^{\mu} = k_{1}^{\mu},k_{2}^{\mu},p_{1}^{\mu},p_{2}^{\mu}$.
For a given set of indices $\{ s_i,n_j \}$ of the input 
amplitude, we find eight identities involving integrals with up to an
additional power on a denominator and an additional power of a scalar 
product at the numerator. Sub-topologies with $t-1$, $t-2$, ... coming
from simplifications between scalar products and corresponding
denominators also appear. As in the auxiliary diagram case, 
discrete symmetries of various topologies generate additional 
identities to be combined with the ibp identities \cite{Bon1}.

Concerning our problem, we can reduce the calculation of all the scalar 
integrals related to the topologies shown in Figs.~\ref{fig2}, 
\ref{fig3} and \ref{fig4} --- and therefore the calculation of the 
Feynman diagrams of Figs. \ref{fig1} and \ref{fig1bis} --- to that of 
the $22$ MI's in Fig. \ref{fig5}. Four of them are the product of
two one-loop MI's. The picture of the diagram alone 
denotes the scalar integral --- only denominators --- while the 
picture of the diagram accompanied by a scalar product denotes the 
scalar integral with that scalar product at the numerator.

%%%%%%%%%%%%%%%%%%%%%% Master Integrals %%%%%%%%%%%%%%%%%%%%%
\bfig
\bc
%%%%%%%%
\subfigure[]{
\begin{fmfgraph*}(20,20)
\fmfleft{i1,i2}
\fmfright{o}
\fmf{photon}{i1,v1}
\fmf{photon}{i2,v2}
\fmf{plain}{v5,o}
\fmf{photon,tension=.3}{v2,v3}
\fmf{photon,tension=.3}{v3,v5}
\fmf{photon,tension=.3}{v1,v4}
\fmf{photon,tension=.3}{v4,v5}
\fmf{photon,tension=0}{v2,v4}
\fmf{photon,tension=0}{v1,v3}
\end{fmfgraph*} }
%
%%%%%%%%%%%%%%%%
%
\subfigure[]{
\begin{fmfgraph*}(20,20)
\fmfleft{i1,i2}
\fmfright{o}
\fmf{photon}{i1,v1}
\fmf{photon}{i2,v2}
\fmf{plain}{v5,o}
\fmf{photon,tension=.3}{v2,v3}
\fmf{photon,tension=.3}{v3,v5}
\fmf{photon,tension=.3}{v1,v4}
\fmf{photon,tension=.3}{v4,v5}
\fmf{plain,tension=0}{v2,v4}
\fmf{photon,tension=0}{v1,v3}
\end{fmfgraph*} }
%
%%%%%%%%%%%%%%%%
%
\subfigure[]{
\begin{fmfgraph*}(20,20)
\fmfleft{i1,i2}
\fmfright{o}
\fmfforce{0.8w,0.5h}{v4}
\fmf{photon}{i1,v1}
\fmf{photon}{i2,v2}
\fmf{plain}{v4,o}
\fmf{photon,tension=.4}{v2,v3}
\fmf{photon,tension=.2}{v3,v4}
\fmf{photon,tension=.15}{v1,v4}
\fmf{photon,tension=0}{v2,v1}
\fmf{plain,tension=0}{v1,v3}
\end{fmfgraph*} }
%
%%%%%%%%%%%%%%%%%%%%%%%
%
\subfigure[]{
\begin{fmfgraph*}(20,20)
\fmfleft{i1,i2}
\fmfright{o}
\fmfforce{0.8w,0.5h}{v4}
\fmf{photon}{i1,v1}
\fmf{photon}{i2,v2}
\fmf{plain}{v4,o}
\fmf{photon,tension=.4}{v2,v3}
\fmf{photon,tension=.2}{v3,v4}
\fmf{photon,tension=.15}{v1,v4}
\fmf{plain,tension=0}{v2,v1}
\fmf{photon,tension=0}{v1,v3}
\end{fmfgraph*} }
%
%%%%%%%%%%%%%%%%%%%%%%%
%
\subfigure[]{
\begin{fmfgraph*}(20,20)
\fmfleft{i1,i2}
\fmfright{o}
\fmfforce{0.8w,0.5h}{v4}
\fmf{photon}{i1,v1}
\fmf{photon}{i2,v2}
\fmf{plain}{v4,o}
\fmf{plain,tension=.4}{v2,v3}
\fmf{photon,tension=.2}{v3,v4}
\fmf{photon,tension=.15}{v1,v4}
\fmf{photon,tension=0}{v2,v1}
\fmf{photon,tension=0}{v1,v3}
\end{fmfgraph*} } \\
%
%%%%%%%%%%%%%%%%%%%%%%%
%
\subfigure[]{
\begin{fmfgraph*}(20,20)
\fmfleft{i1,i2}
\fmfright{o}
\fmfforce{0.8w,0.5h}{v4}
\fmf{photon}{i1,v1}
\fmf{photon}{i2,v2}
\fmf{plain}{v4,o}
\fmflabel{$(k_{1} \cdot k_{2})$}{o}
\fmf{plain,tension=.4}{v2,v3}
\fmf{photon,tension=.2}{v3,v4}
\fmf{photon,tension=.15}{v1,v4}
\fmf{photon,tension=0}{v2,v1}
\fmf{photon,tension=0}{v1,v3}
\end{fmfgraph*} }
%
%%%%%%%%%%%%%%%%%%%%%%%
%
\hspace{20mm}
\subfigure[]{
\begin{fmfgraph*}(20,20)
\fmfleft{i1,i2}
\fmfright{o}
\fmfforce{0.2w,0.9h}{v2}
\fmfforce{0.2w,0.1h}{v1}
\fmfforce{0.2w,0.5h}{v3}
\fmfforce{0.8w,0.5h}{v4}
\fmf{photon}{i1,v1}
\fmf{photon}{i2,v2}
\fmf{plain}{v4,o}
\fmf{photon,tension=0}{v1,v3}
\fmf{photon,tension=0}{v3,v4}
\fmf{photon,tension=0}{v2,v4}
\fmf{plain,tension=0}{v2,v3}
\fmf{photon,tension=0}{v1,v4}
\end{fmfgraph*} }
%
%%%%%%%%%%%%%%%%%%%%%%%
%
\subfigure[]{
\begin{fmfgraph*}(20,20)
\fmfleft{i1,i2}
\fmfright{o}
\fmf{photon}{i1,v1}
\fmf{photon}{i2,v2}
\fmf{plain}{v4,o}
\fmf{photon,tension=.3}{v2,v3}
\fmf{photon,tension=.3}{v1,v3}
\fmf{plain,tension=0}{v2,v1}
\fmf{photon,tension=.2,left}{v3,v4}
\fmf{photon,tension=.2,right}{v3,v4}
\end{fmfgraph*} }
%
%%%%%%%%%%%%%%%%%%%%%%%
%
\subfigure[]{
\begin{fmfgraph*}(20,20)
\fmfforce{0.2w,0.5h}{v1}
\fmfforce{0.5w,0.8h}{v2}
\fmfforce{0.5w,0.2h}{v3}
\fmfforce{0.8w,0.5h}{v4}
\fmfleft{i}
\fmfright{o}
\fmf{plain}{i,v1}
\fmf{plain}{v4,o}
\fmf{photon,tension=.2,left=.4}{v1,v2}
\fmf{photon,tension=.2,right=.4}{v1,v3}
\fmf{photon,tension=.2,left=.4}{v2,v4}
\fmf{photon,tension=.2,right=.4}{v3,v4}
\fmf{plain,tension=0}{v2,v3}
\end{fmfgraph*} } \\
%
%%%%%%%%%%%%%%%%%%%%%%%%
%
\subfigure[]{
\begin{fmfgraph*}(20,20)
\fmfleft{i1,i2}
\fmfright{o}
\fmfforce{0.2w,0.9h}{v2}
\fmfforce{0.2w,0.1h}{v1}
\fmfforce{0.2w,0.55h}{v3}
\fmfforce{0.2w,0.15h}{v5}
\fmfforce{0.8w,0.5h}{v4}
\fmf{photon}{i1,v1}
\fmf{photon}{i2,v2}
\fmf{plain}{v4,o}
\fmf{plain}{v2,v3}
\fmf{photon,left}{v3,v5}
\fmf{photon,right}{v3,v5}
\fmf{photon}{v1,v4}
\fmf{photon}{v2,v4}
\end{fmfgraph*} }
%
%%%%%%%%%%%%%%%%%%%%%%%
%
\subfigure[]{
\begin{fmfgraph*}(20,20)
\fmfleft{i1,i2}
\fmfright{o}
\fmf{photon}{i1,v1}
\fmf{photon}{i2,v2}
\fmf{plain}{v3,o}
\fmf{photon,tension=.3}{v2,v3}
\fmf{photon,tension=.3}{v1,v3}
\fmf{plain,tension=0,right=.5}{v2,v1}
\fmf{photon,tension=0,right=.5}{v1,v2}
\end{fmfgraph*} }
%
%%%%%%%%%%%%%%%%%%%%%%%
%
\subfigure[]{
\begin{fmfgraph*}(20,20)
\fmfleft{i1,i2}
\fmfright{o}
\fmf{photon}{i1,v1}
\fmf{photon}{i2,v2}
\fmf{plain}{v3,o}
\fmflabel{$(p_{2} \cdot k_{1})$}{o}
\fmf{photon,tension=.3}{v2,v3}
\fmf{photon,tension=.3}{v1,v3}
\fmf{plain,tension=0,right=.5}{v2,v1}
\fmf{photon,tension=0,right=.5}{v1,v2}
\end{fmfgraph*} }
%
%%%%%%%%%%%%%%%%%%%%%%%
%
\hspace*{20mm}
\subfigure[]{
\begin{fmfgraph*}(20,20)
\fmfleft{i1,i2}
\fmfright{o}
\fmf{photon}{i1,v1}
\fmf{photon}{i2,v2}
\fmf{plain}{v3,o}
\fmf{photon,tension=.3}{v2,v3}
\fmf{photon,tension=.3}{v1,v3}
\fmf{photon,tension=0,right=.5}{v2,v1}
\fmf{photon,tension=0,right=.5}{v1,v2}
\end{fmfgraph*} } \\
%
%%%%%%%%%%%%%%%%%%%%%%%
%
\subfigure[]{
\begin{fmfgraph*}(20,20)
\fmfleft{i1,i2}
\fmfright{o}
\fmf{photon}{i1,v1}
\fmf{photon}{i2,v2}
\fmf{plain}{v3,o}
\fmf{photon,tension=.3}{v2,v3}
\fmf{photon,tension=.3}{v1,v3}
\fmf{plain,tension=0}{v2,v1}
\fmf{photon,tension=0,left=.5}{v1,v3}
\end{fmfgraph*} } 
%
%%%%%%%%%%%%%%%%%%%%%%%
%
\subfigure[]{
\begin{fmfgraph*}(20,20)
\fmfleft{i}
\fmfright{o}
\fmf{plain}{i,v1}
\fmf{plain}{v3,o}
\fmf{photon,tension=.2,left}{v1,v2}
\fmf{photon,tension=.2,right}{v1,v2}
\fmf{photon,tension=.2,left}{v2,v3}
\fmf{photon,tension=.2,right}{v2,v3}
\end{fmfgraph*} }
%
%%%%%%%%%%%%%%%%%%%%%%%
%
\subfigure[]{
\begin{fmfgraph*}(20,20)
\fmfleft{i}
\fmfright{o}
\fmf{plain}{i,v1}
\fmf{plain}{v3,o}
\fmf{plain,tension=.2,left}{v1,v2}
\fmf{photon,tension=.2,right}{v1,v2}
\fmf{photon,tension=.2,left}{v2,v3}
\fmf{photon,tension=.2,right}{v2,v3}
\end{fmfgraph*} } 
%
%%%%%%%%%%%%%%%%%%%%%%%
%
\subfigure[]{
\begin{fmfgraph*}(20,20)
\fmfleft{i}
\fmfright{o}
\fmfforce{0.2w,0.5h}{v1}
\fmfforce{0.5w,0.8h}{v2}
\fmfforce{0.8w,0.5h}{v3}
\fmf{plain}{i,v1}
\fmf{plain}{v3,o}
\fmf{photon,right}{v1,v3}
\fmf{plain,left=.4}{v1,v2}
\fmf{photon,left=.4}{v2,v3}
\fmf{photon,right=.6}{v2,v3}
\end{fmfgraph*} } 
%
%%%%%%%%%%%%%%%%%%%%%%%
%
\subfigure[]{
\begin{fmfgraph*}(20,20)
\fmfleft{i}
\fmfright{o}
\fmf{plain}{i,v1}
\fmf{plain}{v2,o}
\fmf{photon,tension=.15,left}{v1,v2}
\fmf{photon,tension=.15}{v1,v2}
\fmf{photon,tension=.15,right}{v1,v2}
\end{fmfgraph*} } \\
%
%%%%%%%%%%%%%%%%%%%%%%%
%
\subfigure[]{
\begin{fmfgraph*}(20,20)
\fmfleft{i}
\fmfright{o}
\fmf{plain}{i,v1}
\fmf{plain}{v2,o}
\fmf{plain,tension=.15,left}{v1,v2}
\fmf{photon,tension=.15}{v1,v2}
\fmf{photon,tension=.15,right}{v1,v2}
\end{fmfgraph*} } 
%
%%%%%%%%%%%%%%%%%%%%%%%
%
\subfigure[]{
\begin{fmfgraph*}(20,20)
\fmfleft{i}
\fmfright{o}
\fmf{plain}{i,v1}
\fmf{plain}{v2,o}
\fmflabel{$(k_{1} \cdot k_{2})$}{o}
\fmf{plain,tension=.15,left}{v1,v2}
\fmf{photon,tension=.15}{v1,v2}
\fmf{photon,tension=.15,right}{v1,v2}
\end{fmfgraph*} }
%
%%%%%%%%%%%%%%%%%%%%%%%
%
\hspace*{20mm}
\subfigure[]{
\begin{fmfgraph*}(20,20)
\fmfleft{i}
\fmfright{o}
\fmf{photon}{i,v1}
\fmf{photon}{v2,o}
\fmf{plain,tension=.15,left}{v1,v2}
\fmf{photon,tension=.15}{v1,v2}
\fmf{photon,tension=.15,right}{v1,v2}
\end{fmfgraph*} } 
%
%%%%%%%%%%%%%%%%%%%%%%%
%
\subfigure[]{
\begin{fmfgraph*}(20,20)
\fmfleft{i}
\fmfright{o}
\fmf{plain}{i,v1}
\fmf{plain}{v2,o}
\fmf{photon,tension=.22,left}{v1,v2}
\fmf{photon,tension=.22,right}{v1,v2}
\fmf{plain,right=45}{v2,v2}
\end{fmfgraph*} } 
%
%%%%%%%%%%%%%%%%%%%%%%%
%
%%%%%%%%%%%%%%%%%%%%%%%%
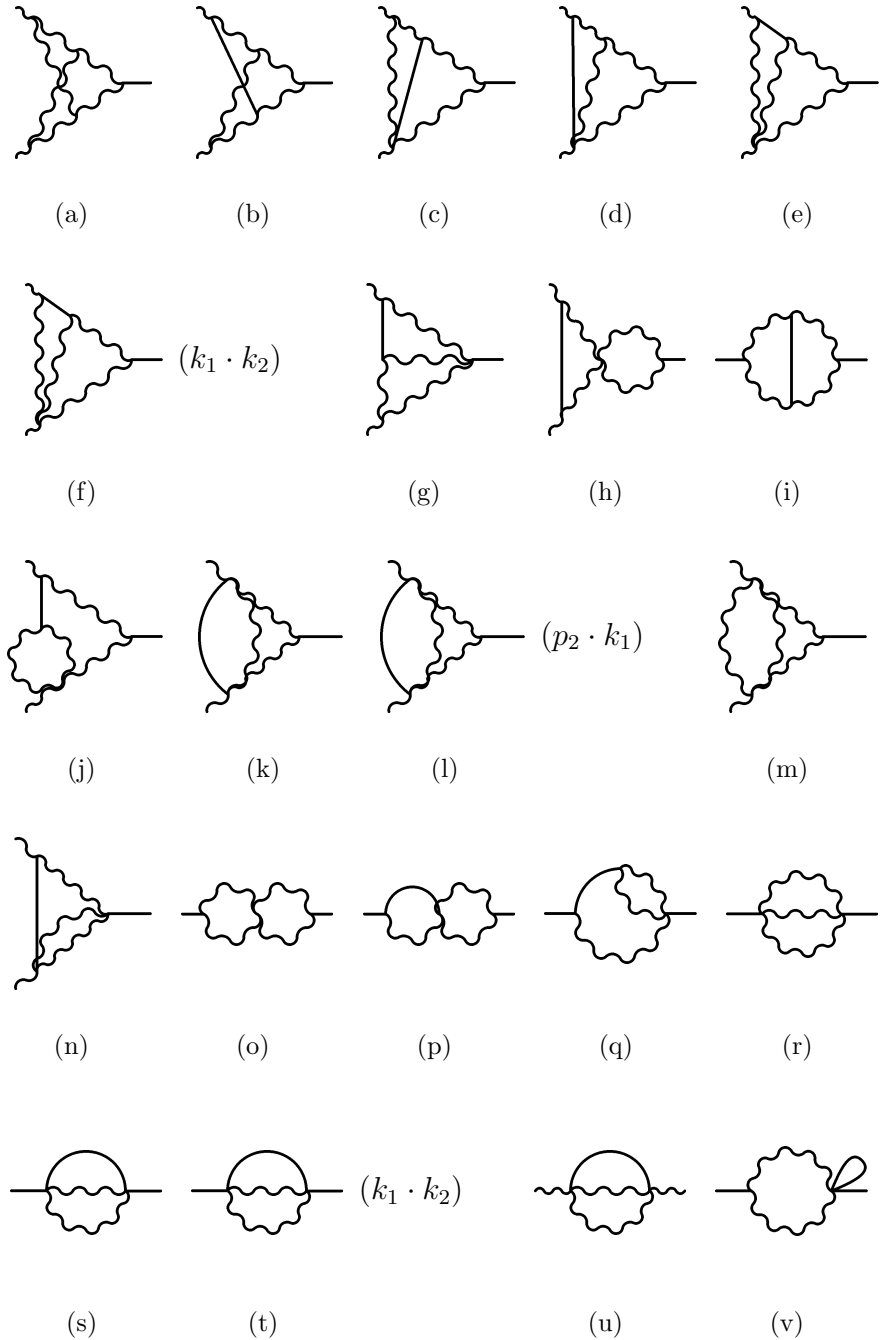
\caption{\label{fig5} The set of 22 Master Integrals. Four of them are 
the product of one-loop master integrals.}
\ec
\efig
%%%%%%%%%%%%%%%%%%%%%%%%%%%%%%%%%%%%%%%%%%%%%%%%%%%%%%%%%%%%%%%%%%%%%

The algorithms explained in this section were implemented in a 
chain of programs written in the algebraic computer language {\tt FORM}
\cite{FORM} and, in the case of the auxiliary-denominator scheme,
also in {\tt Mathematica} \cite{Mathe}. For the solution of the
linear system of ibp identities, we used {\tt SOLVE} \cite{SOLVE}.
This program uses the language {\tt C} as a shell to manage 
a user-provided basic code written in {\tt FORM}. 
External calls to {\tt Maple} \cite{Maple} are done
in order to simplify the coefficients coming in intermediate steps.

\section{Evaluation of the MI's: the differential equations method
\label{diffeqs}}

Let us consider a general amplitude 
$\mathrm{Topo}\left( n_{1},n_{2},n_{3},n_{4},n_{5},n_{6};n_{7} \right)$
in the framework of the auxiliary diagram. As shown in 
the previous section, it can be decomposed, by means of the ibp 
identities, into a superposition of a set of primitive
amplitudes called master integrals (MI's). 
The latter cannot be computed with the
ibp identities, which, in general, only allow for a reduction of the number
of independent amplitudes to be individually computed.
The MI's have been computed with a variety of techniques over the years:
Feynman parameters, dispersion relations, small momentum expansions, large
momentum expansions, just to mention the more common ones. A technique
developed over the last few years which is very efficient in our case is
that of the differential equations in the external kinematical
invariants \cite{Kotikov1,Kotikov2,Kotikov3,Rem1,Rem2,Rem3,Bon1}. Let 
us then review the basics of this method. Once again, let us consider a
specific example: the computation of the MI's in Fig. \ref{fig5} (k) in 
the auxiliary diagram scheme. Let us define:
\begin{eqnarray}
F_{1}(s,a, \epsilon) & = & \mathrm{Topo}(1,0,0,1,1,1;~0) , \\
F_{2}(s,a, \epsilon) & = & \mathrm{Topo}(1,0,0,1,1,1;-1) ,
\end{eqnarray}
The above two MI's appear in the decomposition of the six-denominator 
amplitude presented in the previous section (Fig. \ref{fig1} (b)). They
are functions of $s$ and $a$, as well as of the space-time
dimension $D=4-2\epsilon$. Let us take the derivative of $F_i$ with 
respect to $s$ at fixed $a$ and with respect to $a$ at fixed $s$: 
\bea
& & s \frac{\partial }{\partial s} F_{i} ( s,a, \epsilon) 
= p_{1}^{\mu } \frac{\partial }{\partial p_{1}^{\mu }}
F_{i}(s,a,\epsilon) = 
p_{2}^{\mu }\frac{\partial }{\partial p_{2}^{\mu }}F_{i}(s,a,\epsilon) 
\, , 
\label{p1deriv} \\ 
& & a \frac{\partial }{\partial a}F_{i}(s,a,\epsilon) \, .
\eea
The derivatives with respect to the external momentum component 
$p_{1}^{\mu }$ or $p_{2}^{\mu }$ and with respect to the mass squared 
$a$ can be taken inside the loop integral, because of the properties of
Dimensional Regularization. The resulting amplitudes can be reduced to 
combinations of MI's according to the methods described in the previous 
section, so that:
\begin{eqnarray}
s \frac{\partial }{\partial s} F_{i}(s,a,\epsilon) & = & 
\sum_{j}C_{ij}(s,a,\epsilon) \, F_{j}(s,a,\epsilon) , 
\label{eqdiff1} \\
a\frac{\partial }{\partial a}F_{i}(s,a,\epsilon) & = & 
\sum_{j}K_{ij}(s,a,\epsilon) \, F_{j}(s,a,\epsilon).
\label{eqdiff2} 
\end{eqnarray}
Note the double role played by the ibp identities in the calculation: they
allow for the reduction of independent scalar amplitudes into combinations of master
integrals and they also allow for the generation of the differential
equations to be solved for the MI's themselves.

Eqs.~(\ref{eqdiff1}) and (\ref{eqdiff2}) can be simplified changing variables 
from $s$ and $a$ to $x=-s/a$ and $a.$  This triangular change of 
variables implies that: 
\begin{eqnarray}
x \frac{\partial }{\partial x}F_{i} (x,a,\epsilon) & = & 
s \frac{\partial }{\partial s} F_{i}(s,a,\epsilon) \, , \\
a \frac{\partial }{\partial a}F_{i}(x,a,\epsilon)  & = & 
s\frac{\partial }{\partial s} F_{i}(s,a,\epsilon) + a 
\frac{\partial }{\partial a} F_{i}(s,a,\epsilon) = 
\frac{d_{i}}{2} \, F_{i}(x,a,\epsilon) .
\end{eqnarray}
The second equation holds because of the dimensional relation 
\be
F_{i} (x,a,\epsilon) = a^{d_{i}/2} f_{i} (x,\epsilon) ,
\ee
where $d_{i}$ is the mass dimension of the MI: 
$d_{i}/2=D-\sum_{k=1}^{7}n_{k}^{\left( i\right) }.$ 
In the new variables, the scale $a$ represents an over-all scale of the
amplitudes and its evolution equation is consequently trivial. We 
therefore concentrate on the $x$-evolution equation only and we set $a=1$
from now on in this section: 
\be
\frac{\partial }{\partial x}F_{i} (x,\epsilon)
=\sum_{j=1}^{N_{F}} A_{ij} ( x,\epsilon ) \, F_{j} (x,\epsilon) .
\ee
In our case, by performing the above steps, one obtains the following 
linear system of coupled ordinary differential equations with variable 
coefficients:
\bea
\frac{dF_{1}}{dx}(x,\epsilon) & = &-\frac{1}{x}F_{1}(x,\epsilon)- 
\left( 2-3\epsilon \right) \left( \frac{1}{x}+\frac{1}{1-x}\right) 
F_{2}(x,\epsilon) \nn\\
& & - \left( 1-\epsilon \right) \left( \frac{1}{x}+\frac{1}{1-x}\right) 
\left[ F_{3}(\epsilon)+F_{4}(x,\epsilon) \right] \, , 
\label{sys1} \\
\frac{dF_{2}}{dx}(x,\epsilon) & = & - \left[ ( 1-\epsilon ) \frac{1}{x}
+ ( 2-3\epsilon ) \frac{1}{1-x} \right] F_{2}(x,\epsilon) \nn\\
& & - ( 1-\epsilon ) \left( \frac{1}{x} + \frac{1}{1-x} \right) 
F_{3}(\epsilon) - ( 1-\epsilon ) \frac{1}{1-x} F_{4}(x,\epsilon) \, ,
\label{sys2}
\eea
where we defined:
\begin{eqnarray}
F_3(\epsilon) & = & \mathrm{Topo}(1,0,0,1,1,0;~0)  \, , \\
F_4(x,\epsilon) & = & \mathrm{Topo}(1,0,0,0,1,1;0).  
\end{eqnarray}
$F_{3}$ is represented in Fig. \ref{fig5} (u); 
it is a sunrise diagram with one massive line, 
evaluated at a light-cone external 
momentum, equivalent to the null momentum. $F_{3}$ is then effectively
a vacuum  amplitude, so we dropped the dependence on $x$. It is obtained
from Fig. \ref{fig5} (k) by shrinking one of the internal lines having 
an endpoint in the hard vertex. $F_{4}$ is represented in Fig. 
\ref{fig5} (v) and is the product of a massless bubble and a tadpole; 
it is obtained
from Fig. \ref{fig5} (k) by shrinking the massless line in the bubble. 
Note that there are no other non vanishing sub-topologies. We consider 
the sub-topologies $F_3$ and $F_4$ as known amplitudes\footnote{Their
expressions are given respectively in Eqs. (\ref{effe31}--\ref{effe35}) 
(where we put $F_{3}=F^{(4)}$) ans Eqs. (\ref{effe41}--\ref{effe45}) 
(where $F_{4}=F^{(5)}$).}, so that the system of Eqs. (\ref{sys1}) and 
(\ref{sys2}) is schematically written as:
\begin{equation}
\frac{\partial }{\partial x}F_{i} ( x,\epsilon )
=\sum_{j=1}^{n_F} A_{ij} ( x,\epsilon ) \, F_{j} ( x,\epsilon ) 
+ \Omega _{i} ( x;\epsilon ) ,  
\label{nonomog}
\end{equation}
where the $\Omega _{i}( x,\epsilon)$ are known functions constituting 
the non-homogeneous part of the system. Let us note that the 
coefficients of the homogeneous terms have a regular expansion around 
$\epsilon=0$, involving, in this case, only two terms:
\begin{equation} 
A_{ij}(x,\epsilon) =A_{ij}^{(0)} (x) + \epsilon A_{ij}^{(1)} (x) \, .
\end{equation}
A great simplification is offered by the fact that the system is 
triangular in our basis, because the equation for 
$dF_2/dx$ depends only on $F_2$ and not on $F_1$: 
$A_{21}(x,\epsilon)=0$. Let us then consider this equation first. We 
substitute the explicit expression for the expansion of $\Omega_2$ 
in powers of $\epsilon$ and the symbolic expansion for $F_2$, up to the
desired order $n$:
\bea
\Omega_{2}(x,\epsilon) & = & \sum_{j=-2}^{n} \epsilon^j 
\Omega_{2}^{(j)}
(x) + {\mathcal O} \left( \epsilon^{n+1} \right) \, , \\
F_{2}(x,\epsilon) & = & \sum_{j=-2}^{n} \epsilon^j F_{2}^{(j)}
(x) + {\mathcal O} \left( \epsilon^{n+1} \right) \, .
\eea

The double pole --- the leading singularity --- obeys a differential 
equation of the form
\begin{equation}
\frac{dF_2^{(-2)}}{dx}(x)  =  A_{22}^{(0)}(x)  F_2^{(-2)}(x) +
\Omega^{(-2)}(x) 
\end{equation}
where
\begin{equation}
A_{22}^{(0)}(x) = -\frac{1}{x}-\frac{2}{1-x} \, .
\end{equation}
The associated homogeneous equation,
\begin{equation}
\frac{ d{\omega}_{2}}{dx}(x)  =  A_{22}^{(0)}(x) \, {\omega}_2 (x)
\end{equation}
is solved by separation of variables to give:
\begin{equation}
\omega_2(x) = \exp\left[ \int^{x} A_{22}^{(0)}(t) dt \right] = 
\frac{(1-x)^2}{x},
\end{equation}
with an overall constant omitted in the last member.
The double pole coefficient is then computed with the method of 
of the variation of the constants:
\begin{equation}
F_2^{(-2)}(x)= \omega_2(x)  \int^{x} K_2(t)  \Omega^{(-2)}(t) dt  \, , 
\end{equation}
where the primitive involves an arbitrary constant 
to be fixed imposing
initial conditions, and where the kernel $K_2$ is given by
\begin{equation}
K_2(t)=\frac{1}{\omega_2(t)}=-\frac{1}{1-t}+\frac{1}{(1-t)^2} \, .
\end{equation}

Higher orders are computed through the recursive relation in $\epsilon$:
\begin{equation}
\frac{dF_2^{(i)}(x)}{dx} = A_{22}^{(0)}(x) F_2^{(i)}(x) 
+ A_{22}^{(1)}(x) F_2^{(i-1)}(x) +\Omega^{(i)}(x). 
\end{equation}
The solution of the above equation is again obtained with the method
of variation of constants:
\begin{equation}
F_2^{(i)}(x) = \omega_2(x)  \int^{x} K_2(t)  \tilde{\Omega}^{(i)}(t) dt  \, , 
\end{equation}
where 
\begin{equation}
\tilde{\Omega}^{(i)}(t) = \Omega^{(i)}(t) + A_{22}^{(1)}(t) F_2^{(i-1)}(t)
\end{equation}
is the inhomegeneous term redefined to include the amplitude of previous order 
in the $\epsilon$ expansion. Note that the kernel $K_2$ is the same for each order 
in $\epsilon$.

The full evaluation of $F_2$ requires, as we already said, an initial 
condition on the $x$ evolution.
The latter can often be derived by studying the behaviour of the MI 
close to a threshold or a pseudothreshold. In our case, let us consider 
the point $x=1$: it is a pseudothreshold as it corresponds to $s=-m^2<0$
and therefore is not a singular point for the amplitude, 
so that:
\begin{equation}
\lim_{x \rightarrow 1} (1-x) \frac{dF_2(x,\epsilon)}{dx} = 0.
\end{equation}
By imposing the above condition to the differential equation, an 
initial value for the amplitude in $x=1$ is obtained:
\begin{equation}
(2-3\epsilon)F_2(1;\epsilon) +
(1-\epsilon)F_3(\epsilon)+(1-\epsilon)F_4(1;\epsilon)=0. 
\end{equation}
Once $F_{2}$ has been explicitly computed, one replaces the result in 
the first equation and solves it for $F_1$ in the same way. 

In general, however, the system is not triangular and the above method 
is not directly applicable. One can try a transformation to a new set 
of MI's,
\begin{equation}
F_1,~~F_2~\rightarrow~F_{1}',~~F_{2}',  
\end{equation}
in order to have the system in a triangular form. A triangularization 
to order $\epsilon$ is actually sufficient, i.e. one for which 
$A_{21}^{(0)}=0$ but $A_{21}^{(1)} \ne 0$\footnote{In more complicated 
cases, it may happen that a triangular form of the system  
cannot be reached, and the coupled system of $n_{F}$ equations
must be solved. This is equivalent to the solution of a
differential equation of order $n_F$ for anyone of the MI's, 
for which no general algorithm exists. In our problem, the number
of MI's for a given topology is at most $n_{F}=2$, corresponding to a
second-order differential equation. Moreover, it turned out that the
solution of the associated homogeneous system was rather simple,
so that the complete solution could be found with the method of the
variation of arbitrary constants.}.

The MI's can be expressed in terms of the one-dimensional harmonic 
polylogarithms (HPL's), 
the latter having the following recursive integral definition
\begin{eqnarray}
H(1;x)&=&-\log(1-x)  \\
H(0,x)&=&\log(x)    \\ 
H(-1;x)&=&\log(1+x)
\end{eqnarray}
and
\begin{equation}
H(a,\vec{w};x)=\int_{0}^{x} f(a;y) H(\vec{w};y) dy,
\end{equation}
where the basic weight functions are defined as:
\begin{eqnarray}
f(1;x)  & = &\frac{1}{1-x}  \\ 
f(0;x)  & = &\frac{1}{x}    \\
f(-1;x) & = &\frac{1}{1+x}.
\end{eqnarray}

For a complete discussion of HPL's and their properties we refer to
\cite{Polylog}; for their numerical evaluation see \cite{Polylog3}.

As an "empirical" rule, we found that it was sufficient to stop the
$\epsilon$ expansion beyond the order involving weight $w=4$ HPL's.

The massless MI's cannot be computed with the differential equation method
and a more traditional method as that of Feynman parameters or
dispersion relations has to be used.
As discussed in the introduction, they are indeed of the form:
\begin{equation}
{\rm (massless~amp.)}=
a^{4-n_d}\left(\frac{\mu^2}{a}\right)^{2\epsilon}G(\epsilon) 
x^{4-n_d-2\epsilon}
\end{equation}
Since the $x$-dependence is factorized, the $x$ evolution gives 
only a trivial, dimensional, information, analogous to the $a$ evolution. 
The non-trivial quantity to compute is $G(\epsilon)$. In other words,
all the information is contained in the initial data --- for example the 
amplitude at $x=1$ --- that can be extracted only by
direct integration.

For a graphical summary of the whole method
discussed in sections \ref{Master} and \ref{diffeqs},
see the flowchart in Fig. \ref{fig6}.
\bfig
\bc
\begin{picture}(386,170)(-6,0)
\SetWidth{1.5}
\SetPFont{Helvetica}{12}
\GBox(0,420)(150,460){.8}
\PText(18,443)(0)[l]{Tensorial Amplitude}
\GBox(0,340)(150,380){.8}
%\PText(38,483)(0)[l]{Independent}
\PText(22,363)(0)[l]{Scalar Amplitudes}
\GBox(0,240)(150,300){.8}
\PText(38,283)(0)[l]{Independent}
\PText(22,263)(0)[l]{Scalar Amplitudes}
\GBox(0,160)(150,200){.5}
\PText(30,183)(0)[l]{Master Integrals}
\GBox(0,80)(150,120){.8}
\PText(18,103)(0)[l]{System of Diff. Eqs.}
%\PText(30,203)(0)[l]{Eqs. for the MI's}
%
\GBox(0,0)(255,40){.7}
\PText(5,23)(0)[l]{Solution of the System in Laurent series of}
\Text(85,7.2)[r]{${\Large \epsilon}$}
\ArrowLine(95,420)(95,380)
\ArrowLine(95,340)(95,300)
\ArrowLine(95,240)(95,200)
\ArrowLine(95,160)(95,120)
\ArrowLine(95,80)(95,40)
\SetWidth{0.5}
\SetPFont{Helvetica}{10}
\DashLine(100,400)(220,400){5}
\DashLine(100,320)(220,320){5}
\DashLine(100,220)(220,220){5}
\DashLine(100,140)(220,140){5}
\GBox(220,385)(390,415){1}
\PText(226,403)(0)[l]{Projection on invariant form factors}
\GBox(220,305)(390,335){1}
\PText(245,323)(0)[l]{Auxiliary diagram or shifts}
\GBox(220,205)(390,235){1}
\PText(223,223)(0)[l]{IBP's and general symmetry relations}
\GBox(220,125)(390,155){1}
\PText(230,143)(0)[l]{Generation of Diff. Eqs. (+ IBP's)}
\end{picture}
\ec
\vspace*{5mm}
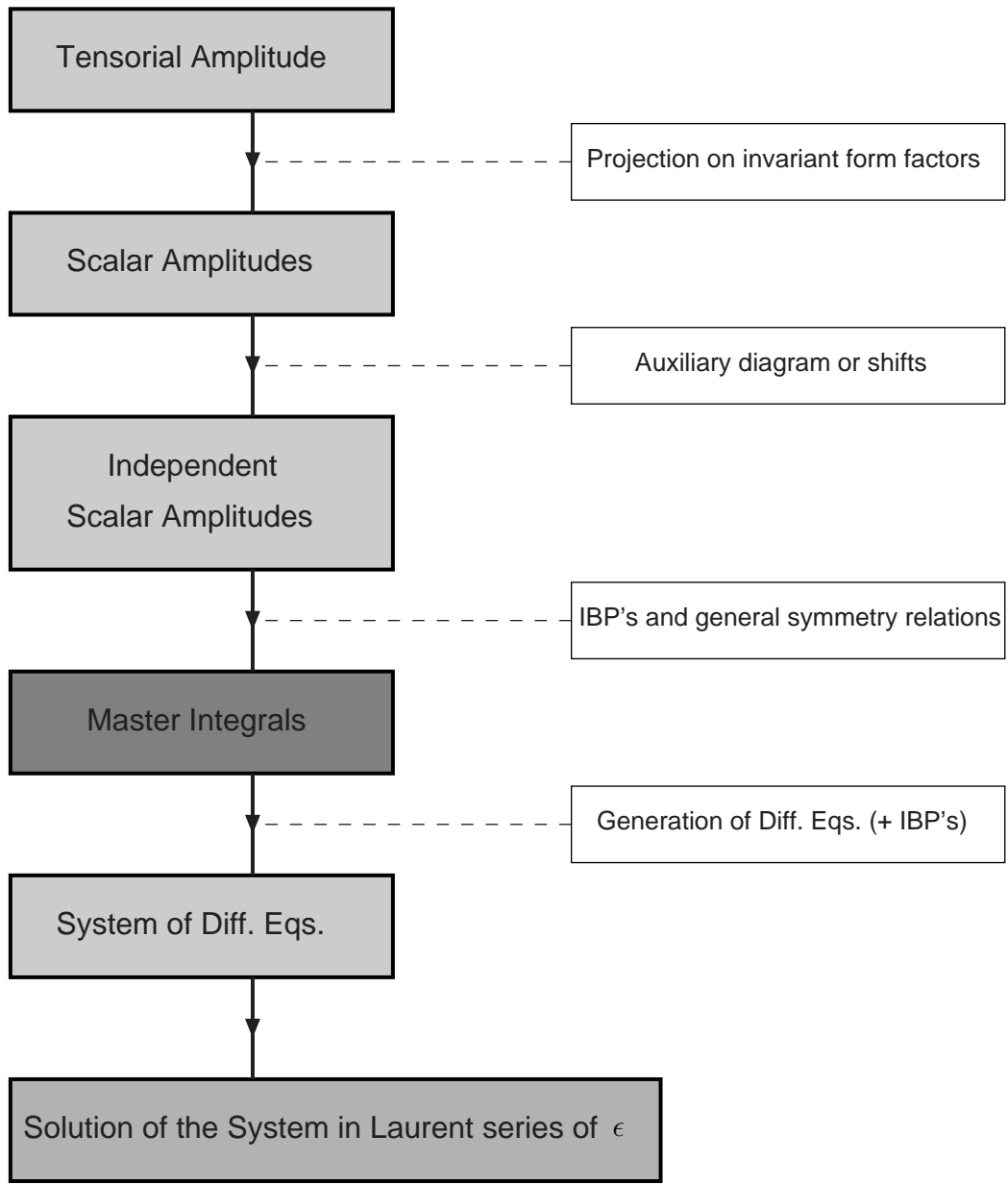
\caption{\label{fig6} Flowchart of the method used for the reduction 
to the MI's and their evaluation.}
\efig
\end{fmffile}
\begin{fmffile}{UnaMassa2}

\newcommand{\be}{\begin{equation}}
\newcommand{\ee}{\end{equation}}
\newcommand{\nn}{\nonumber}
\newcommand{\bea}{\begin{eqnarray}}
\newcommand{\eea}{\end{eqnarray}}
\newcommand{\bfig}{\begin{figure}}
\newcommand{\efig}{\end{figure}}
\newcommand{\bc}{\begin{center}}
\newcommand{\ec}{\end{center}}

\section{Results for the master integrals \label{Results}}

In this section we present the results of our computation of the MI's 
involving up to five denominators included, which constitute a  
necessary input for the calculation of the two-loop vertex diagrams 
in Figs.~\ref{fig1} and \ref{fig1bis}. They are expanded in a Laurent 
series in 
\begin{equation}
\epsilon = 2-D/2, 
\end{equation}
up to the required order in $\epsilon$. The coefficients of the series 
are expressed in terms of HPL's \cite{Polylog,Polylog3} of the variable 
$x$, defined as :
\be
x = \frac{q^{2}}{a} \, ,
\ee
where $ q = p_{1}+p_{2} $, and\footnote
{Note that with our definition of the scalar product $s=-q^2$.} $ a = m^2$ . 
We denote by $\mu$ the mass scale of the Dimensional Regularization (DR)
--- the so-called unit of mass. The explicit values of the denominators 
${\mathcal D}_{i}$ are given in appendix \ref{app1}.
We work in Minkowski space and we normalize the loop measure as:
\be
\left\{ d^{D} k \right\} = \frac{d^{D}k}{i\pi^{\frac{D}{2}} \Gamma 
\left( 3 - \frac{D}{2} \right) } = \frac{d^{(4-2\epsilon)}k}{i
\pi^{(2-\epsilon)} \Gamma 
\left( 1+\epsilon \right) }  .
\ee
This definition makes the expression of the one-loop tadpole 
--- the simplest of all loop diagrams --- particularly simple 
(see appendix \ref{app2}).

\subsection{Topology $t = 3$ \label{3den}}

\bea
\parbox{20mm}{\begin{fmfgraph*}(15,15)
\fmfleft{i}
\fmfright{o}
\fmf{plain}{i,v1}
\fmf{plain}{v2,o}
\fmf{photon,tension=.15,left}{v1,v2}
\fmf{photon,tension=.15}{v1,v2}
\fmf{photon,tension=.15,right}{v1,v2}
\end{fmfgraph*} }  & = & \mu^{2(4-D)} 
\int \{ d^{D}k_{1} \} \{ d^{D}k_{2} \}
\frac{1}{{\mathcal D}_{1} {\mathcal D}_{2} {\mathcal D}_{11} } \\
& = & \left( \frac{\mu^{2}}{a} \right) ^{2 \epsilon} 
\sum_{i=-2}^{2} \epsilon^{i} F^{(1)}_{i} + {\mathcal O} \left( 
\epsilon^{3} \right) , 
\eea
where:
\bea
\frac{F^{(1)}_{-2}}{a} & = & 0 \, , \\
\frac{F^{(1)}_{-1}}{a} & = & - \frac{x}{4} \, , \\
\frac{F^{(1)}_{0}}{a} & = & - x \biggl[ \frac{13}{8} - \frac{1}{2} H(0,x)
\biggr] \, , \\
\frac{F^{(1)}_{1}}{a} & = & - \frac{x}{4} \biggl[ \frac{115}{4} - 2
\zeta(2) - 13 H(0,x) + 4 H(0,0,x) \biggr] \, , \\
\frac{F^{(1)}_{2}}{a} & = & - \frac{x}{2} \biggl[ \frac{865}{16} - 
\frac{13 \zeta(2)}{2} - 5 \zeta(3) - \left( \frac{115}{4} - 2 \zeta(2) 
\right) H(0,x) \nn\\
& & + 13 H(0,0,x) - 4 H(0,0,0,x)  \biggr] \, .
\eea
The above amplitude is easily computed with Feynman parameters or with 
an iteration of the general formula for a massless bubble given in \cite{Chet}.
\bea
\parbox{23mm}{\begin{fmfgraph*}(15,15)
\fmfleft{i}
\fmfright{o}
\fmf{plain}{i,v1}
\fmf{plain}{v2,o}
\fmf{plain,tension=.15,left}{v1,v2}
\fmf{photon,tension=.15}{v1,v2}
\fmf{photon,tension=.15,right}{v1,v2}
\end{fmfgraph*} }  \quad \quad \quad & = & \mu^{2(4-D)} 
\int \{ d^{D}k_{1} \} \{ d^{D}k_{2} \}
\frac{1}{{\mathcal D}_{2} {\mathcal D}_{11} {\mathcal D}_{12} } \\
& = & \left( \frac{\mu^{2}}{a} \right) ^{2 \epsilon} 
\sum_{i=-2}^{2} \epsilon^{i} F^{(2)}_{i} + {\mathcal O} \left( 
\epsilon^{3} 
\right) , \\
\parbox{35mm}{\begin{fmfgraph*}(15,15)
\fmfleft{i}
\fmfright{o}
\fmf{plain}{i,v1}
\fmf{plain}{v2,o}
\fmflabel{$(k_{1} \cdot k_{2})$}{o}
\fmf{plain,tension=.15,left}{v1,v2}
\fmf{photon,tension=.15}{v1,v2}
\fmf{photon,tension=.15,right}{v1,v2}
\end{fmfgraph*} }  & = & \mu^{2(4-D)} 
\int \{ d^{D}k_{1} \} \{ d^{D}k_{2} \}
\frac{k_{1} \cdot k_{2}}{{\mathcal D}_{2} {\mathcal D}_{11} {\mathcal D}_{12} } \\
& = & \left( \frac{\mu^{2}}{a} \right) ^{2 \epsilon} 
\sum_{i=-2}^{2} \epsilon^{i} F^{(3)}_{i} + {\mathcal O} \left( 
\epsilon^{3} 
\right) , 
\eea
where:
\bea
\frac{F^{(2)}_{-2}}{a} & = & - \frac{1}{2} \, , \\
\frac{F^{(2)}_{-1}}{a} & = & - \frac{3}{2} - \frac{x}{4} \, , \\
\frac{F^{(2)}_{0}}{a} & = & - 3 - \zeta(2) - \frac{13}{8} x 
- \frac{1}{2} \left[ \frac{1}{x} - x \right] H(-1,x) + H(0,-1,x) \, , \\
\frac{F^{(2)}_{1}}{a} & = & - \frac{15}{4} \! - \! 3 \zeta(2) \! +
\! \zeta(3) \! - \! \frac{x}{16} \bigl[ 115 \! + \! 8 
\zeta(2) \bigr] - \frac{1}{4} \left( \frac{1}{x} - x \right) \bigl[ 
13 H(-1,x) \nn\\
& & - 8 H(-1,-1,x) \bigr] + \frac{1}{2} \left[ 6 + 2x - \frac{1}{x}
\right] H(0,-1,x) \nn\\
& & - 4 H(0,-1,-1,x) + H(0,0,-1,x) \, , \\
\frac{F^{(2)}_{2}}{a} & = & \frac{21}{8} - 6 \zeta(2) - 
\frac{9 \zeta^{2}(2)}{5} + 3 \zeta(3)
- \left[ \frac{865}{32} + \frac{13 \zeta(2)}{4} - 
\frac{\zeta(3)}{2} \right] x \nn\\
& & - \frac{1}{8} \bigl[ 115 + 8 \zeta(2) \bigr] \left[ \frac{1}{x} -x
\right] H(-1,x) + 13 \left[ \frac{1}{x} -x \right] H(-1,-1,x)
\nn\\
& & \frac{1}{4} \biggl[ 24 + 8 \zeta(2) + 26 x - \frac{13}{x} 
\biggr] H(0,-1,x) - \left[ \frac{1}{x} -x \right] \times
\nn\\
& & \times \bigl[ 8 H(-1,-1,-1,x)- 3 H(-1,0,-1,x) \bigr] +
\biggl[ 3 + x - \frac{1}{2x} \biggr] \times \nn\\
& & \times \bigl[ H(0,0,-1,x) - 4 H(0,-1,-1,x)
\bigr] + 16 H(0,-1,-1,-1,x) \nn\\
& & - \! 6 H(0,-1,0,-1,x) \! - \! 4 H(0,0,-1,-1,x)
\nn\\
& & + H(0,0,0,-1,x) \bigr] \, , \\
%%%%%%
%%%%%%
%%%%%%
\frac{F^{(3)}_{-2}}{a^{2}} & = & - \frac{1}{4} \\
\frac{F^{(3)}_{-1}}{a^{2}} & = & - \frac{3}{4} - \frac{x}{4} -
\frac{x^{2}}{24} \, , \\
\frac{F^{(3)}_{0}}{a^{2}} & = & - \frac{19}{12} - \frac{\zeta(2)}{2} -
\frac{35x}{24} - \frac{13x^{2}}{48} + \frac{1}{12} \left[ 3 - 
\frac{2}{x} + 6x + x^{2} \right] H(-1,x) \nn\\
& & + \frac{1}{2} H(0,-1,x) \, , \\
\frac{F^{(3)}_{1}}{a^{2}} & = & - \frac{5}{2} - \frac{3 \zeta(2)}{2} + 
\frac{\zeta(3)}{2} - \left[ \frac{95}{16} + \frac{\zeta(2)}{2} 
\right] x - \biggl[ \frac{115}{96} + 
\frac{\zeta(2)}{12} \biggr] x^{2} \nn\\
& & + \biggl[ \frac{35}{24} + \frac{37x}{12} + \frac{13x^{2}}{24} - 
\frac{13}{12x} \biggr] H(-1,x) + \biggl[ \frac{7}{4} + x +
\frac{x^{2}}{6} \nn\\
& & - \frac{1}{6x} \biggr] H(0,-1,x) - \biggl[ 1 + 2x + \frac{x^{2}}{3} 
- \frac{2}{3x}  \biggr] H(-1,-1,x) 
\nn\\
& & - 2 H(0,-1,-1,x) + \frac{1}{2} H(0,0,-1,x) \, , \\
\frac{F^{(3)}_{2}}{a^{2}} & = & - \frac{41}{24} \! - \frac{19 \zeta(2)}{6}
- \frac{9 \zeta^{2}(2)}{10} + \frac{3 \zeta(3)}{2} - \! 
\frac{1}{12} \! \biggl[ 35 \zeta(2) \! - \! 6 \zeta(3) \nn\\
& & + \! \frac{2015}{8} \biggr] x - \frac{1}{24} \left[ 13 \zeta(2) 
- 2 \zeta(3) + \frac{865}{8} \right] x^{2} + \frac{1}{2} \biggl[ 
\frac{95}{8} + \zeta(2) \nn\\
& & + \biggl( \frac{105}{4} + 2 \zeta(2) \biggr) x 
+ \biggl( \frac{115}{24} +  \frac{\zeta(2)}{3} \biggr) x^{2} - 
\left( \frac{115}{12} +  
\frac{2 \zeta(2)}{3} \right) \frac{1}{x} \biggr] \times \nn\\
& & \times H(-1,x) + \biggl[ \frac{37}{8} + \zeta(2) + 6 x + 
\frac{13 x^{2}}{12} - \frac{13}{12 x} \biggr] H(0,-1,x) \nn\\
& & - \frac{1}{6} \biggl[ 35 + 74x + 13 x^{2} - \frac{26}{x} \biggr] 
H(-1,-1,x) - \frac{1}{6} \biggl[ 3 + 6x \nn\\
& & + x^{2} - \frac{2}{x} \biggr] \bigl[ 3 H(-1,0,-1,x) 
-8 H(-1,-1,-1,x) \bigr] \nn\\
& & + \frac{1}{12} \biggl[ 21 + 12x + 2 x^{2} - \frac{2}{x} \biggr] 
\bigl[ H(0,0,-1,x) - 4 H(0,-1,-1,x) \bigr] \nn\\
& & + \frac{1}{2} \bigl[ H(0,0,0,-1,x) - 6H(0,-1,0,-1,x) 
- 4 H(0,0,-1,-1,x) \nn\\
& & + 16 H(0,-1,-1,-1,x) \bigr]
\, .
\eea

\bea
\parbox{20mm}{\begin{fmfgraph*}(15,15)
\fmfleft{i}
\fmfright{o}
\fmf{photon}{i,v1}
\fmf{photon}{v2,o}
\fmf{plain,tension=.15,left}{v1,v2}
\fmf{photon,tension=.15}{v1,v2}
\fmf{photon,tension=.15,right}{v1,v2}
\end{fmfgraph*} } & = & \mu^{2(4-D)} 
\int \{ d^{D}k_{1} \} \{ d^{D}k_{2} \}
\frac{1}{{\mathcal D}_{2} {\mathcal D}_{8} {\mathcal D}_{12} } \\
& = & \left( \frac{\mu^{2}}{a} \right) ^{2 \epsilon} 
\sum_{i=-2}^{2} \epsilon^{i} F^{(4)}_{i} + {\mathcal O} \left( 
\epsilon^{3} 
\right) , 
\eea
where:
\bea
\frac{F^{(4)}_{-2}}{a} & = & - \frac{1}{2} \, , 
\label{effe31} \\
\frac{F^{(4)}_{-1}}{a} & = & - \frac{3}{2} \, , 
\label{effe32} \\
\frac{F^{(4)}_{0}}{a} & = & - \frac{7}{2} - \zeta(2) \, , 
\label{effe33} \\
\frac{F^{(4)}_{1}}{a} & = & - \frac{15}{2} - 3 \zeta(2)
+ \zeta(3) \, , 
\label{effe34} \\
\frac{F^{(4)}_{2}}{a} & = & - \frac{31}{2} - 7 \zeta(2)
- \frac{ 9 \zeta^{2}(2)}{5} + 3 \zeta(3) 
\label{effe35} 
\, .
\eea
The above amplitude is easily computed with Feynman parameters.
\bea
\parbox{20mm}{\begin{fmfgraph*}(15,15)
\fmfleft{i}
\fmfright{o}
\fmf{plain}{i,v1}
\fmf{plain}{v2,o}
\fmf{photon,tension=.22,left}{v1,v2}
\fmf{photon,tension=.22,right}{v1,v2}
\fmf{plain,right=45}{v2,v2}
\end{fmfgraph*} }  & = & \mu^{2(4-D)} 
\int \{ d^{D}k_{1} \} \{ d^{D}k_{2} \} 
\frac{1}{{\mathcal D}_{1} {\mathcal D}_{9} {\mathcal D}_{13} } \\
& = & \left( \frac{\mu^{2}}{a} \right) ^{2 \epsilon} 
\sum_{i=-2}^{2} \epsilon^{i} F^{(5)}_{i} + {\mathcal O} \left( 
\epsilon^{3} \right) , 
\eea
where:
\bea
\frac{F^{(5)}_{-2}}{a} & = & - 1 \, , 
\label{effe41} \\
\frac{F^{(5)}_{-1}}{a} & = & -3 + H(0,x) \, , 
\label{effe42} \\
\frac{F^{(5)}_{0}}{a} & = & - 7 + \zeta(2) +
3 H(0,x) - H(0,0,x) \, , 
\label{effe43} \\
\frac{F^{(5)}_{1}}{a} & = & - 15 + 3 \zeta(2) + 2 \zeta(3) + (7- \zeta(2)) 
H(0,x) - 3 H(0,0,x) \nn\\
& & + H(0,0,0,x) \, , 
\label{effe44} \\
\frac{F^{(5)}_{2}}{a} & = & - 31 + 7 \zeta(2) + \frac{9}{10} \zeta^{2}(2)
+ 6 \zeta(3) + (15 - 3 \zeta(2)- 2 \zeta(3)) H(0,x) \nn\\
& & - ( 7 - \zeta(2) ) H(0,0,x) + 3 H(0,0,0,x) - H(0,0,0,0,x) \, .
\label{effe45} 
\eea

\subsection{Topology $t=4$ \label{4den}}

\bea
\parbox{20mm}{\begin{fmfgraph*}(15,15)
\fmfleft{i}
\fmfright{o}
\fmfforce{0.2w,0.5h}{v1}
\fmfforce{0.5w,0.8h}{v2}
\fmfforce{0.8w,0.5h}{v3}
\fmf{plain}{i,v1}
\fmf{plain}{v3,o}
\fmf{photon,right}{v1,v3}
\fmf{plain,left=.4}{v1,v2}
\fmf{photon,left=.4}{v2,v3}
\fmf{photon,right=.6}{v2,v3}
\end{fmfgraph*} }  & = & \mu^{2(4-D)} 
\int \{ d^{D}k_{1} \} \{ d^{D}k_{2} \} 
\frac{1}{{\mathcal D}_{2} {\mathcal D}_{3} {\mathcal D}_{9} 
{\mathcal D}_{12} } \\
& = & \left( \frac{\mu^{2}}{a} \right) ^{2 \epsilon} 
\sum_{i=-2}^{2} \epsilon^{i} F^{(6)}_{i} + {\mathcal O} \left( 
\epsilon^{3} \right) , 
\eea
where:
\bea
F^{(6)}_{-2} & = & \frac{1}{2} \, , \\
F^{(6)}_{-1} & = & \frac{5}{2} - \biggl[ 1 + \frac{1}{x} \biggr] H(-1,x)  
\, , \\
F^{(6)}_{0} & = & \frac{19}{2} + \zeta(2) - H(0,x) - \biggl[ 1 + \frac{1}{x} 
\biggr] \bigl[ 4 H(-1,x) - 2 H(-1,-1,x) \nn\\
& & - H(-1,0,x) + H(0,-1,x) \bigr] \, , \\
F^{(6)}_{1} & = & \frac{65}{2} + 3 \zeta(2) - \zeta(3) - 7 H(0,x) - 12
\biggl[ 1 + \frac{1}{x} \biggr] H(-1,x) + 2 H(0,0,x) \nn\\
& & + \biggl[ 1 + \frac{1}{x} \biggr] \bigl[  4 H(-1,0,x)
- 4 H(0,-1,x) + 8 H(-1,-1,x) \nn\\
& & - 4 H(-1,-1,-1,x) - 2 H(-1,-1,0,x) + 2 H(-1,0,-1,x) \nn\\
& & - 2 H(-1,0,0,x) + 2 H(0,-1,-1,x) + H(0,-1,0,x) \nn\\
& & - H(0,0,-1,x) \bigr] \, , \\
F^{(6)}_{2} & = & \frac{211}{2} + 5 \zeta(2) + \frac{9 \zeta^{2}(2)}{5} - 9
\zeta(3) - \bigl( 33 - 2 \zeta(2) \bigr) H(0,x) \nn\\
& & - 2 \bigl( 16 - 3 \zeta(3) \bigr) \biggl[ 1 + \frac{1}{x} \biggr] 
H(-1,x)+ 14 H(0,0,x) - 4 H(0,0,0,x) \nn\\
& & + \biggl[ 1 + \frac{1}{x} \biggr] 
\bigl[ 2 ( 6 - \zeta(2)) H(-1,0,x) - 12 H(0,-1,x) + 24 H(-1,-1,x) \nn\\
& & - 8 H(-1,-1,0,x) + 8 H(-1,0,-1,x) - 8 H(-1,0,0,x) \nn\\
& & + 8 H(0,-1,-1,x) + 4 H(0,-1,0,x) - 4 H(0,0,-1,x) \nn\\
& & - 16 H(-1,-1,-1,x) + 8 H(-1,-1,-1,-1,x) \nn\\
& & + 4 H(-1,-1,-1,0,x) - 4 H(-1,-1,0,-1,x) \nn\\
& & + 4 H(-1,-1,0,0,x) - 4 H(-1,0,-1,-1,x) - 2 H(-1,0,-1,0,x) \nn\\
& & + 2 H(-1,0,0,-1,x) + 4 H(-1,0,0,0,x) - 4 H(0,-1,-1,-1,x) \nn\\
& &  - 2 H(0,-1,-1,0,x) + 2 H(0,-1,0,-1,x) - 2 H(0,-1,0,0,x) \nn\\
& & + 2 H(0,0,-1,-1,x) + H(0,0,-1,0,x) - H(0,0,0,-1,x) \bigr]
\, .
\eea

\bea
\parbox{20mm}{\begin{fmfgraph*}(15,15)
\fmfleft{i}
\fmfright{o}
\fmf{plain}{i,v1}
\fmf{plain}{v3,o}
\fmf{photon,tension=.2,left}{v1,v2}
\fmf{photon,tension=.2,right}{v1,v2}
\fmf{photon,tension=.2,left}{v2,v3}
\fmf{photon,tension=.2,right}{v2,v3}
\end{fmfgraph*} }  & = & \mu^{2(4-D)} 
\int \{ d^{D}k_{1} \} \{ d^{D}k_{2} \}
\frac{1}{{\mathcal D}_{1} {\mathcal D}_{2} {\mathcal D}_{9} 
{\mathcal D}_{10} } \\
& = & \left( \frac{\mu^{2}}{a} \right) ^{2 \epsilon} 
\sum_{i=-2}^{2} \epsilon^{i} F^{(7)}_{i} + {\mathcal O} \left( 
\epsilon^{3} \right) , 
\eea
where:
\bea
F^{(7)}_{-2} & = & 1 \, , \\
F^{(7)}_{-1} & = & 4 - 2 H(0,x) \, , \\
F^{(7)}_{0} & = & 12 - 2 \zeta(2)  - 8 H(0,x) + 4 H(0,0,x) \, , \\
F^{(7)}_{1} & = & 32 - 8 \zeta(2) - 4 \zeta(3) - 4
\bigl[ 6 - \zeta(2) \bigr] H(0,x) + 16 H(0,0,x) \nn\\
& & - 8 H(0,0,0,x) \, , \\
F^{(7)}_{2} & = & 80 - 24 \zeta(2) - \frac{4 \zeta^{2}(2)}{5}
- 16 \zeta(3) - 8 \bigl[ 8 - 2 \zeta(2) - \zeta(3) \bigr] H(0,x) \nn\\
& & + 8 \bigl[ 6 - \zeta(2) \bigr] H(0,0,x) - 32 H(0,0,0,x) + 
16 H(0,0,0,0,x) \, .
\eea

\bea
\parbox{20mm}{\begin{fmfgraph*}(15,15)
\fmfleft{i}
\fmfright{o}
\fmf{plain}{i,v1}
\fmf{plain}{v3,o}
\fmf{plain,tension=.2,left}{v1,v2}
\fmf{photon,tension=.2,right}{v1,v2}
\fmf{photon,tension=.2,left}{v2,v3}
\fmf{photon,tension=.2,right}{v2,v3}
\end{fmfgraph*} }  & = & \mu^{2(4-D)} 
\int \{ d^{D}k_{1} \} \{ d^{D}k_{2} \}
\frac{1}{{\mathcal D}_{2} {\mathcal D}_{10} {\mathcal D}_{9} 
{\mathcal D}_{12} } \\
& = & \left( \frac{\mu^{2}}{a} \right) ^{2 \epsilon} 
\sum_{i=-2}^{2} \epsilon^{i} F^{(8)}_{i} + {\mathcal O} \left( 
\epsilon^{3} \right) , 
\eea
where:
\bea
F^{(8)}_{-2} & = & 1 \, , \\
F^{(8)}_{-1} & = & 4 -  H(0,x) - \biggl[ 1 + \frac{1}{x} \biggr] H(-1,x) 
\, , \\
F^{(8)}_{0} & = & 12 - \zeta(2)  - 4 H(0,x) - 4 \biggl[ 1 + \frac{1}{x} 
\biggr] H(-1,x) + H(0,0,x) \nn\\
& & + \biggl[ 1 + \frac{1}{x} \biggr] \bigl[ H(-1,0,x)
+ 2 H(-1,1,x) \bigr] \, , \\
F^{(8)}_{1} & = & 32 - 4 \zeta(2) - 2 \zeta(3) - \bigl[ 12 - \zeta(2) \bigr] 
H(0,x) - \bigl[ 12 - \zeta(2) \bigr] \biggl[ 1 + \frac{1}{x} \biggr] 
\times \nn\\
& & \times H(-1,x) + 4 H(0,0,x) + \biggl[ 1 + \frac{1}{x} \biggr] 
\bigl[ 4 H(-1,0,x) + 8 H(-1,-1,x) \bigr] \nn\\
& & - H(0,0,0,x) - \biggl[ 1 + \frac{1}{x} 
\biggr] \bigl[ H(-1,0,0,x) + 2 H(-1,-1,0,x) \nn\\
& & + 4 H(-1,-1,-1,x) \bigr] \, , \\
F^{(8)}_{2} & = & 80 - 12 \zeta(2) - \frac{9 \zeta^{2}(2)}{10} - 8 \zeta(3)
- \bigl[ 32 - 4 \zeta(2) - 2 \zeta(3) \bigr] \biggl[ H(0,x) \nn\\ 
& & + \biggl( 1 + \frac{1}{x} \biggr) H(-1,x) \biggr] 
+ \bigl[ 12 - \zeta(2) \bigr] \biggl\{ H(0,0,x) + 
\biggl( 1 + \frac{1}{x} \biggr) \times \nn\\ 
& & \times \bigl[ 2 H(-1,0,x) + 2 H(-1,-1,x) \bigr] 
\biggr\} - 4 H(0,0,0,x) \nn\\
& & - \! \biggl[ 1 \! + \! \frac{1}{x} \biggr] \bigl[ 8 \! H(-1,-1,0,x) 
+ \! 4 H(-1,0,0,x) \! + \! 16 H(-1,-1,-1,x) \nn\\
& & - H(-1,0,0,0,x) - 4 H(-1,-1,-1,0,x) - 2 H(-1,-1,0,0,x) \nn\\
& &  - 8 H(-1,-1,-1,-1,x) \bigr] + H(0,0,0,0,x) 
 \, .
\eea

\bea
\parbox{23mm}{\begin{fmfgraph*}(15,15)
\fmfleft{i1,i2}
\fmfright{o}
\fmf{photon}{i1,v1}
\fmf{photon}{i2,v2}
\fmf{plain}{v3,o}
\fmf{photon,tension=.3}{v2,v3}
\fmf{photon,tension=.3}{v1,v3}
\fmf{plain,tension=0,right=.5}{v2,v1}
\fmf{photon,tension=0,right=.5}{v1,v2}
\end{fmfgraph*} } \quad \quad \quad  & = & \mu^{2(4-D)} 
\int \{ d^{D}k_{1} \} \{ d^{D}k_{2} \}
\frac{1}{{\mathcal D}_{2} {\mathcal D}_{7} {\mathcal D}_{8} 
{\mathcal D}_{12} } \\
& = & \left( \frac{\mu^{2}}{a} \right) ^{2 \epsilon} 
\sum_{i=-2}^{1} \epsilon^{i} F^{(9)}_{i} + {\mathcal O} \left( 
\epsilon^{2} \right) , \\
\parbox{35mm}{\begin{fmfgraph*}(15,15)
\fmfleft{i1,i2}
\fmfright{o}
\fmf{photon}{i1,v1}
\fmf{photon}{i2,v2}
\fmf{plain}{v3,o}
\fmflabel{$(p_{2} \cdot k_{1})$}{o}
\fmf{photon,tension=.3}{v2,v3}
\fmf{photon,tension=.3}{v1,v3}
\fmf{plain,tension=0,right=.5}{v2,v1}
\fmf{photon,tension=0,right=.5}{v1,v2}
\end{fmfgraph*} }  & = & \mu^{2(4-D)} 
\int \{ d^{D}k_{1} \} \{ d^{D}k_{2} \}
\frac{p_{2} \cdot k_{1}}{{\mathcal D}_{2} {\mathcal D}_{7} {\mathcal D}_{8} 
{\mathcal D}_{12} } \\
& = & \left( \frac{\mu^{2}}{a} \right) ^{2 \epsilon} 
\sum_{i=-2}^{1} \epsilon^{i} F^{(10)}_{i} + {\mathcal O} \left( 
\epsilon^{2} \right) ,
\eea
where:
\bea
F^{(9)}_{-2} & = & \frac{1}{2} \, , \\
F^{(9)}_{-1} & = & \frac{5}{2} - H(0,x) \, , \\
F^{(9)}_{0} & = & \frac{19}{2} - 2 \zeta(2) - 5 H(0,x) + H(0,0,x)
- \left[ 1 - \frac{1}{x} \right] H(1,0,x) \nn\\
& & + \frac{1}{x} H(0,1,0,x) \, , \\
F^{(9)}_{1} & = & \frac{65}{2} - 10 \zeta(2) - \zeta(3) - \bigl[ 19 - \zeta(2)
\bigr] H(0,x) - 3 \zeta(2) \biggl[ 1 - \frac{1}{x} \biggr] H(1,x) \nn\\
& & + 5 H(0,0,x) + \frac{3 \zeta(2)}{x} H(0,1,x) - 5 \biggl[ 1 - 
\frac{1}{x} \biggr] H(1,0,x) \nn\\ 
& & - H(0,0,0,x) + \frac{2}{x} H(0,1,0,x) - \biggl[ 1 - \frac{1}{x} 
\biggr] \bigl[ H(0,1,0,x) \nn\\
& & - H(1,0,0,x) + 3 H(1,1,0,x) \bigr] + \frac{1}{x} \bigl[
H(0,0,1,0,x) \nn\\
& & - H(0,1,0,0,x) + 3 H(0,1,1,0,x) \bigr] \, , \\
\frac{F^{(10)}_{-2}}{a} & = & \frac{x}{16}  \, , \\
\frac{F^{(10)}_{-1}}{a} & = & \frac{x}{32} \bigl[ 9 - 4 H(0,x) \bigr] \, , \\
\frac{F^{(10)}_{0}}{a} & = & \frac{5}{8} + \frac{x}{64} \bigl[ 63 - 16
\zeta(2) \bigr] - \frac{3}{16} \bigl[ 2 + 3x \bigr] H(0,x) + \frac{x}{8}
H(0,0,x) \nn\\
& & - \frac{1}{8} \biggl[ x - \frac{1}{x} \biggr] H(1,0,x) +
\frac{1}{4x} H(0,1,0,x) \, , \\
\frac{F^{(10)}_{1}}{a} & = & \frac{57}{16} + \frac{405 x}{128} - \frac{9}{8}
\bigl( 1+x \bigr) \zeta(2) - \frac{x}{8} \zeta(3) - \frac{1}{32} \bigl[
70 + 63x \nn\\
& & - 4 \zeta(2)x \bigr] H(0,x) - \frac{3 \zeta(2)}{8} \biggl[ x - 
\frac{1}{x} \biggr] H(1,x) + \frac{3}{16} \bigl[ 2 + 3x \bigr] H(0,0,x) 
\nn\\
& & - \frac{1}{16} \biggl[ 2 + 9x - \frac{11}{x} \biggr] H(1,0,x) 
+ \frac{3 \zeta(2)}{4x} H(0,1,x) - \frac{1}{8} H(0,0,0,x) \nn\\
& & +\frac{1}{2x} H(0,1,0,x) - \frac{1}{8} \biggl[ x - \frac{1}{x} 
\biggr] \bigl[ H(0,1,0,x) - H(1,0,0,x) \nn\\
& & + 3 H(1,1,0,x) \bigr] + \frac{1}{4x} \bigl[ H(0,0,1,0,x) - 
H(0,1,0,0,x) \nn\\
& & + 3 H(0,1,1,0,x) \bigr] \, .
\eea

\bea
\parbox{20mm}{\begin{fmfgraph*}(15,15)
\fmfleft{i1,i2}
\fmfright{o}
\fmf{photon}{i1,v1}
\fmf{photon}{i2,v2}
\fmf{plain}{v3,o}
\fmf{photon,tension=.3}{v2,v3}
\fmf{photon,tension=.3}{v1,v3}
\fmf{photon,tension=0,right=.5}{v2,v1}
\fmf{photon,tension=0,right=.5}{v1,v2}
\end{fmfgraph*} }  & = & \mu^{2(4-D)} 
\int \{ d^{D}k_{1} \} \{ d^{D}k_{2} \}
\frac{1}{{\mathcal D}_{1} {\mathcal D}_{2} {\mathcal D}_{7} 
{\mathcal D}_{8} } 
\label{4denmassless} \\
& = & \left( \frac{\mu^{2}}{a} \right) ^{2 \epsilon} 
\sum_{i=-2}^{2} \epsilon^{i} F^{(11)}_{i} + {\mathcal O} \left( 
\epsilon^{3} \right) , 
\eea
where:
\bea
F^{(11)}_{-2} & = & \frac{1}{2} \, , \\
F^{(11)}_{-1} & = & \frac{5}{2} - H(0,x) \, , \\
F^{(11)}_{0} & = & \frac{19}{2} - 5 H(0,x) + 2 H(0,0,x) \, , \\
F^{(11)}_{1} & = & \frac{65}{2} - 4 \zeta(3) - 19 H(0,x) + 10 H(0,0,x) 
- 4 H(0,0,0,x) \, , \\
F^{(11)}_{2} & = & \frac{211}{2} - \frac{12}{5} \zeta^{2}(2) - 20 \zeta(3) - [
65 - 8 \zeta(3) ] H(0,x) + 38 H(0,0,x) \nn\\
& & - 20 H(0,0,0,x) + 8 H(0,0,0,0,x) 
\, .
\eea
As discussed in section~3, the above amplitude, being massless, 
cannot be calculated by means of the differential equations technique. 
It is easily computed with Feynman parameters.
\bea
\parbox{20mm}{\begin{fmfgraph*}(15,15)
\fmfleft{i1,i2}
\fmfright{o}
\fmf{photon}{i1,v1}
\fmf{photon}{i2,v2}
\fmf{plain}{v3,o}
\fmf{photon,tension=.3}{v2,v3}
\fmf{photon,tension=.3}{v1,v3}
\fmf{plain,tension=0}{v2,v1}
\fmf{photon,tension=0,left=.5}{v1,v3}
\end{fmfgraph*} }  & = & \mu^{2(4-D)} 
\int \{ d^{D}k_{1} \} \{ d^{D}k_{2} \}
\frac{1}{{\mathcal D}_{2} {\mathcal D}_{4} {\mathcal D}_{8} 
{\mathcal D}_{12} } \\
& = & \left( \frac{\mu^{2}}{a} \right) ^{2 \epsilon} 
\sum_{i=-2}^{2} \epsilon^{i} F^{(12)}_{i} + {\mathcal O} \left( 
\epsilon^{3} \right) , 
\eea
where:
\bea
F^{(12)}_{-2} & = & \frac{1}{2} \, , \\
F^{(12)}_{-1} & = & \frac{3}{2} \, , \\
F^{(12)}_{0} & = & \frac{5}{2} + \zeta(2) + H(0,x)
+ \left[ 1 - \frac{1}{x} \right] H(1,0,x) \, , \\
F^{(12)}_{1} & = & - \frac{1}{2} + 5 \zeta(2) - \zeta(3)
+ 7 H(0,x) + 2\zeta(2) \left[ 1 - \frac{1}{x} 
\right] H(1,x) \nn\\
& & - 2 H(0,0,x) + \biggl[ 1 - \frac{1}{x} \biggr] \bigl[ 4 H(1,0,x) + 
H(0,1,0,x) \nn\\
& & - 2 H(1,0,0,x) + 2 H(1,1,0,x) \bigr] \, , \\
F^{(12)}_{2} & = & - \frac{51}{2} + 19 \zeta(2) + 
\frac{9 \zeta^{2}(2)}{5} + \zeta(3) 
+ \bigl[ 33 - 2 \zeta(2) \bigr] H(0,x) + 4 \bigl[ \zeta(3) \nn\\
& & + 2 \zeta(2) \bigr] \left[ 1 - 
\frac{1}{x} \right] H(1,x) - 14 H(0,0,x) 
+ 2 \zeta(2) \left[ 1 - 
\frac{1}{x} \right] H(0,1,x) 
\nn\\
& & + 2 \bigl[ 6 - \zeta(2) \bigr] \left[ 1 - 
\frac{1}{x} \right] H(1,0,x) + 4 \zeta(2) \left[ 1 - 
\frac{1}{x} \right] H(1,1,x) \nn\\
& &  + 4 H(0,0,0,x) + \left[ 1 - \frac{1}{x} 
\right] \bigl[ 4 H(0,1,0,x) - 8 H(1,0,0,x) \nn\\
& & + 8 H(1,1,0,x) + H(0,0,1,0,x) - 2 H(0,1,0,0,x) \nn\\
& & + 2 H(0,1,1,0,x) + 4 H(1,0,0,0,x) + 2 H(1,0,1,0,x) \nn\\
& & - 4 H(1,1,0,0,x) + 4 H(1,1,1,0,x) \bigr]
\, .
\eea

\subsection{Topology $t=5$ \label{5den}}

\begin{eqnarray}
\parbox{20mm}{\begin{fmfgraph*}(15,15)
\fmfleft{i1,i2}
\fmfright{o}
\fmfforce{0.8w,0.5h}{v4}
\fmf{photon}{i1,v1}
\fmf{photon}{i2,v2}
\fmf{plain}{v4,o}
\fmf{photon,tension=.4}{v2,v3}
\fmf{photon,tension=.2}{v3,v4}
\fmf{photon,tension=.15}{v1,v4}
\fmf{photon,tension=0}{v2,v1}
\fmf{plain,tension=0}{v1,v3}
\end{fmfgraph*} } & = & \mu^{2(4-D)} 
\int \{ d^{D}k_{1} \} \{ d^{D}k_{2} \}
\frac{1}{{\mathcal D}_{1} {\mathcal D}_{4} {\mathcal D}_{7} 
{\mathcal D}_{8} {\mathcal D}_{13} } \\
& = & \left( \frac{\mu^{2}}{a} \right) ^{2 \epsilon} 
\sum_{i=-1}^{0} \epsilon^{i} F^{(13)}_{i} + {\mathcal O} \left( 
\epsilon \right) , 
\eea
where:
\bea
a F^{(13)}_{-1} & = & \frac{1}{x} \bigl[ H(0,0,-1,x) + H(0,1,0,x) \bigr] \, , 
\\
a F^{(13)}_{0} & = & \frac{1}{x} \bigl[ 3 \zeta(2) H(0,1,x) - 4 H(0,0,-1,-1,x)
+ H(0,0,0,-1,x) \nn\\
& & + H(0,0,1,0,x) - H(0,1,0,0,x) + 3 H(0,1,1,0,x) \bigr] \,  .
\eea

\begin{eqnarray}
\parbox{20mm}{\begin{fmfgraph*}(15,15)
\fmfleft{i1,i2}
\fmfright{o}
\fmfforce{0.8w,0.5h}{v4}
\fmf{photon}{i1,v1}
\fmf{photon}{i2,v2}
\fmf{plain}{v4,o}
\fmf{photon,tension=.4}{v2,v3}
\fmf{photon,tension=.2}{v3,v4}
\fmf{photon,tension=.15}{v1,v4}
\fmf{plain,tension=0}{v2,v1}
\fmf{photon,tension=0}{v1,v3}
\end{fmfgraph*} } & = & \mu^{2(4-D)} 
\int \{ d^{D}k_{1} \} \{ d^{D}k_{2} \}
\frac{1}{{\mathcal D}_{2} {\mathcal D}_{4} {\mathcal D}_{7} 
{\mathcal D}_{8} {\mathcal D}_{12} } \\
& = & \left( \frac{\mu^{2}}{a} \right) ^{2 \epsilon} 
F^{(14)}_{0} + {\mathcal O} \left( 
\epsilon \right) , 
\eea
where:
\bea
a F^{(14)}_{0} & = & \frac{1}{x} \bigl[ \zeta(2) H(0,1,x) + H(0,1,0,0,x)
+ H(0,1,1,0,x) \bigr] \, .
\eea

\begin{eqnarray}
\parbox{23mm}{\begin{fmfgraph*}(15,15)
\fmfleft{i1,i2}
\fmfright{o}
\fmfforce{0.8w,0.5h}{v4}
\fmf{photon}{i1,v1}
\fmf{photon}{i2,v2}
\fmf{plain}{v4,o}
\fmf{plain,tension=.4}{v2,v3}
\fmf{photon,tension=.2}{v3,v4}
\fmf{photon,tension=.15}{v1,v4}
\fmf{photon,tension=0}{v2,v1}
\fmf{photon,tension=0}{v1,v3}
\end{fmfgraph*} } \quad \quad \quad & = & \mu^{2(4-D)} 
\int \{ d^{D}k_{1} \} \{ d^{D}k_{2} \}
\frac{1}{{\mathcal D}_{1} {\mathcal D}_{2} {\mathcal D}_{7} 
{\mathcal D}_{8} {\mathcal D}_{15} } \\
& = & \left( \frac{\mu^{2}}{a} \right) ^{2 \epsilon} 
F^{(15)}_{0} + {\mathcal O} \left( 
\epsilon \right) , \\
\parbox{35mm}{\begin{fmfgraph*}(15,15)
\fmfleft{i1,i2}
\fmfright{o}
\fmfforce{0.8w,0.5h}{v4}
\fmf{photon}{i1,v1}
\fmf{photon}{i2,v2}
\fmf{plain}{v4,o}
\fmflabel{$(k_{1} \cdot k_{2})$}{o}
\fmf{plain,tension=.4}{v2,v3}
\fmf{photon,tension=.2}{v3,v4}
\fmf{photon,tension=.15}{v1,v4}
\fmf{photon,tension=0}{v2,v1}
\fmf{photon,tension=0}{v1,v3}
\end{fmfgraph*} } & = & \mu^{2(4-D)} 
\int \{ d^{D}k_{1} \} \{ d^{D}k_{2} \}
\frac{k_1 \cdot k_2}{{\mathcal D}_{1} {\mathcal D}_{2} {\mathcal D}_{7} 
{\mathcal D}_{8} {\mathcal D}_{15}} \\
& = & \left( \frac{\mu^{2}}{a} \right) ^{2 \epsilon} 
\sum_{i=-2}^{1} \epsilon^{i} F^{(16)}_{i} + {\mathcal O} \left( 
\epsilon^{2} \right) ,
\eea
where:
\bea
a F^{(15)}_{0} & = & \frac{1}{x} \bigl[ 2 \zeta(2) H(0, \! -1,x) \! - \! 
H(0, \! -1,0, \! -1,x) \! + \! H(0, \! -1,0,0,x) \bigr] \, , \\
%a Q_{1} & = & \frac{1}{x} \bigl[ - 3 \zeta(3) H(0,-1,x) - 4 \zeta(2) 
%H(0,-1,-1,x) \nn\\
%& & + 2 \zeta(2) H(0,0,-1,x) - \zeta(2) H(0,-1,0,x) \nn\\
%& & + 2 H(0,-1,-1,0,-1,x) - 2 H(0,-1,-1,0,0,x) \nn\\
%& & + 4 H(0,-1,0,-1,-1,x) - H(0,-1,0,0,-1,x) \nn\\
%& & - 3 H(0,-1,0,0,0,x) - H(0,0,-1,0,-1,x) \nn\\
%& & + H(0,0,-1,0,0,x) \bigr] \, \\
aF^{(16)}_{-2} & = & \frac{1}{2} \, , \\
aF^{(16)}_{-1} & = & 2 - H(0,x) \, , \\
aF^{(16)}_{0} & = & 7 - 3 \zeta(2) - \frac{5}{2} H(0,x) - \bigl( 2 - \zeta(2) 
\bigr) \biggl[ 1 + \frac{1}{x} \biggr] H(-1,x) 
\nn\\
& & + \frac{1}{2} H(0,0,x) \! + \! \frac{1}{2} \biggl[ 3 \! - \! 
\frac{1}{x} \biggr] H(0, \! -1,x) \! + \! \frac{1}{2} \biggl[ 1 \! + \! 
\frac{1}{x} \biggr] \bigl[ H( \! -1,0,0,x) \nn\\
& & - H(-1,0,-1,x) \bigr] \, , \\
aF^{(16)}_{1} & = & 24 - \frac{17}{2} \zeta(2) + \frac{\zeta(3)}{2} -
\frac{1}{2} \bigl[ 7 - 3 \zeta(2) \bigr] H(0,x) - \biggl( 13 - 2
\zeta(2) \nn\\
& & + \frac{3}{2} \zeta(3) \biggr) \biggl[ 1 + \frac{1}{x} \biggr]
H(-1,x) - \frac{\zeta(2)}{2} \biggl[ 1 + \frac{1}{x} \biggr] 
H(-1,0,x) \nn\\
& & + 5 H(0,-1,x) + \biggl( \zeta(2) - \frac{7}{2} \biggl) \biggl[ 1 + 
\frac{1}{x} \biggr] H(0,-1,x) \nn\\
& & + 2 \bigl( 4 - \zeta(2) \bigr) \biggl[ 1 + \frac{1}{x} \biggr] 
H(-1,-1,x) + \frac{1}{2} H(0,0,0,x) \nn\\
& & - 8 H(0, \! -1, \! -1,x) \! + \! 2 H(0,0, \! -1,x) \! + \! 
\frac{1}{2} \biggl[ 1 \! + \! \frac{1}{x} \biggr] \bigr[ 4 \! H(0, \! 
-1, \! -1,x) \nn\\
& & - H(0,0,-1,x) - 2 H(-1,0,-1,x) + 2 H(-1,0,0,x) \nn\\
& & + 2 H( \! -1, \! -1,0, \! -1,x) \! - \! 2 H( \! -1, \! -1,0,0,x) \! 
+ \! 4 H( \! -1,0, \! -1, \! -1,x) \nn\\
& & - H(-1,0,0,-1,x) - 3 H(-1,0,0,0,x) - H(0,-1,0,-1,x) \nn\\
& & + H(0,-1,0,0,x) \bigr]
\, .
\eea

The above topology is the only 5-denominator one having 2 MI's.

\begin{eqnarray}
\parbox{20mm}{\begin{fmfgraph*}(15,15)
\fmfleft{i1,i2}
\fmfright{o}
\fmfforce{0.2w,0.9h}{v2}
\fmfforce{0.2w,0.1h}{v1}
\fmfforce{0.2w,0.5h}{v3}
\fmfforce{0.8w,0.5h}{v4}
\fmf{photon}{i1,v1}
\fmf{photon}{i2,v2}
\fmf{plain}{v4,o}
\fmf{photon,tension=0}{v1,v3}
\fmf{photon,tension=0}{v3,v4}
\fmf{photon,tension=0}{v2,v4}
\fmf{plain,tension=0}{v2,v3}
\fmf{photon,tension=0}{v1,v4}
\end{fmfgraph*} } & = & \mu^{2(4-D)} 
\int \{ d^{D}k_{1} \} \{ d^{D}k_{2} \}
\frac{1}{{\mathcal D}_{1} {\mathcal D}_{2} {\mathcal D}_{6} 
{\mathcal D}_{11} {\mathcal D}_{15}} \\
& = & \left( \frac{\mu^{2}}{a} \right) ^{2 \epsilon} 
\sum_{i=-1}^{0} \epsilon^{i} F^{(17)}_{i} + {\mathcal O} \left( 
\epsilon \right) , 
\eea
where:
\bea
a F^{(17)}_{-1} & = & \frac{1}{x} H(0,1,0,x) \, , \\
a F^{(17)}_{0} & = & \frac{1}{x} \bigl[ 2 \zeta(2) H(0,1,x) + H(0,0,1,0,x) 
- 2 H(0,1,0,0,x) \nn\\
& & + 2 H(0,1,1,0,x) \bigr] \, . 
\eea

\begin{eqnarray}
\parbox{20mm}{\begin{fmfgraph*}(15,15)
\fmfleft{i1,i2}
\fmfright{o}
\fmf{photon}{i1,v1}
\fmf{photon}{i2,v2}
\fmf{plain}{v4,o}
\fmf{photon,tension=.3}{v2,v3}
\fmf{photon,tension=.3}{v1,v3}
\fmf{plain,tension=0}{v2,v1}
\fmf{photon,tension=.2,left}{v3,v4}
\fmf{photon,tension=.2,right}{v3,v4}
\end{fmfgraph*} }  & = & \mu^{2(4-D)} 
\int \{ d^{D}k_{1} \} \{ d^{D}k_{2} \}
\frac{1}{{\mathcal D}_{2} {\mathcal D}_{4} {\mathcal D}_{5} 
{\mathcal D}_{10} {\mathcal D}_{12}} \\
& = & \left( \frac{\mu^{2}}{a} \right) ^{2 \epsilon} 
\sum_{i=-1}^{1} \epsilon^{i} F^{(18)}_{i} + {\mathcal O} \left( 
\epsilon^{2} \right) , 
\end{eqnarray}
where:
\bea
a F^{(18)}_{-1} & = &  - \frac{1}{x} H(1,0,x) \, , \\
a F^{(18)}_{0} & = &  - \frac{1}{x} \bigl[ \zeta(2) H(1,x) \! + \! 2 H(1,0,x) 
\! - \! 3 H(1,0,0,x) \! + \! H(1,1,0,x) \bigr] , \\
a F^{(18)}_{1} & = & - \frac{1}{x} \bigl[ 2( \zeta(2) \! + \! \zeta(3) ) H(1,x)
\! + \! \zeta(2) H(1,1,x) \! + \! ( 4 \! - \! 3 \zeta(2) ) H(1,0,x) 
\nn\\
& &  - 6 H(1,0,0,x) + 2 H(1,1,0,x) + 7 H(1,0,0,0,x) \nn\\
& & - 3 H(1,1,0,0,x) + H(1,1,1,0,x) \bigr] 
\, .
\eea

The above amplitude has a simple ultraviolet pole coming from the massless
bubble.

\begin{eqnarray}
\parbox{20mm}{\begin{fmfgraph*}(15,15)
\fmfforce{0.2w,0.5h}{v1}
\fmfforce{0.5w,0.8h}{v2}
\fmfforce{0.5w,0.2h}{v3}
\fmfforce{0.8w,0.5h}{v4}
\fmfleft{i}
\fmfright{o}
\fmf{plain}{i,v1}
\fmf{plain}{v4,o}
\fmf{photon,tension=.2,left=.4}{v1,v2}
\fmf{photon,tension=.2,right=.4}{v1,v3}
\fmf{photon,tension=.2,left=.4}{v2,v4}
\fmf{photon,tension=.2,right=.4}{v3,v4}
\fmf{plain,tension=0}{v2,v3}
\end{fmfgraph*} } & = & \mu^{2(4-D)} 
\int \{ d^{D}k_{1} \} \{ d^{D}k_{2} \}
\frac{1}{{\mathcal D}_{1} {\mathcal D}_{3} {\mathcal D}_{9} 
{\mathcal D}_{11} {\mathcal D}_{13}} \\
& = & \left( \frac{\mu^{2}}{a} \right) ^{2 \epsilon} 
\sum_{i=0}^{2} \epsilon^{i} F^{(19)}_{i} + {\mathcal O} \left( 
\epsilon^{2} \right) , 
\eea
where:
\bea
a F^{(19)}_{0} & = & \frac{2}{x} \bigl[ \zeta(2) H(1,x) - H(0,0,-1,x) +
H(1,0,0,x) \nn\\
& & - 2 H(1,0,-1,x) \bigr] \, , \\
a F^{(19)}_{1} & = & \frac{2}{x} \bigl[ ( 2 \zeta(2) - \zeta(3) ) H(1,x) 
+ \zeta(2) ( H(1,0,x) + H(1,1,x) ) \nn\\
& &  - 2 H(0,0,-1,x) - 4 H(1,0,-1,x) + 2 H(1,0,0,x) \nn\\
& &  + 4 H(0,0,-1,-1,x) - H(0,0,0,-1,x) + 8 H(1,0,-1,-1,x) \nn\\
& &  - 3 H(1,0,0,-1,x) - 3 H(1,0,0,0,x) - 2 H(1,1,0,-1,x) \nn\\
& &  + H(1,1,0,0,x) \bigr] \,  .
\eea

The finite part of the above amplitude is also given in \cite{fleischer}.

\begin{eqnarray}
\parbox{20mm}{\begin{fmfgraph*}(15,15)
\fmfleft{i1,i2}
\fmfright{o}
\fmfforce{0.2w,0.9h}{v2}
\fmfforce{0.2w,0.1h}{v1}
\fmfforce{0.2w,0.55h}{v3}
\fmfforce{0.2w,0.15h}{v5}
\fmfforce{0.8w,0.5h}{v4}
\fmf{photon}{i1,v1}
\fmf{photon}{i2,v2}
\fmf{plain}{v4,o}
\fmf{plain}{v2,v3}
\fmf{photon,left}{v3,v5}
\fmf{photon,right}{v3,v5}
\fmf{photon}{v1,v4}
\fmf{photon}{v2,v4}
\end{fmfgraph*} }  & = & \mu^{2(4-D)} 
\int \{ d^{D}k_{1} \} \{ d^{D}k_{2} \}
\frac{1}{{\mathcal D}_{2} {\mathcal D}_{3} {\mathcal D}_{4} 
{\mathcal D}_{5} {\mathcal D}_{12}} \\
& = & \left( \frac{\mu^{2}}{a} \right) ^{2 \epsilon} 
\sum_{i=-1}^{1} \epsilon^{i} F^{(20)}_{i} + {\mathcal O} \left( 
\epsilon^{2} \right) , 
\eea
where:
\bea
a F^{(20)}_{-1} & = & - \frac{1}{x} H(1,0,x) \, , \\
a F^{(20)}_{0} & = & - \frac{1}{x} \bigl[ \zeta(2) H(1,x) + H(0,1,0,x) + 2
H(1,0,x) + H(1,1,0,x) \nn\\
& & - 2 H(1,0,0,x) \bigr]   \, , \\
a F^{(20)}_{1} & = & - \frac{1}{x} \bigl[ ( 2 \zeta(2) + 3 \zeta(3) ) H(1,x) 
+ \zeta(2) H(0,1,x) + 4 H(1,0,x) \nn\\
& & + \zeta(2) H(1,1,x) + H(0,0,1,0,x) + 2 H(0,1,0,x) \nn\\
& & - 2 H(0,1,0,0,x) + H(0,1,1,0,x) - 4 H(1,0,0,x) \nn\\
& & + 4 H(1,0,0,0,x) + H(1,0,1,0,x) + 2 H(1,1,0,x) \nn\\
& & - 2 H(1,1,0,0,x) + H(1,1,1,0,x) \bigr]  \, .
\eea
The above amplitude has a simple ultraviolet pole coming from the 
nested bubble.

\section{Six-denominator diagrams \label{6Denominatori}}

In this section we present the results of all the six-denominator 
diagrams with up to one massive line entering the EW form factor.
The ladder and vertex insertion diagrams --- the planar
topologies --- are all reducible, while both the crossed ladders 
--- the non-planar topologies ---
are MI's, i.e. cannot be reduced to simpler 
topologies by using the ibp identities. Even though these diagrams have 
a different ``diagrammatic status'', we present them together because 
they are the most important ones and may be used for reference.
Series representations for the pole and finite parts of the ladder and 
crossed ladder diagrams have been given in \cite{fleischer,fleischer0}. 
The authors compute a large number of terms in a small momentum expansion 
of the diagrams and then fit the results to an assumed basis of functions 
and trascendent constants. In the above paper, formulas for the 
resummation of the series in terms of Nielsen polylogarithms 
\cite{Nielsen,Kolbig} are also reported. By using them, we could check 
that the pole parts of our results agree with theirs. Finally, the 
leading logarithms of all the six denominator amplitudes have been 
computed in \cite{kodaira} by means of a pole approximation for the 
vector boson propagators; we are in agreement with them also.

\subsection{Planar topologies}

\begin{eqnarray}
\parbox{20mm}{\begin{fmfgraph*}(15,15)
\fmfleft{i1,i2}
\fmfright{o}
\fmf{photon}{i1,v1}
\fmf{photon}{i2,v2}
\fmf{plain}{v5,o}
\fmf{photon,tension=.3}{v2,v3}
\fmf{photon,tension=.3}{v3,v5}
\fmf{photon,tension=.3}{v1,v4}
\fmf{photon,tension=.3}{v4,v5}
\fmf{photon,tension=0}{v2,v1}
\fmf{photon,tension=0}{v4,v3}
\end{fmfgraph*} } & = & \mu^{2(4-D)} 
\int \{ d^{D}k_{1} \} \{ d^{D}k_{2} \}
\frac{1}{{\mathcal D}_{1} {\mathcal D}_{2} {\mathcal D}_{4} 
{\mathcal D}_{5} {\mathcal D}_{7} {\mathcal D}_{8}} \\
& = & \left( \frac{\mu^{2}}{a} \right) ^{2 \epsilon} 
\sum_{i=-4}^{0} \epsilon^{i} A_{i} + {\mathcal O} \left( 
\epsilon \right) , 
\eea
where:
\bea
a^{2} A_{-4} & = & \frac{1}{4 x^{2}} \, , \\
a^{2} A_{-3} & = & - \frac{1}{2x^{2}} H(0,x)  \, , \\
a^{2} A_{-2} & = & \frac{1}{x^{2}} \biggl[ \zeta(2) + H(0,0,x) 
\biggr] \, , \\
a^{2} A_{-1} & = & \frac{1}{x^{2}} \biggl[ 5 \zeta(3) - 2 \zeta(2) 
H(0,x) - 2 H(0,0,0,x) \biggr] \, , \\
a^{2} A_{0} & = & \frac{1}{x^{2}} \biggl[ \frac{11 \zeta^{2}(2)}{5} - 10
\zeta(3) H(0,x) + 4 \zeta(2) H(0,0,x) \nn\\
& & + 4 H(0,0,0,0,x) \biggr] \, .
\eea

The above diagram has also been computed in \cite{Gonsalves83} using 
Feynman parameters and in \cite{VanNeerven86} using dispersion 
relations; we computed it by means of the reduction to simpler (massless)
amplitudes, the latter computed with Feynman parameters.
We are in complete agreement with these references.
The leading singularity of this diagram is a pole in $\epsilon$ of 
fourth order or, equivalently, a forth power in $\log(x)$ in the finite 
part for $x \rightarrow 0$. We have indeed a soft and a collinear 
singularity for each loop, each one contributing by a simple pole to the
amplitude.

\begin{eqnarray}
\parbox{20mm}{\begin{fmfgraph*}(15,15)
\fmfleft{i1,i2}
\fmfright{o}
\fmf{photon}{i1,v1}
\fmf{photon}{i2,v2}
\fmf{plain}{v5,o}
\fmf{photon,tension=.3}{v2,v3}
\fmf{photon,tension=.3}{v3,v5}
\fmf{photon,tension=.3}{v1,v4}
\fmf{photon,tension=.3}{v4,v5}
\fmf{plain,tension=0}{v2,v1}
\fmf{photon,tension=0}{v4,v3}
\end{fmfgraph*} } & = & \mu^{2(4-D)} 
\int \{ d^{D}k_{1} \} \{ d^{D}k_{2} \}
\frac{1}{{\mathcal D}_{2} {\mathcal D}_{4} {\mathcal D}_{5} 
{\mathcal D}_{7} {\mathcal D}_{8} {\mathcal D}_{12}} 
\label{koda} \\
& = & \left( \frac{\mu^{2}}{a} \right) ^{2 \epsilon} 
B_{0} + {\mathcal O} \left( 
\epsilon \right) , 
\eea
where:
\bea
a^{2} B_{0} & = & \frac{1}{x^{2}} \bigl[ 6 \zeta(3) H(1,x) - 2 \zeta(2) 
H(1,1,x) + H(1,0,1,0,x) \nn\\
& & - 2 H(1,1,0,0,x) - 2 H(1,1,1,0,x) \bigr] \, .
\eea

As expected on the basis of one's physical intuition, the above diagram 
is infrared finite. The massive vector boson cannot indeed propagate for
large distances and the same is true for the photon, which is 
``trapped'' inside the external loop.

\begin{eqnarray}
\parbox{20mm}{\begin{fmfgraph*}(15,15)
\fmfleft{i1,i2}
\fmfright{o}
\fmf{photon}{i1,v1}
\fmf{photon}{i2,v2}
\fmf{plain}{v5,o}
\fmf{photon,tension=.3}{v2,v3}
\fmf{photon,tension=.3}{v3,v5}
\fmf{photon,tension=.3}{v1,v4}
\fmf{photon,tension=.3}{v4,v5}
\fmf{photon,tension=0}{v2,v1}
\fmf{plain,tension=0}{v4,v3}
\end{fmfgraph*} } & = & \mu^{2(4-D)} 
\int \{ d^{D}k_{1} \} \{ d^{D}k_{2} \}
\frac{1}{{\mathcal D}_{1} {\mathcal D}_{4} {\mathcal D}_{5} 
{\mathcal D}_{7} {\mathcal D}_{8} {\mathcal D}_{13}} \\
& = & \left( \frac{\mu^{2}}{a} \right) ^{2 \epsilon} 
\sum_{i=-2}^{0} \epsilon^{i} C_{i} + {\mathcal O} \left( 
\epsilon \right) , 
\eea
where:
\bea
a^{2} C_{-2} & = & - \frac{1}{x^{2}} H(1,0,x)  \, , \\
a^{2} C_{-1} & = & - \frac{1}{x^{2}} \bigl[ \zeta(2) H(1,x) 
+ 4 H(1,0,-1,x) + 3 H(1,1,0,x) \nn\\
& & - 3 H(1,0,0,x) \bigr] \, , \\
a^{2} C_{0} & = & - \frac{1}{x^{2}} \bigl[ 2 \zeta(3) H(1,x) + 7
\zeta(2) H(1,1,x) - 3 \zeta(2) H(1,0,x) \nn\\
& &  - 16 H(1,0,-1,-1,x) + 6 H(1,0,0,-1,x) + 7 H(1,0,0,0,x) \nn\\
& & + 3 H(1,0,1,0,x) + 4 H(1,1,0,-1,x) - 5 H(1,1,0,0,x) \nn\\
& & + 9 H(1,1,1,0,x) \bigr] \, .
\eea

Unlike Eq. (\ref{koda}), in the above diagram the photon is 
not trapped by the massive line and can propagate to asymptotic times. 
As a consequence, we have a double infrared pole, having a similar dynamical origin to 
the one in the massless one-loop vertex.

\begin{eqnarray}
\parbox{20mm}{\begin{fmfgraph*}(15,15)
\fmfleft{i1,i2}
\fmfright{o}
\fmf{photon}{i1,v1}
\fmf{photon}{i2,v2}
\fmf{plain}{v5,o}
\fmf{plain,tension=.3}{v2,v3}
\fmf{photon,tension=.3}{v3,v5}
\fmf{photon,tension=.3}{v1,v4}
\fmf{photon,tension=.3}{v4,v5}
\fmf{photon,tension=0}{v2,v1}
\fmf{photon,tension=0}{v4,v3}
\end{fmfgraph*} } & = & \mu^{2(4-D)} 
\int \{ d^{D}k_{1} \} \{ d^{D}k_{2} \}
\frac{1}{{\mathcal D}_{1} {\mathcal D}_{2} {\mathcal D}_{5} 
{\mathcal D}_{7} {\mathcal D}_{8} {\mathcal D}_{15}} \\
& = & \left( \frac{\mu^{2}}{a} \right) ^{2 \epsilon} 
\sum_{i=-2}^{0} \epsilon^{i} E_{i} + {\mathcal O} \left( 
\epsilon \right) , 
\eea
where:
\bea
a^{2} E_{-2} & = & \frac{1}{2} \biggl[ \frac{1}{x} - \frac{1}{(1+x)} 
\biggr] \bigl[ \zeta(2) + H(0,-1,x) \bigr] \, , \\
a^{2} E_{-1} & = & \frac{1}{2} \biggl[ \frac{1}{x} - \frac{1}{(1+x)} 
\biggr] \bigl[ \zeta(3) - 2 \zeta(2) ( H(0,x) + H(-1,x) ) \nn\\
& & - 2 H(-1,0,-1,x) - 2 H(0,-1,0,x) + H(0,0,-1,x) \bigr]  \, , \\
a^{2} E_{0} & = & \biggl[ \frac{1}{x} - \frac{1}{(1+x)} 
\biggr] \biggl\{ - \frac{\zeta^{2}(2)}{10} - \zeta(3) \bigl[ H(0,x) + 
H(-1,x) \bigr] \nn\\
& & + \! \zeta(2) \bigl[ 2 H(0,0,x) \! + \! 2 H( \! -1, \! -1,x) \! + \!
2 H( \! -1,0,x) \! - \! 3 H(0, \! -1,x) \bigr] \nn\\
& & - 4 H(0,-1,-1,-1,x) + 2 H(0,-1,0,-1,x) + H(0,-1,0,0,x) \nn\\
& & + \frac{1}{2} H(0,0,0,-1,x) - H(0,0,-1,0,x) + 2 H(0,-1,-1,0,x) \nn\\
& & + 2 H(-1,0,-1,0,x) - H(-1,0,0,-1,x) \nn\\
& & + 2 H(-1,-1,0,-1,x) \biggl\} 
\, .
\eea

\begin{eqnarray}
\parbox{20mm}{\begin{fmfgraph*}(15,15)
\fmfleft{i1,i2}
\fmfright{o}
\fmfforce{0.2w,0.9h}{v2}
\fmfforce{0.2w,0.1h}{v1}
\fmfforce{0.2w,0.5h}{v3}
\fmfforce{0.8w,0.5h}{v5}
\fmf{photon}{i1,v1}
\fmf{photon}{i2,v2}
\fmf{plain}{v5,o}
\fmf{photon,tension=0}{v2,v5}
\fmf{photon,tension=0}{v3,v4}
\fmf{photon,tension=.4}{v1,v4}
\fmf{photon,tension=.4}{v4,v5}
\fmf{photon,tension=0}{v1,v3}
\fmf{photon,tension=0}{v2,v3}
\end{fmfgraph*} }  & = & \mu^{2(4-D)} 
\int \{ d^{D}k_{1} \} \{ d^{D}k_{2} \}
\frac{1}{{\mathcal D}_{1} {\mathcal D}_{2} {\mathcal D}_{3} 
{\mathcal D}_{4} {\mathcal D}_{5} {\mathcal D}_{6}} \\
& = & \left( \frac{\mu^{2}}{a} \right) ^{2 \epsilon} 
\sum_{i=-3}^{0} \epsilon^{i} G_{i} + {\mathcal O} \left( 
\epsilon \right) , 
\eea
where:
\bea
a^{2} G_{-3} & = & \frac{3}{4x^{2}}   \, , \\
a^{2} G_{-2} & = & - \frac{3}{2x^{2}} \bigl[ 1 + H(0,x) \bigr] \, , \\
a^{2} G_{-1} & = & \frac{3}{x^{2}} \bigl[ 1 - \zeta(2) + H(0,x) +
H(0,0,x) \bigr] \, , \\
a^{2} G_{0} & = & - \frac{3}{x^{2}} \bigl[ 2 - 2 \zeta(2) + 3 \zeta(3) 
+ 2 ( 1 - \zeta(2)) H(0,x) + 2 H(0,0,x) \nn\\
& & + 2 H(0,0,0,x) \bigr]   \, .
\eea

The above diagram has also been computed in \cite{Gonsalves83} using 
Feynman parameters and in \cite{VanNeerven86} using dispersion 
relations; we have computed it by means of the reduction 
to simpler (massless) amplitudes, the latter computed with Feynman parameters 
in section 3. We are in complete agreement with the previous results.
The massless vertex correction has at most a triple pole in $\epsilon$ 
and is therefore less singular than the massless ladder or crossed 
ladder (see later). As well known, the vertex correction 
is indeed neglected in the double 
logarithmic analysis of QED form factors.

\begin{eqnarray}
\parbox{20mm}{\begin{fmfgraph*}(15,15)
\fmfleft{i1,i2}
\fmfright{o}
\fmfforce{0.2w,0.9h}{v2}
\fmfforce{0.2w,0.1h}{v1}
\fmfforce{0.2w,0.5h}{v3}
\fmfforce{0.8w,0.5h}{v5}
\fmf{photon}{i1,v1}
\fmf{photon}{i2,v2}
\fmf{plain}{v5,o}
\fmf{photon,tension=0}{v2,v5}
\fmf{photon,tension=0}{v3,v4}
\fmf{plain,tension=.4}{v1,v4}
\fmf{photon,tension=.4}{v4,v5}
\fmf{photon,tension=0}{v1,v3}
\fmf{photon,tension=0}{v2,v3}
\end{fmfgraph*} }   & = & \mu^{2(4-D)} 
\int \{ d^{D}k_{1} \} \{ d^{D}k_{2} \}
\frac{1}{{\mathcal D}_{1} {\mathcal D}_{2} {\mathcal D}_{3} 
{\mathcal D}_{4} {\mathcal D}_{5} {\mathcal D}_{17}} \\
& = & \left( \frac{\mu^{2}}{a} \right) ^{2 \epsilon} 
\sum_{i=-3}^{0} \epsilon^{i} I_{i} + {\mathcal O} \left( 
\epsilon \right) , 
\eea
where:
\bea
a^{2} I_{-3} & = & - \frac{3}{4x}   \, , \\
a^{2} I_{-2} & = & - \frac{1}{2x^{2}} \bigl[ (1 + x) H(-1,x) - x H(0,x)
\bigr] \, , \\
a^{2} I_{-1} & = & \frac{1}{x^{2}} \bigl[ \zeta(2) x - 2 x H(0,-1,x) + 2 (
(1+x) H(-1,-1,x) \bigr] \, , \\
a^{2} I_{0} & = & \frac{1}{x^{2}} \Bigl\{ \zeta(2) (1+x) H(-1,x) - 
\zeta(2) x H(0,x) - x  H(0,0,0,x) \nn\\
& & + 2x [ 4 H(0,-1,-1,x) - H(0,0,-1,x) ] + (1+x) [ H(-1,0,0,x) \nn\\
& & + 2 H(-1,0,-1,x) - 8 H(-1,-1,-1,x) ] \Bigl\} \, .
\eea

\begin{eqnarray}
\parbox{20mm}{\begin{fmfgraph*}(15,15)
\fmfleft{i1,i2}
\fmfright{o}
\fmfforce{0.2w,0.9h}{v2}
\fmfforce{0.2w,0.1h}{v1}
\fmfforce{0.2w,0.5h}{v3}
\fmfforce{0.8w,0.5h}{v5}
\fmf{photon}{i1,v1}
\fmf{photon}{i2,v2}
\fmf{plain}{v5,o}
\fmf{photon,tension=0}{v2,v5}
\fmf{photon,tension=0}{v3,v4}
\fmf{photon,tension=.4}{v1,v4}
\fmf{photon,tension=.4}{v4,v5}
\fmf{photon,tension=0}{v1,v3}
\fmf{plain,tension=0}{v2,v3}
\end{fmfgraph*} }  & = & \mu^{2(4-D)} 
\int \{ d^{D}k_{1} \} \{ d^{D}k_{2} \}
\frac{1}{{\mathcal D}_{2} {\mathcal D}_{3} {\mathcal D}_{4} 
{\mathcal D}_{5} {\mathcal D}_{6} {\mathcal D}_{12}} \\
& = & \left( \frac{\mu^{2}}{a} \right) ^{2 \epsilon} 
\sum_{i=-2}^{0} \epsilon^{i} J_{i} + {\mathcal O} \left( 
\epsilon \right) , 
\eea
where:
\bea
a^{2} J_{-2} & = & \frac{\zeta(2)}{2x} \, , \\
a^{2} J_{-1} & = & \frac{1}{2x} \bigl[ \zeta(3) - 2 \zeta(2) \bigl( H(0,x)
+ H(1,x) \bigr) - 2 H(0,1,0,x) \nn\\
& & - 2 H(1,1,0,x) \bigr] \, , \\
a^{2} J_{0} & = & - \frac{1}{x} \biggl\{ \frac{\zeta^{2}(2)}{10} +
\zeta(3) \bigl[ H(0,x) + H(1,x) \bigr] - \zeta(2) \bigl[ 2 H(0,0,x)
\nn\\
& &  - 3 H(0,1,x) + 2 H(1,0,x) - 3 H(1,1,x) - H(0,0,1,0,x)  \nn\\
& & + 2 H(0,1,0,0,x) - 3 H(0,1,1,0,x) - H(1,0,1,0,x)  \nn\\
& & + 2 H(1,1,0,0,x) - 3 H(1,1,1,0,x) \biggr\} \, .
\eea

\begin{eqnarray}
\parbox{20mm}{\begin{fmfgraph*}(15,15)
\fmfleft{i1,i2}
\fmfright{o}
\fmfforce{0.2w,0.93h}{v2}
\fmfforce{0.2w,0.07h}{v1}
\fmfforce{0.8w,0.5h}{v5}
\fmfforce{0.2w,0.4h}{v9}
\fmfforce{0.5w,0.45h}{v10}
\fmfforce{0.2w,0.5h}{v11}
\fmf{photon}{i1,v1}
\fmf{photon}{i2,v2}
\fmf{plain}{v5,o}
\fmf{photon}{v2,v3}
\fmf{photon,tension=.25,right}{v3,v4}
\fmf{photon,tension=.25}{v3,v4}
\fmf{photon}{v4,v5}
\fmf{photon}{v1,v5}
\fmf{photon}{v1,v2}
\end{fmfgraph*} }  & = & \mu^{2(4-D)} 
\int \{ d^{D}k_{1} \} \{ d^{D}k_{2} \}
\frac{1}{{\mathcal D}_{1} {\mathcal D}_{2} {\mathcal D}_{4}^{2} 
{\mathcal D}_{5} {\mathcal D}_{7}} \\
& = & \left( \frac{\mu^{2}}{a} \right) ^{2 \epsilon} 
\sum_{i=-2}^{0} \epsilon^{i} K_{i} + {\mathcal O} \left( 
\epsilon \right) , 
\eea
where:
\bea
a^2 K_{-2} & = & \frac{3}{2x^2} \, , \\
a^2 K_{-1} & = & \frac{3}{2x^2} \bigl[ 1 - 2 H(0,x) \bigr] \, , \\
a^2 K_{0} & = & \frac{3}{2x^2} \bigl[ 3 - 2 \zeta(2) - 2 H(0,x) + 4 H(0,0,x)
\bigr] \, .
\eea

\begin{eqnarray}
\parbox{20mm}{\begin{fmfgraph*}(15,15)
\fmfleft{i1,i2}
\fmfright{o}
\fmfforce{0.2w,0.93h}{v2}
\fmfforce{0.2w,0.07h}{v1}
\fmfforce{0.8w,0.5h}{v5}
\fmfforce{0.2w,0.4h}{v9}
\fmfforce{0.5w,0.45h}{v10}
\fmfforce{0.2w,0.5h}{v11}
\fmf{photon}{i1,v1}
\fmf{photon}{i2,v2}
\fmf{plain}{v5,o}
\fmf{photon}{v2,v3}
\fmf{plain,tension=.25,right}{v3,v4}
\fmf{photon,tension=.25}{v3,v4}
\fmf{photon}{v4,v5}
\fmf{photon}{v1,v5}
\fmf{photon}{v1,v2}
\end{fmfgraph*} }  & = & \mu^{2(4-D)} 
\int \{ d^{D}k_{1} \} \{ d^{D}k_{2} \}
\frac{1}{{\mathcal D}_{1} {\mathcal D}_{4}^{2} {\mathcal D}_{5} 
{\mathcal D}_{7} {\mathcal D}_{13}} \\
& = & \left( \frac{\mu^{2}}{a} \right) ^{2 \epsilon} 
\sum_{i=-2}^{0} \epsilon^{i} L_{i} + {\mathcal O} \left( 
\epsilon \right) , 
\eea
where:
\bea
a^2 L_{-2} & = & - \frac{1}{2x} \biggl[ 1 - \frac{4}{x} \biggr] \, , \\
a^2 L_{-1} & = &  \frac{1}{2x} \biggl[ \frac{5}{x} + \biggl( 1 - 
\frac{4}{x} \biggr) H(0,x) - \biggl( 1 + \frac{2}{x} + \frac{1}{x^2} 
\biggr) H(-1,x) \biggr] \, , \\
a^2 L_{0} & = & \frac{1}{4x} \biggl[ \frac{17}{x} + 2 \zeta(2) 
\biggl( 1 - \frac{4}{x} \biggr) + \biggl( 3 - \frac{8}{x} \biggr) 
H(0,x) - 3 \biggl( 1 + \frac{4}{x} \nn\\
& & + \frac{3}{x^2} \biggr) H(-1,x) - 2 \biggl( 1 + \frac{4}{x} \biggr) 
H(0,0,x) - 6  H(0,-1,x) \nn\\
& & + 8 \biggl( 1 + \frac{2}{x} + \frac{1}{x^2} \biggr) H(-1,-1,x)
 \, .
\eea

\begin{eqnarray}
\parbox{20mm}{\begin{fmfgraph*}(15,15)
\fmfleft{i1,i2}
\fmfright{o}
\fmfforce{0.2w,0.93h}{v2}
\fmfforce{0.2w,0.07h}{v1}
\fmfforce{0.8w,0.5h}{v5}
\fmfforce{0.2w,0.4h}{v9}
\fmfforce{0.5w,0.45h}{v10}
\fmfforce{0.2w,0.5h}{v11}
\fmf{photon}{i1,v1}
\fmf{photon}{i2,v2}
\fmf{plain}{v5,o}
\fmf{photon}{v2,v3}
\fmf{photon,tension=.25,right}{v3,v4}
\fmf{photon,tension=.25}{v3,v4}
\fmf{photon}{v4,v5}
\fmf{photon}{v1,v5}
\fmf{plain}{v2,v1}
\end{fmfgraph*} } & = & \mu^{2(4-D)} 
\int \{ d^{D}k_{1} \} \{ d^{D}k_{2} \}
\frac{1}{{\mathcal D}_{2} {\mathcal D}_{4}^{2} {\mathcal D}_{5} 
{\mathcal D}_{7} {\mathcal D}_{12}} \\
& = & \left( \frac{\mu^{2}}{a} \right) ^{2 \epsilon} 
\sum_{i=-2}^{0} \epsilon^{i} M_{i} + {\mathcal O} \left( 
\epsilon \right) , 
\eea
where:
\bea
a^2 M_{-2} & = & - \frac{1}{2x}  \, , \\
a^2 M_{-1} & = &  - \frac{1}{2x} + \biggl( \frac{1}{x} + \frac{1}{(1-x)}
\biggr) H(0,x)  \, , \\
a^2 M_{0} & = & - \frac{1}{2x} \bigl( 7 - 2 \zeta(2) \bigr) + 
\biggl[ \frac{1}{x} + \frac{1}{(1-x)} \biggr] \bigl[ 2 \zeta(2) + H(0,x)
- 2 H(0,0,x)
\nn\\
& & + 2 H(1,0,x) \bigr] \, .
\eea

\begin{eqnarray}
\parbox{20mm}{\begin{fmfgraph*}(15,15)
\fmfleft{i1,i2}
\fmfright{o}
\fmfforce{0.2w,0.93h}{v2}
\fmfforce{0.2w,0.07h}{v1}
\fmfforce{0.2w,0.3h}{v3}
\fmfforce{0.2w,0.7h}{v4}
\fmfforce{0.8w,0.5h}{v5}
\fmf{photon}{i1,v1}
\fmf{photon}{i2,v2}
\fmf{plain}{v5,o}
\fmf{photon}{v2,v5}
\fmf{photon}{v1,v3}
\fmf{photon}{v2,v4}
\fmf{photon}{v1,v5}
\fmf{photon,right}{v4,v3}
\fmf{photon,right}{v3,v4}
\end{fmfgraph*} }  & = & \mu^{2(4-D)} 
\int \{ d^{D}k_{1} \} \{ d^{D}k_{2} \}
\frac{1}{{\mathcal D}_{1}^{2} {\mathcal D}_{2} {\mathcal D}_{3} 
{\mathcal D}_{4} {\mathcal D}_{5}} \\
& = & \left( \frac{\mu^{2}}{a} \right) ^{2 \epsilon} 
\sum_{i=-2}^{0} \epsilon^{i} N_{i} + {\mathcal O} \left( 
\epsilon \right) , 
\eea
where:
\bea
a^2 N_{-2} & = & - \frac{3}{4x^2}  \, , \\
a^2 N_{-1} & = &  \frac{3}{2x^2} H(0,x)  \, , \\
a^2 N_{0} & = & - \frac{3}{x^2} \bigl( 1 + H(0,0,x) \bigr) \, .
\eea

\begin{eqnarray}
\parbox{20mm}{\begin{fmfgraph*}(15,15)
\fmfleft{i1,i2}
\fmfright{o}
\fmfforce{0.2w,0.93h}{v2}
\fmfforce{0.2w,0.07h}{v1}
\fmfforce{0.2w,0.3h}{v3}
\fmfforce{0.2w,0.7h}{v4}
\fmfforce{0.8w,0.5h}{v5}
\fmf{photon}{i1,v1}
\fmf{photon}{i2,v2}
\fmf{plain}{v5,o}
\fmf{photon}{v2,v5}
\fmf{photon}{v1,v3}
\fmf{plain}{v2,v4}
\fmf{photon}{v1,v5}
\fmf{photon,right}{v4,v3}
\fmf{photon,right}{v3,v4}
\end{fmfgraph*} }  & = & \mu^{2(4-D)} 
\int \{ d^{D}k_{1} \} \{ d^{D}k_{2} \}
\frac{1}{{\mathcal D}_{1} {\mathcal D}_{2} {\mathcal D}_{3} 
{\mathcal D}_{4} {\mathcal D}_{5} {\mathcal D}_{12}} \\
& = & \left( \frac{\mu^{2}}{a} \right) ^{2 \epsilon} 
\sum_{i=-3}^{0} \epsilon^{i} O_{i} + {\mathcal O} \left( 
\epsilon \right) , 
\eea
where:
\bea
a^2 O_{-3} & = & \frac{1}{4x}  \, , \\
a^2 O_{-2} & = & \frac{1}{2x} \bigl[ 1 - H(0,x) \bigr]  \, , \\
a^2 O_{-1} & = &  \frac{1}{x} \bigl[ 1 - H(0,x) + H(0,0,x) - H(1,0,x) \bigr]
\, , \\
a^2 O_{0} & = & \frac{1}{x} \bigl[ 2 - 2 \zeta(3) - 2 H(0,x) + \zeta(2)
H(1,x) + 2 H(0,0,x) \nn\\
& & + 2 H(1,0,x) - 2 H(0,0,0,x) + H(0,1,0,x) - 2 H(1,0,0,x) \nn\\
& & + H(1,1,0,x)
\bigr] \, .
\eea

\begin{eqnarray}
\parbox{20mm}{\begin{fmfgraph*}(15,15)
\fmfleft{i1,i2}
\fmfright{o}
\fmfforce{0.2w,0.9h}{v2}
\fmfforce{0.2w,0.1h}{v1}
\fmfforce{0.2w,0.5h}{v3}
\fmfforce{0.8w,0.5h}{v5}
\fmf{photon}{i1,v1}
\fmf{photon}{i2,v2}
\fmf{plain}{v5,o}
\fmf{photon,tension=0}{v2,v5}
%\fmf{photon,tension=0}{v3,v4}
\fmf{plain,left=90}{v3,v3}
\fmf{photon,tension=.4}{v1,v4}
\fmf{photon,tension=.4}{v4,v5}
\fmf{photon,tension=0}{v1,v3}
\fmf{photon,tension=0}{v2,v3}
\end{fmfgraph*} }  & = & \mu^{2(4-D)} 
\int \{ d^{D}k_{1} \} \{ d^{D}k_{2} \}
\frac{1}{{\mathcal D}_{1}^{2} {\mathcal D}_{4} {\mathcal D}_{5} 
{\mathcal D}_{13}} \\
& = & \left( \frac{\mu^{2}}{a} \right) ^{2 \epsilon} 
\sum_{i=-2}^{0} \epsilon^{i} P_{i} + {\mathcal O} \left( 
\epsilon \right) , 
\eea
where:
\bea
a P_{-2} & = & \frac{2}{x^2}   \, , \\
a P_{-1} & = &  - \frac{2}{x^2} H(0,x) \, , \\
a P_{0} & = & \frac{2}{x^2} \bigl[ 1 - \zeta(2) + H(0,0,x)
\bigr] \, .
\eea
The above amplitude is actually a 5 denominator one; we include it 
here just for practical convenience.

\subsection{Crossed ladders \label{6den}}

The massless crossed ladder is a MI, i.e. it cannot be reduced to simpler
(massless) topologies by means of ibp identites. Its direct computation
is rather involved so we report the result as taken from 
ref. \cite{Gonsalves83}. We checked Gonsalves 
result for the crossed ladder with Eq.~(3.4) in \cite{VanNeerven86}.
If we include in the latter equation the dimensional factor
$x^{-2-2 \epsilon}$ (in our notation), a complete agreement between the 
two computations is found\footnote{Note that the definition of 
$\epsilon$ in the above cited Van Neerven paper
is different from ours: $\epsilon_{VN} = -2 \epsilon $. }.

\begin{eqnarray}
\parbox{20mm}{\begin{fmfgraph*}(15,15)
\fmfleft{i1,i2}
\fmfright{o}
\fmf{photon}{i1,v1}
\fmf{photon}{i2,v2}
\fmf{plain}{v5,o}
\fmf{photon,tension=.3}{v2,v3}
\fmf{photon,tension=.3}{v3,v5}
\fmf{photon,tension=.3}{v1,v4}
\fmf{photon,tension=.3}{v4,v5}
\fmf{photon,tension=0}{v2,v4}
\fmf{photon,tension=0}{v1,v3}
\end{fmfgraph*} }  & = & \mu^{2(4-D)} 
\int \{ d^{D}k_{1} \} \{ d^{D}k_{2} \}
\frac{1}{{\mathcal D}_{1} {\mathcal D}_{2} {\mathcal D}_{3} 
{\mathcal D}_{4} {\mathcal D}_{6} {\mathcal D}_{11}} \\
& = & \left( \frac{\mu^{2}}{a} \right) ^{2 \epsilon} 
\sum_{i=-4}^{0} \epsilon^{i} Q_{i} + {\mathcal O} \left( 
\epsilon \right) , 
\eea
where:
\bea
a^{2} Q_{-4} & = & \frac{1}{x^2}  \, , \\
a^{2} Q_{-3} & = & - \frac{2}{x^2} H(0,x) \, , \\
a^{2} Q_{-2} & = & - \frac{1}{x^2} \bigl[ 7 \zeta(2) - 
                            4 H(0,0,x) \bigr] \, , \\
a^{2} Q_{-1} & = &  - \frac{1}{x^2} \bigl[ 27 \zeta(3) - 14
\zeta(2) H(0,x) + 8 H(0,0,0,x) \bigr] \, , \\
a^{2} Q_{0} & = & - \frac{1}{x^2} \biggl[ 
\frac{57 \zeta^{2}(2)}{5} - 54 \zeta(3) H(0,x) + 28 \zeta(2) H(0,0,x)
\nn\\
& & - 16 H(0,0,0,0,x) \biggr] \, .
\eea
Note that the above diagram is the only one having 
a fourth order (infrared) pole in $\epsilon$ together,
as we have seen, with the planar ladder.

\begin{eqnarray}
\parbox{20mm}{\begin{fmfgraph*}(15,15)
\fmfleft{i1,i2}
\fmfright{o}
\fmf{photon}{i1,v1}
\fmf{photon}{i2,v2}
\fmf{plain}{v5,o}
\fmf{photon,tension=.3}{v2,v3}
\fmf{photon,tension=.3}{v3,v5}
\fmf{photon,tension=.3}{v1,v4}
\fmf{photon,tension=.3}{v4,v5}
\fmf{plain,tension=0}{v2,v4}
\fmf{photon,tension=0}{v1,v3}
\end{fmfgraph*} }  & = & \mu^{2(4-D)} 
\int \{ d^{D}k_{1} \} \{ d^{D}k_{2} \}
\frac{1}{{\mathcal D}_{1} {\mathcal D}_{2} {\mathcal D}_{3} 
{\mathcal D}_{6} {\mathcal D}_{11} {\mathcal D}_{15}} \\
& = & \left( \frac{\mu^{2}}{a} \right) ^{2 \epsilon} 
\sum_{i=-2}^{0} \epsilon^{i} R_{i} + {\mathcal O} \left( 
\epsilon \right) , 
\eea
where:
\bea
a^{2} R_{-2} & = & \frac{1}{2} \biggl[ \frac{1}{x} + \frac{1}{(1-x)} 
\biggr] \bigl[ \zeta(2) - 2 H(0,-1,x) \bigr] \, , \\
a^{2} R_{-1} & = & \frac{1}{2} \biggl[ \frac{1}{x} + \frac{1}{(1-x)} 
\biggr] \bigl[ \zeta(3) - 2 \zeta(2) ( H(0,x) - 2 H(1,x) ) \nn\\
& & + 8 H(0,-1,-1,x) - 4 H(0,0,-1,x) + 2 H(0,1,0,x) \nn\\
& & - 8 H(1,0,-1,x) \bigr]  \, , \\
a^{2} R_{0} & = & \biggl[ \frac{1}{x} + \frac{1}{(1-x)} 
\biggr] \biggl\{ - \frac{\zeta^{2}(2)}{10} - \zeta(3) \bigl[ H(0,x) - 
2 H(1,x) \bigr] \nn\\
& & + \zeta(2) \bigl[ 2 H(0,0,x) + 8 H(1,1,x) - 4 H(1,0,x)- 6 H(0,-1,x) 
\nn\\
& & + 7 H(0,1,x) \bigr]- 16 H(0,-1,-1,-1,x) + 8 H(0,-1,0,-1,x) \nn\\
& & - 2 H(0,-1,0,0,x) + 8 H(0,0,-1,-1,x) - 2 H(0,0,0,-1,x) \nn\\
& & + H(0,0,1,0,x) - 8 H(0,1,0,-1,x) - H(0,1,0,0,x) \nn\\
& & + 3 H(0,1,1,0,x) + 16 H(1,0,-1,-1,x) - 8 H(1,0,0,-1,x) \nn\\
& & + 4 H(1,0,1,0,x) - 16 H(1,1,0,-1,x) \biggl\} \, .
\eea
The differential equation for the above amplitude is a rather
lengthy expression, as it involves fourteen subtopologies.

\section{Large momentum expansion, $|s| \gg m^2$ \label{GrandiP}}

In this section we give the asymptotic expansions of
the six denominator scalar integrals, i.e. the expansion in powers
of $1/x$ and $L=\log(x)$ up to the order $1/x^{4}$ included. 
As already noted in the introduction, these 
expansions are relevant for the study of the structure of the
infrared logarithms coming from multiple emission.
The exact expression of the massless diagrams involves only 
an overall power of $x$ and powers of $L$, so their expansion coincides 
with the former and is not duplicated.
The coefficients not explicitly given are zero.
\be
\parbox{20mm}{\begin{fmfgraph*}(15,15)
\fmfleft{i1,i2}
\fmfright{o}
\fmf{photon}{i1,v1}
\fmf{photon}{i2,v2}
\fmf{plain}{v5,o}
\fmf{photon,tension=.3}{v2,v3}
\fmf{photon,tension=.3}{v3,v5}
\fmf{photon,tension=.3}{v1,v4}
\fmf{photon,tension=.3}{v4,v5}
\fmf{plain,tension=0}{v2,v1}
\fmf{photon,tension=0}{v4,v3}
\end{fmfgraph*} } = \left( \frac{\mu^{2}}{a} \right) ^{2 \epsilon}
\sum_{j=2}^{4} \frac{1}{x^{j}} B_{-j}^{0} 
+ {\mathcal O} \left(\frac{1}{x^{5}} \right) \, , 
\ee
where:
\bea
a^{2} B^{0}_{-2} & = & \frac{37 \zeta^{2}(2)}{5} 
                       - 4 \zeta(3) L 
		       + 2 \zeta(2) L^{2} 
		       + \frac{1}{24} L^{4} \, , \\
a^{2} B^{0}_{-3} & = & - 10 
                       - 2 \zeta(2) 
		       + 4 \zeta(3) 
		       - 4 \bigl[ 1 + \zeta(2) \bigr] L 
		       - \frac{1}{2} L^{2} 
		       - \frac{1}{6} L^{3} \, , \\
a^{2} B^{0}_{-4} & = & - \frac{43}{8} 
                       + \frac{\zeta(2)}{2} 
		       + 2 \zeta(3) 
		       - \biggl[ \frac{5}{2} 
		       + 2 \zeta(2) 
                         \biggr] L 
		       - \frac{1}{8} L^{2} 
		       - \frac{1}{12} L^{3} \, .
\eea
\be
\parbox{20mm}{\begin{fmfgraph*}(15,15)
\fmfleft{i1,i2}
\fmfright{o}
\fmf{photon}{i1,v1}
\fmf{photon}{i2,v2}
\fmf{plain}{v5,o}
\fmf{photon,tension=.3}{v2,v3}
\fmf{photon,tension=.3}{v3,v5}
\fmf{photon,tension=.3}{v1,v4}
\fmf{photon,tension=.3}{v4,v5}
\fmf{photon,tension=0}{v2,v1}
\fmf{plain,tension=0}{v4,v3}
\end{fmfgraph*} } = \left( \frac{\mu^{2}}{a} \right) ^{2 \epsilon}
\sum_{j=2}^{4} \frac{1}{x^{j}} \biggl[
\sum_{i=-2}^{0} \epsilon^{i} C_{-j}^{i} \biggr]  
+ {\mathcal O} \left( \frac{1}{x^{5}}\right) \, , 
\ee
where:
\bea
a^{2} C^{-2}_{-2} & = & 2 \zeta(2) + \frac{1}{2} L^{2}  \, , \\
a^{2} C^{-1}_{-2} & = & 3 \zeta(3) - \zeta(2) L - \frac{1}{3} L^{2} 
\, , \\
a^{2} C^{0}_{-2} & = & \frac{61 \zeta^{2}(2)}{10} - 11 \zeta(3) L + 2
\zeta(2) L^{2} + \frac{1}{6} L^{4}
\, , \\
a^{2} C^{-2}_{-3} & = & - 1 - L  \, , \\
a^{2} C^{-1}_{-3} & = & \zeta(2) - L + L^{2} \, , \\
a^{2} C^{0}_{-3} & = & - 26 + 3 \zeta(2) 
+ 11 \zeta(3) - 2 \bigl[ 3 + 2 \zeta(2) 
\bigr] L + 3 L^{2} - \frac{2}{3} L^{3} \, , \\
a^{2} C^{-2}_{-4} & = & - \frac{1}{4} - \frac{1}{2} L  \, , \\
a^{2} C^{-1}_{-4} & = & - \frac{5}{4} + \frac{\zeta(2)}{2} 
- \frac{7}{4} L + \frac{1}{2} L^{2} \, , \\
a^{2} C^{0}_{-4} & = & - \frac{117}{8} + \frac{17 \zeta(2)}{4} 
+ \frac{11 \zeta(3)}{2} - \biggl[ \frac{5}{2} + 2 \zeta(2) 
\biggr] L + \frac{13}{4} L^{2} - \frac{1}{3} L^{3} \, , 
\eea
The expansion above is in agreement with that obtained in \cite{smirnov} 
by using a kinematic region decomposition of the integrand.
\begin{equation}
\parbox{20mm}{\begin{fmfgraph*}(15,15)
\fmfleft{i1,i2}
\fmfright{o}
\fmf{photon}{i1,v1}
\fmf{photon}{i2,v2}
\fmf{plain}{v5,o}
\fmf{plain,tension=.3}{v2,v3}
\fmf{photon,tension=.3}{v3,v5}
\fmf{photon,tension=.3}{v1,v4}
\fmf{photon,tension=.3}{v4,v5}
\fmf{photon,tension=0}{v2,v1}
\fmf{photon,tension=0}{v4,v3}
\end{fmfgraph*} } = \left( \frac{\mu^{2}}{a} \right) ^{2 \epsilon}
\sum_{j=2}^{4} \frac{1}{x^{j}} \biggl[ \sum_{i=-2}^{0} \epsilon^{i}
 E_{-j}^{i}  \biggr] 
+ {\mathcal O} \left(\frac{1}{x^{5}}\right) \, , 
\end{equation}
where:
\bea
a^{2} E^{-2}_{-2} & = & \zeta(2) 
                       + \frac{1}{4} L^{2} \, , \\
a^{2} E^{-1}_{-2} & = & \frac{5 \zeta(3)}{2} 
			- \frac{3 \zeta(2)}{2} L  
                        - \frac{1}{4} L^{3}   \, , \\
a^{2} E^{0}_{-2} & = & \frac{7 \zeta^{2}(2)}{4} 
		       - 8 \zeta(3) L
                       + \frac{7 \zeta(2)}{4} L^{2}
                       + \frac{7}{48} L^{4}  \, , \\
a^{2} E^{-2}_{-3} & = & - \frac{1}{2} 
                       - \zeta(2) 
                       - \frac{1}{4} L^{2} \, , \\
a^{2} E^{-1}_{-3} & = & \frac{1}{2} 
                        - 2 \zeta(2)
                        - \frac{5 \zeta(3)}{2} 
			+ \frac{3 \zeta(2)}{2} L  
                        - \frac{1}{2} L^{2}  
                        + \frac{1}{4} L^{3} \, , \\
a^{2} E^{0}_{-3} & = & - \frac{17}{2} 
                       + 2 \zeta(2) 
                       - \frac{7 \zeta^{2}(2)}{4} 
		       - 5 \zeta(3) 
                       - 4 L 
                       + 3 \zeta(2) L
                       + 8 \zeta(3) L \nn\\
& & 
                       - \frac{7 \zeta(2)}{4} L^{2}
                       + \frac{1}{2} L^{3}
                       - \frac{7}{48} L^{4} \, , \\
a^{2} E^{-2}_{-4} & = & \frac{5}{8} 
                       + \zeta(2) 
                       + \frac{1}{4} L^{2} \, , \\
a^{2} E^{-1}_{-4} & = & - \frac{1}{16} 
                        + 3 \zeta(2)
                        + \frac{5 \zeta(3)}{2} 
			- \frac{3 \zeta(2)}{2} L  
                        + \frac{3}{4} L^{2}  
                        - \frac{1}{4} L^{3} \, , \\
a^{2} E^{0}_{-4} & = & \frac{297}{32} 
                       - \frac{\zeta(2)}{2} 
                       + \frac{7 \zeta^{2}(2)}{4} 
		       + \frac{15 \zeta(3)}{2}
                       + \frac{11}{2} L 
                       - \frac{9 \zeta(2)}{2} L 
                       - 8 \zeta(3) L \nn\\
& & 
                       + \frac{1}{2} L^{2}
                       + \frac{7 \zeta(2)}{4} L^{2}
                       - \frac{3}{4} L^{3}
                       + \frac{7}{48} L^{4} \, ,
\eea
\be
\parbox{20mm}{\begin{fmfgraph*}(15,15)
\fmfleft{i1,i2}
\fmfright{o}
\fmfforce{0.2w,0.9h}{v2}
\fmfforce{0.2w,0.1h}{v1}
\fmfforce{0.2w,0.5h}{v3}
\fmfforce{0.8w,0.5h}{v5}
\fmf{photon}{i1,v1}
\fmf{photon}{i2,v2}
\fmf{plain}{v5,o}
\fmf{photon,tension=0}{v2,v5}
\fmf{photon,tension=0}{v3,v4}
\fmf{plain,tension=.4}{v1,v4}
\fmf{photon,tension=.4}{v4,v5}
\fmf{photon,tension=0}{v1,v3}
\fmf{photon,tension=0}{v2,v3}
\end{fmfgraph*} } = \left( \frac{\mu^{2}}{a} \right) ^{2 \epsilon}
\sum_{j=1}^{4} \frac{1}{x^{j}} \biggl[ \sum_{i=-3}^{0} \epsilon^{i}
I_{-j}^{i} \biggr]  
+ {\mathcal O} \left(\frac{1}{x^{5}}\right) \, , 
\ee
where:
\bea
a^{2} I^{-3}_{-1} & = & - \frac{3}{4} \, , \\
a^{2} I^{-1}_{-1} & = & - \zeta(2) \, , \\
a^{2} I^{0}_{-1} & = & 4 \zeta(3) \, , \\
a^{2} I^{-2}_{-2} & = & - \frac{1}{2} - \frac{1}{2} L \, , \\
a^{2} I^{-1}_{-2} & = & 2 + 2 L + L^{2}  \, , \\
a^{2} I^{0}_{-2} & = & - 5 
                       + 3 \zeta(2)
                       - 4 \zeta(3)
                       - 5 L
                       + 3 \zeta(2) L
                       - \frac{5}{2} L^{2} 
                       - \frac{5}{6} L^{3} \, , \\
a^{2} I^{-2}_{-3} & = & - \frac{1}{4} \, , \\
a^{2} I^{-1}_{-3} & = & \frac{1}{2} + L  \, , \\
a^{2} I^{0}_{-3} & = &   \frac{21}{8} 
                       + \frac{3 \zeta(2)}{2}
                       + \frac{1}{4} L
                       - \frac{5}{4} L^{2} \, , \\
a^{2} I^{-2}_{-4} & = & \frac{1}{12} \, , \\
a^{2} I^{-1}_{-4} & = & \frac{2}{9} - \frac{1}{3} L \, , \\
a^{2} I^{0}_{-4} & = & - \frac{211}{216} 
                       - \frac{\zeta(2)}{2}
                       - \frac{47}{36} L
                       + \frac{5}{12} L^{2} \, , 
\eea
\be
\parbox{20mm}{\begin{fmfgraph*}(15,15)
\fmfleft{i1,i2}
\fmfright{o}
\fmfforce{0.2w,0.9h}{v2}
\fmfforce{0.2w,0.1h}{v1}
\fmfforce{0.2w,0.5h}{v3}
\fmfforce{0.8w,0.5h}{v5}
\fmf{photon}{i1,v1}
\fmf{photon}{i2,v2}
\fmf{plain}{v5,o}
\fmf{photon,tension=0}{v2,v5}
\fmf{photon,tension=0}{v3,v4}
\fmf{photon,tension=.4}{v1,v4}
\fmf{photon,tension=.4}{v4,v5}
\fmf{photon,tension=0}{v1,v3}
\fmf{plain,tension=0}{v2,v3}
\end{fmfgraph*} } = \left( \frac{\mu^{2}}{a} \right) ^{2 \epsilon}
\sum_{j=1}^{4} \frac{1}{x^{j}} \biggl[ \sum_{i=-2}^{0} \epsilon^{i}
 J_{-j}^{i} \biggr]  
+ {\mathcal O} \left(\frac{1}{x^{5}}\right) \, , 
\ee
where:
\bea
a^{2} J^{-2}_{-1} & = & \frac{\zeta(2)}{2} \, , \\
a^{2} J^{-1}_{-1} & = & - \frac{\zeta(3)}{2}  \, , \\
a^{2} J^{0}_{-1} & = & \frac{6 \zeta^{2}(2)}{5} \, , \\
a^{2} J^{-1}_{-2} & = & 1  
                       + \zeta(2)
                       +  L
                       + \frac{1}{2} L^{2}   \, , \\
a^{2} J^{0}_{-2} & = & - 4 
                       + \zeta(2)
                       - 4 \zeta(3)
                       - 4 L
                       + \zeta(2) L
                       - 2 L^{2}
                       - \frac{2}{3} L^{3} \, , \\
a^{2} J^{-1}_{-3} & = & - \frac{5}{8}  
                       + \frac{\zeta(2)}{2}
                       - \frac{1}{4} L
                       + \frac{1}{4} L^{2}   \, , \\
a^{2} J^{0}_{-3} & = & \frac{13}{8} 
                       + \frac{7 \zeta(2)}{4}
                       - 2 \zeta(3)
                       + \frac{9}{4} L
                       + \frac{\zeta(2)}{2} L
                       + \frac{3}{4} L^{2}
                       - \frac{1}{3} L^{3}  \, , \\
a^{2} J^{-1}_{-4} & = & - \frac{59}{108}  
                       + \frac{\zeta(2)}{3}
                       - \frac{7}{18} L
                       + \frac{1}{6} L^{2}  \, , \\
a^{2} J^{0}_{-4} & = & \frac{67}{648} 
                       + \frac{29 \zeta(2)}{18}
                       - \frac{4 \zeta(3)}{3}
                       + \frac{155}{108} L
                       + \frac{\zeta(2)}{3} L
                       + \frac{37}{36} L^{2}
                       - \frac{2}{9} L^{3} \, ,
\eea
\be
\parbox{20mm}{\begin{fmfgraph*}(15,15)
\fmfleft{i1,i2}
\fmfright{o}
\fmfforce{0.2w,0.93h}{v2}
\fmfforce{0.2w,0.07h}{v1}
\fmfforce{0.8w,0.5h}{v5}
\fmfforce{0.2w,0.4h}{v9}
\fmfforce{0.5w,0.45h}{v10}
\fmfforce{0.2w,0.5h}{v11}
\fmf{photon}{i1,v1}
\fmf{photon}{i2,v2}
\fmf{plain}{v5,o}
\fmf{photon}{v2,v3}
\fmf{plain,tension=.25,right}{v3,v4}
\fmf{photon,tension=.25}{v3,v4}
\fmf{photon}{v4,v5}
\fmf{photon}{v1,v5}
\fmf{photon}{v1,v2}
\end{fmfgraph*} } = \left( \frac{\mu^{2}}{a} \right) ^{2 \epsilon}
\sum_{j=1}^{4} \frac{1}{x^{j}} \biggl[ \sum_{i=-2}^{0} \epsilon^{i}
 L_{-j}^{i} \biggr]  
+ {\mathcal O} \left(\frac{1}{x^{5}}\right) \, , 
\ee
where:
\bea
a^{2} L^{-2}_{-1} & = & - \frac{1}{2} \, , \\
a^{2} L^{0}_{-1} & = & - \zeta(2) \, , \\
a^{2} L^{-2}_{-2} & = & 2 \, , \\
a^{2} L^{-1}_{-2} & = & 2  
                       - 3  L  \, , \\
a^{2} L^{0}_{-2} & = & 5 
                       - 2 \zeta(2)
                       - 3 L
                       + 3 L^{2} \, , \\
a^{2} L^{-1}_{-3} & = & - \frac{3}{4} 
                       - \frac{1}{2} L  \, , \\
a^{2} L^{0}_{-3} & = & - 2
                       + \frac{3}{4} L
                       + L^{2} \, , \\
a^{2} L^{-1}_{-4} & = & - \frac{1}{6}  \, , \\
a^{2} L^{0}_{-4} & = & \frac{1}{6} 
                       + \frac{2}{3} L \, , 
\eea
\be
\parbox{20mm}{\begin{fmfgraph*}(15,15)
\fmfleft{i1,i2}
\fmfright{o}
\fmfforce{0.2w,0.93h}{v2}
\fmfforce{0.2w,0.07h}{v1}
\fmfforce{0.8w,0.5h}{v5}
\fmfforce{0.2w,0.4h}{v9}
\fmfforce{0.5w,0.45h}{v10}
\fmfforce{0.2w,0.5h}{v11}
\fmf{photon}{i1,v1}
\fmf{photon}{i2,v2}
\fmf{plain}{v5,o}
\fmf{photon}{v2,v3}
\fmf{photon,tension=.25,right}{v3,v4}
\fmf{photon,tension=.25}{v3,v4}
\fmf{photon}{v4,v5}
\fmf{photon}{v1,v5}
\fmf{plain}{v2,v1}
\end{fmfgraph*} } = \left( \frac{\mu^{2}}{a} \right) ^{2 \epsilon}
\sum_{j=1}^{4} \frac{1}{x^{j}} \biggl[ \sum_{i=-2}^{0} \epsilon^{i}
M_{-j}^{i} \biggr]  
+ {\mathcal O} \left(\frac{1}{x^{5}}\right) \, , 
\ee
where:
\bea
a^{2} M^{-2}_{-1} & = & - \frac{1}{2} \, , \\
a^{2} M^{-1}_{-1} & = & - \frac{1}{2} \, , \\
a^{2} M^{0}_{-1} & = & - \frac{3}{2}
                       - \zeta(2) \, , \\
a^{2} M^{-1}_{-2} & = & - L  \, , \\
a^{2} M^{0}_{-2} & = & 2 \zeta(2)
                       - L
                       + 2 L^{2} \, , \\
a^{2} M^{-1}_{-3} & = & - L  \, , \\
a^{2} M^{0}_{-3} & = & 2 
                       + 2 \zeta(2)
                       - 3 L
                       + 2 L^{2}  \, , \\
a^{2} M^{-1}_{-4} & = & - L  \, , \\
a^{2} M^{0}_{-4} & = & \frac{5}{2} 
                       + 2 \zeta(2)
                       - 4 L
                       + 2 L^{2} \, ,
\eea
\be
\parbox{20mm}{\begin{fmfgraph*}(15,15)
\fmfleft{i1,i2}
\fmfright{o}
\fmfforce{0.2w,0.93h}{v2}
\fmfforce{0.2w,0.07h}{v1}
\fmfforce{0.2w,0.3h}{v3}
\fmfforce{0.2w,0.7h}{v4}
\fmfforce{0.8w,0.5h}{v5}
\fmf{photon}{i1,v1}
\fmf{photon}{i2,v2}
\fmf{plain}{v5,o}
\fmf{photon}{v2,v5}
\fmf{photon}{v1,v3}
\fmf{plain}{v2,v4}
\fmf{photon}{v1,v5}
\fmf{photon,right}{v4,v3}
\fmf{photon,right}{v3,v4}
\end{fmfgraph*} } = \left( \frac{\mu^{2}}{a} \right) ^{2 \epsilon}
\sum_{j=1}^{4} \frac{1}{x^{j}} \biggl[ \sum_{i=-3}^{0} \epsilon^{i}
O_{-j}^{i} \biggr]  
+ {\mathcal O} \left(\frac{1}{x^{5}}\right) \, , 
\ee
where:
\bea
a^{2} O^{-3}_{-1} & = & \frac{1}{4} \, , \\
a^{2} O^{-2}_{-1} & = &   \frac{1}{2}
                        - \frac{1}{2} L \, , \\
a^{2} O^{-1}_{-1} & = & 1
                        - 2 \zeta(2)
                        - L  \, , \\
a^{2} O^{0}_{-1} & = & 2
                       - 4 \zeta(2)
                       - \zeta(3)
                       - 2 L 
                       - \zeta(2) L \, , \\
a^{2} O^{-1}_{-2} & = &   1
                        + L  \, , \\
a^{2} O^{0}_{-2} & = & - 1
                       - \zeta(2)
                       -  L 
                       - \frac{3}{2} L^{2} \, , \\
a^{2} O^{-1}_{-3} & = &   \frac{1}{4}
                        + \frac{1}{2} L  \, , \\
a^{2} O^{0}_{-3} & = & \frac{7}{8} 
                       - \frac{\zeta(2)}{2}
                       + \frac{3}{4} L 
                       - \frac{3}{4} L^{2} \, , \\
a^{2} O^{-1}_{-4} & = &   \frac{1}{9}
                        + \frac{1}{3} L  \, , \\
a^{2} O^{0}_{-4} & = & \frac{25}{36} 
                       - \frac{\zeta(2)}{3}
                       + \frac{5}{6} L 
                       - \frac{1}{2} L^{2} \, ,
\eea
\be
\parbox{20mm}{\begin{fmfgraph*}(15,15)
\fmfleft{i1,i2}
\fmfright{o}
\fmf{photon}{i1,v1}
\fmf{photon}{i2,v2}
\fmf{plain}{v5,o}
\fmf{photon,tension=.3}{v2,v3}
\fmf{photon,tension=.3}{v3,v5}
\fmf{photon,tension=.3}{v1,v4}
\fmf{photon,tension=.3}{v4,v5}
\fmf{plain,tension=0}{v2,v4}
\fmf{photon,tension=0}{v1,v3}
\end{fmfgraph*} } = \left( \frac{\mu^{2}}{a} \right) ^{2 \epsilon}
\sum_{j=2}^{4} \frac{1}{x^{j}}  \biggl[ \sum_{i=-2}^{0} \epsilon^{i}
 R_{-j}^{i} \biggr] + {\mathcal O} \left( \frac{1}{x^{5}} \right) \, , 
\ee
where:
\bea
a^{2} R^{-2}_{-2} & = & \frac{\zeta(2)}{2} 
                        + \frac{1}{2} L^{2} \, , \\
a^{2} R^{-1}_{-2} & = & - \frac{21 \zeta(3)}{2}
                        + 3 \zeta(2) L  
                        - \frac{5}{6} L^{3} \, , \\
a^{2} R^{0}_{-2} & = & - \frac{163 \zeta^{2}(2)}{10} 
                       + 36 \zeta(3) L
                       - \frac{19 \zeta(2)}{2} L^{2} 
                       + \frac{13}{24} L^{4} \, , \\
a^{2} R^{-2}_{-3} & = & - 1 
                        + \frac{\zeta(2)}{2} 
                        + \frac{1}{2} L^{2} \, , \\
a^{2} R^{-1}_{-3} & = & 8
                        + 2 \zeta(2)
			- \frac{21 \zeta(3)}{2} 
			+ 9 L
                        + 3 \zeta(2) L
                        + 2 L^{2} 
                        - \frac{5}{6} L^{3}  \, , \\
a^{2} R^{0}_{-3} & = &  34
                       + 15 \zeta(2)
		       - \frac{163}{10} \zeta^{2}(2)
		       - 42 \zeta(3)
                       - 2 L 
                       + 12 \zeta(2) L
                       + 36 \zeta(3) L \nn\\
& & 
                       - 13 L^{2}
                       - \frac{19 \zeta(2)}{2} L^{2} 
                       - \frac{10}{3} L^{3}
                       + \frac{13}{24} L^{4} \, , \\
a^{2} R^{-2}_{-4} & = & - \frac{3}{4}
                        + \frac{\zeta(2)}{2}
                        + \frac{1}{2} L^{2} \, , \\
a^{2} R^{-1}_{-4} & = & \frac{15}{2} 
                        + 3 \zeta(2)
                        - \frac{21 \zeta(3)}{2} 
                        + \frac{37}{4} L 
                        + 3 \zeta(2)  L 
                        + 3 L^{2}
                        - \frac{5}{6} L^{3}  \, , \\
a^{2} R^{0}_{-4} & = &  \frac{453}{8} 
                       + \frac{87 \zeta(2)}{4}  
                       - \frac{163 \zeta^{2}(2)}{10} 
		       - 63 \zeta(3)
                       + \frac{31}{2} L 
                       + 18 \zeta(2) L
                       + 36 \zeta(3) L \nn\\
& & 
                       - \frac{39}{4} L^{2}
                       - \frac{19 \zeta(2)}{2} L^{2} 
                       - 5 L^{3}
                       + \frac{13}{24} L^{4} \, .
\eea

\section{Small momentum expansions, $|s| \ll m^2$ \label{PiccoliP} }

In this section we present the small momentum expansions of the 
six-denomi\-na\-tor diagrams, i.e. expansions in powers of $x$ and 
$L=\log(x)$ up to first order in $x$ included. As in the previous 
section, we do not consider the massless scalar integrals since their 
expansion is trivial.
The coefficients not explicitly given are zero.
\be
\parbox{20mm}{\begin{fmfgraph*}(15,15)
\fmfleft{i1,i2}
\fmfright{o}
\fmf{photon}{i1,v1}
\fmf{photon}{i2,v2}
\fmf{plain}{v5,o}
\fmf{photon,tension=.3}{v2,v3}
\fmf{photon,tension=.3}{v3,v5}
\fmf{photon,tension=.3}{v1,v4}
\fmf{photon,tension=.3}{v4,v5}
\fmf{plain,tension=0}{v2,v1}
\fmf{photon,tension=0}{v4,v3}
\end{fmfgraph*} } = \left( \frac{\mu^{2}}{a} \right) ^{2 \epsilon}
\sum_{j=-1}^{1}
x^{j} B_{j}^{0} 
+ {\mathcal O} (x^2) \, , 
\ee
where:
\bea
a^{2} B^{0}_{-1} & = & 6 \zeta(3)  \, , \\
a^{2} B^{0}_{0} & = & - 3
                       - \zeta(2) 
                       + 3 \zeta(3) 
		       + 2 L
                       - \frac{1}{2} L^{2} \, , \\
a^{2} B^{0}_{1} & = & - \frac{17}{12} 
                       - \zeta(2) 
                       + 2 \zeta(3)
                       + \frac{5}{4} L 
                       - \frac{1}{2} L^{2} \, , 
\eea
\be
\parbox{20mm}{\begin{fmfgraph*}(15,15)
\fmfleft{i1,i2}
\fmfright{o}
\fmf{photon}{i1,v1}
\fmf{photon}{i2,v2}
\fmf{plain}{v5,o}
\fmf{photon,tension=.3}{v2,v3}
\fmf{photon,tension=.3}{v3,v5}
\fmf{photon,tension=.3}{v1,v4}
\fmf{photon,tension=.3}{v4,v5}
\fmf{photon,tension=0}{v2,v1}
\fmf{plain,tension=0}{v4,v3}
\end{fmfgraph*} } = \left( \frac{\mu^{2}}{a} \right) ^{2 \epsilon}
\sum_{j=-1}^{1} x^{j} \biggl[ 
\sum_{i=-2}^{0} \epsilon^{i} 
C_{j}^{i} \biggr] 
+ {\mathcal O} (x^2) \, , 
\ee
where:
\bea
a^{2} C^{-2}_{-1} & = & 1 - L \, , \\
a^{2} C^{-1}_{-1} & = & 3
                        - \zeta(2)
                        - 3 L
                        + \frac{3}{2} L^{2}  \, , \\
a^{2} C^{0}_{-1} & = & 7
                        - 3 \zeta(2)
                        - 2 \zeta(3)
                        - 7 L
                        + 3 \zeta(2) L
                        + \frac{7}{2} L^{2}
                        - \frac{7}{6} L^{3} \, , \\
a^{2} C^{-2}_{0} & = & \frac{1}{4} 
                       - \frac{1}{2} L \, , \\
a^{2} C^{-1}_{0} & = & \frac{5}{8}
                        - \frac{\zeta(2)}{2}
                        - \frac{9}{4} L
                        + \frac{3}{4} L^{2} \, , \\
a^{2} C^{0}_{0} & = &  \frac{89}{16}
                        - \frac{17 \zeta(2)}{4}
                        - \zeta(3)
                        - \frac{49}{8} L
                        + \frac{3 \zeta(2)}{2} L
                        + \frac{17}{8} L^{2}
                        - \frac{7}{12} L^{3} \, , \\
a^{2} C^{-2}_{1} & = & \frac{1}{9} 
                       - \frac{1}{3} L \, , \\
a^{2} C^{-1}_{1} & = & \frac{31}{36}
                        - \frac{\zeta(2)}{3}
                        - \frac{11}{6} L
                        + \frac{1}{2} L^{2} \, , \\
a^{2} C^{0}_{1} & = & \frac{4709}{648}
                        - \frac{23 \zeta(2)}{6}
                        - \frac{2 \zeta(3)}{3}
                        - \frac{160}{27} L
                        + \zeta(2) L
                        + \frac{59}{36} L^{2}
                        - \frac{7}{18} L^{3} \, , 
\eea
\begin{equation}
\parbox{20mm}{\begin{fmfgraph*}(15,15)
\fmfleft{i1,i2}
\fmfright{o}
\fmf{photon}{i1,v1}
\fmf{photon}{i2,v2}
\fmf{plain}{v5,o}
\fmf{plain,tension=.3}{v2,v3}
\fmf{photon,tension=.3}{v3,v5}
\fmf{photon,tension=.3}{v1,v4}
\fmf{photon,tension=.3}{v4,v5}
\fmf{photon,tension=0}{v2,v1}
\fmf{photon,tension=0}{v4,v3}
\end{fmfgraph*} } = \left( \frac{\mu^{2}}{a} \right) ^{2 \epsilon}
\sum_{j=-1}^{1} x^{j} \biggl[ 
\sum_{i=-2}^{0} \epsilon^{i} 
E_{j}^{i} \biggr] 
+ {\mathcal O} (x^2) \, , 
\end{equation}
where:
\bea
a^{2} E^{-2}_{-1} & = & \frac{\zeta(2)}{2} \, , \\
a^{2} E^{-1}_{-1} & = & \frac{\zeta(3)}{2}
                        - \zeta(2) L \, , \\
a^{2} E^{0}_{-1} & = & - \frac{\zeta^{2}(2)}{10} 
                        - \zeta(3) L
                        + \zeta(2) L^{2} \, , \\
a^{2} E^{-2}_{0} & = & \frac{1}{2} 
                       - \frac{\zeta(2)}{2} \, , \\
a^{2} E^{-1}_{0} & = & \frac{5}{2}
                        - \zeta(2)
                        - \frac{\zeta(3)}{2}
                        -  L
                        + \zeta(2) L \, , \\
a^{2} E^{0}_{0} & = &  \frac{13}{2} \! 
                        - 5 \zeta(2)  \! 
			+ \frac{1}{10} \zeta^{2}(2) \! 
                        - \zeta(3) \! 
                        - 3 L
                        + 2 \zeta(2) L
                        +  \zeta(3) L
                        + \frac{1}{2} L^{2} \, , \\
a^{2} E^{-2}_{1} & = & - \frac{5}{8} 
                       + \frac{\zeta(2)}{2} \, , \\
a^{2} E^{-1}_{1} & = & - \frac{53}{16}
                        + \frac{3 \zeta(2)}{2}
                        + \frac{\zeta(3)}{2}
                        + \frac{5}{4} L
                        - \zeta(2) L , \\
& & 
                        - \zeta(2) L^{2} \, , \\
a^{2} E^{0}_{1} & = & - \frac{333}{32}
                        + \frac{29 \zeta(2)}{4}
			- \frac{\zeta^{2}(2)}{10} 
                        + \frac{3 \zeta(3)}{2}
                        - \frac{39}{8} L
                        - 3 \zeta(2) L
                        - \zeta(3) L \nn\\
& & 
                        - \frac{5}{8} L^{2} 
                        + \zeta(2) L^{2}\, , 
\eea
\be
\parbox{20mm}{\begin{fmfgraph*}(15,15)
\fmfleft{i1,i2}
\fmfright{o}
\fmfforce{0.2w,0.9h}{v2}
\fmfforce{0.2w,0.1h}{v1}
\fmfforce{0.2w,0.5h}{v3}
\fmfforce{0.8w,0.5h}{v5}
\fmf{photon}{i1,v1}
\fmf{photon}{i2,v2}
\fmf{plain}{v5,o}
\fmf{photon,tension=0}{v2,v5}
\fmf{photon,tension=0}{v3,v4}
\fmf{plain,tension=.4}{v1,v4}
\fmf{photon,tension=.4}{v4,v5}
\fmf{photon,tension=0}{v1,v3}
\fmf{photon,tension=0}{v2,v3}
\end{fmfgraph*} } = \left( \frac{\mu^{2}}{a} \right) ^{2 \epsilon}
\sum_{j=-1}^{1} x^{j} \biggl[ 
\sum_{i=-3}^{0} \epsilon^{i} 
I_{j}^{i} \biggr] 
+ {\mathcal O} (x^2) \, , 
\ee
where:
\bea
a^{2} I^{-3}_{-1} & = & - \frac{3}{4} \, , \\
a^{2} I^{-2}_{-1} & = & - \frac{1}{2}
                        + \frac{1}{2} L \, , \\
a^{2} I^{-1}_{-1} & = &  \zeta(2) \, , \\
a^{2} I^{0}_{-1} & = & 1
                        + \zeta(2) 
                        -  L
                        - \zeta(2) L
                        + \frac{1}{2} L^{2}
                        - \frac{1}{6} L^{3} \, , \\
a^{2} I^{-2}_{0} & = & - \frac{1}{4}  \, , \\
a^{2} I^{-1}_{0} & = & - 1 \, , \\
a^{2} I^{0}_{0} & = &  - \frac{1}{8}
                        + \frac{ \zeta(2)}{2}
                        - \frac{3}{4} L
                        + \frac{1}{4} L^{2}  \, , \\
a^{2} I^{-2}_{1} & = &   \frac{1}{12}  \, , \\
a^{2} I^{-1}_{1} & = &  \frac{1}{2} \, , \\
a^{2} I^{0}_{1} & = &  \frac{215}{216}
                        - \frac{  \zeta(2)}{6}
                        + \frac{5}{36} L
                        - \frac{1}{12} L^{2}  \, , 
\eea
\be
\parbox{20mm}{\begin{fmfgraph*}(15,15)
\fmfleft{i1,i2}
\fmfright{o}
\fmfforce{0.2w,0.9h}{v2}
\fmfforce{0.2w,0.1h}{v1}
\fmfforce{0.2w,0.5h}{v3}
\fmfforce{0.8w,0.5h}{v5}
\fmf{photon}{i1,v1}
\fmf{photon}{i2,v2}
\fmf{plain}{v5,o}
\fmf{photon,tension=0}{v2,v5}
\fmf{photon,tension=0}{v3,v4}
\fmf{photon,tension=.4}{v1,v4}
\fmf{photon,tension=.4}{v4,v5}
\fmf{photon,tension=0}{v1,v3}
\fmf{plain,tension=0}{v2,v3}
\end{fmfgraph*} } = \left( \frac{\mu^{2}}{a} \right) ^{2 \epsilon}
\sum_{j=-1}^{1} x^{j} \biggl[
\sum_{i=-2}^{0} \epsilon^{i} 
J_{j}^{i} \biggr] 
+ {\mathcal O} (x^2) \, , 
\ee
where:
\bea
a^{2} J^{-2}_{-1} & = & \frac{\zeta(2)}{2} \, , \\
a^{2} J^{-1}_{-1} & = & \frac{\zeta(3)}{2}
                        - \zeta(2) L \, , \\
a^{2} J^{0}_{-1} & = & - \frac{\zeta^{2}(2)}{10}
                        - \zeta(3) L
                        + \zeta(2) L^{2} \, , \\
a^{2} J^{-1}_{0} & = &   2 
                        - \zeta(2)
                        -  L \, , \\
a^{2} J^{0}_{0} & = &  9
                        - 5 \zeta(2)
                        - \zeta(3)
                        - 5 L
                        + 2 \zeta(2) L
                        + L^{2}  \, , \\
a^{2} J^{-1}_{1} & = &  1 
                        - \frac{\zeta(2)}{2}
                        - \frac{3}{4} L \, , \\
a^{2} J^{0}_{1} & = &  \frac{81}{16}
                        - \frac{11 \zeta(2)}{4}
                        - \frac{ \zeta(3)}{2}
                        - \frac{27}{8} L
                        + \zeta(2) L
                        + \frac{3}{4} L^{2}  \, ,
\eea
\be
\parbox{20mm}{\begin{fmfgraph*}(15,15)
\fmfleft{i1,i2}
\fmfright{o}
\fmfforce{0.2w,0.93h}{v2}
\fmfforce{0.2w,0.07h}{v1}
\fmfforce{0.8w,0.5h}{v5}
\fmfforce{0.2w,0.4h}{v9}
\fmfforce{0.5w,0.45h}{v10}
\fmfforce{0.2w,0.5h}{v11}
\fmf{photon}{i1,v1}
\fmf{photon}{i2,v2}
\fmf{plain}{v5,o}
\fmf{photon}{v2,v3}
\fmf{plain,tension=.25,right}{v3,v4}
\fmf{photon,tension=.25}{v3,v4}
\fmf{photon}{v4,v5}
\fmf{photon}{v1,v5}
\fmf{photon}{v1,v2}
\end{fmfgraph*} } = \left( \frac{\mu^{2}}{a} \right) ^{2 \epsilon}
\sum_{j=-2}^{1} x^{j} \biggl[
\sum_{i=-2}^{0} \epsilon^{i} 
L_{j}^{i} \biggr] 
+ {\mathcal O} (x^2) \, , 
\ee
where:
\bea
a^{2} L^{-2}_{-2} & = & 2 \, , \\
a^{2} L^{-1}_{-2} & = & 2  
                        - 2 L \, , \\
a^{2} L^{0}_{-2} & = & 2 
                        - 2 \zeta(2)
                        - 2 L
                        + L^{2} \, , \\
a^{2} L^{-2}_{-1} & = & - \frac{1}{2}  \, , \\
a^{2} L^{-1}_{-1} & = & - \frac{3}{4}
                        + \frac{1}{2} L \, , \\
a^{2} L^{0}_{-1} & = & - \frac{7}{8}
                        + \frac{\zeta(2)}{2}
                        + \frac{3}{4} L
                        - \frac{1}{4} L^{2} \, , \\
a^{2} L^{-1}_{0} & = &   - \frac{1}{6} \, , \\
a^{2} L^{0}_{0} & = &  - \frac{1}{2}  \, , \\
a^{2} L^{-1}_{1} & = &  \frac{1}{24} \, , \\
a^{2} L^{0}_{1} & = &  \frac{11}{48}  \, ,
\eea
\be
\parbox{20mm}{\begin{fmfgraph*}(15,15)
\fmfleft{i1,i2}
\fmfright{o}
\fmfforce{0.2w,0.93h}{v2}
\fmfforce{0.2w,0.07h}{v1}
\fmfforce{0.8w,0.5h}{v5}
\fmfforce{0.2w,0.4h}{v9}
\fmfforce{0.5w,0.45h}{v10}
\fmfforce{0.2w,0.5h}{v11}
\fmf{photon}{i1,v1}
\fmf{photon}{i2,v2}
\fmf{plain}{v5,o}
\fmf{photon}{v2,v3}
\fmf{photon,tension=.25,right}{v3,v4}
\fmf{photon,tension=.25}{v3,v4}
\fmf{photon}{v4,v5}
\fmf{photon}{v1,v5}
\fmf{plain}{v2,v1}
\end{fmfgraph*} } = \left( \frac{\mu^{2}}{a} \right) ^{2 \epsilon}
\sum_{j=-1}^{1} x^{j} \biggl[
\sum_{i=-2}^{0} \epsilon^{i} 
M_{j}^{i} \biggr] 
+ {\mathcal O} (x^2) \, , 
\ee
where:
\bea
a^{2} M^{-2}_{-1} & = & - \frac{1}{2} \, , \\
a^{2} M^{-1}_{-1} & = & - \frac{1}{2}
                        + L \, , \\
a^{2} M^{0}_{-1} & = & - \frac{3}{2}
                        + \zeta(2)
                        + L
                        - L^{2} \, , \\
a^{2} M^{-1}_{0} & = &    L \, , \\
a^{2} M^{0}_{0} & = &  - 2
                        + 2 \zeta(2)
			+ 3 L
			- L^{2}  \, , \\
a^{2} M^{-1}_{1} & = &   L \, , \\
a^{2} M^{0}_{1} & = &  - \frac{5}{2}
                        + 2 \zeta(2)
			+ 4 L
			- L^{2}  \, ,
\eea
\be
\parbox{20mm}{\begin{fmfgraph*}(15,15)
\fmfleft{i1,i2}
\fmfright{o}
\fmfforce{0.2w,0.93h}{v2}
\fmfforce{0.2w,0.07h}{v1}
\fmfforce{0.2w,0.3h}{v3}
\fmfforce{0.2w,0.7h}{v4}
\fmfforce{0.8w,0.5h}{v5}
\fmf{photon}{i1,v1}
\fmf{photon}{i2,v2}
\fmf{plain}{v5,o}
\fmf{photon}{v2,v5}
\fmf{photon}{v1,v3}
\fmf{plain}{v2,v4}
\fmf{photon}{v1,v5}
\fmf{photon,right}{v4,v3}
\fmf{photon,right}{v3,v4}
\end{fmfgraph*} } = \left( \frac{\mu^{2}}{a} \right) ^{2 \epsilon}
\sum_{j=-1}^{1} x^{j} \biggl[
\sum_{i=-3}^{0} \epsilon^{i} 
O_{j}^{i} \biggr] 
+ {\mathcal O} (x^2) \, , 
\ee
where:
\bea
a^{2} O^{-3}_{-1} & = &  \frac{1}{4} \, , \\
a^{2} O^{-2}_{-1} & = & \frac{1}{2}
                        - \frac{1}{2} L \, , \\
a^{2} O^{-1}_{-1} & = & 1
                        - L 
                        + \frac{1}{2} L^{2} \, , \\
a^{2} O^{0}_{-1} & = & 2
                        - 2 \zeta(3) 
                        - 2 L
                        + L^{2}
                        - \frac{1}{3} L^{3} \, , \\
a^{2} O^{-1}_{0} & = &  - 1
                        + L \, , \\
a^{2} O^{0}_{0} & = &  - 6
                        +  \zeta(2) 
                        + 5 L
                        - L^{2}  \, , \\
a^{2} O^{-1}_{1} & = &  - \frac{1}{4} 
                        + \frac{1}{2} L \, , \\
a^{2} O^{0}_{1} & = &  - \frac{7}{4}
                        + \frac{\zeta(2)}{2}
                        + \frac{9}{4} L
                        - \frac{1}{2} L^{2}  \, .
\eea
\be
\parbox{20mm}{\begin{fmfgraph*}(15,15)
\fmfleft{i1,i2}
\fmfright{o}
\fmf{photon}{i1,v1}
\fmf{photon}{i2,v2}
\fmf{plain}{v5,o}
\fmf{photon,tension=.3}{v2,v3}
\fmf{photon,tension=.3}{v3,v5}
\fmf{photon,tension=.3}{v1,v4}
\fmf{photon,tension=.3}{v4,v5}
\fmf{plain,tension=0}{v2,v4}
\fmf{photon,tension=0}{v1,v3}
\end{fmfgraph*} } = \left( \frac{\mu^{2}}{a} \right) ^{2 \epsilon}
\sum_{j=-1}^{1} x^{j}  \biggl[ \sum_{i=-2}^{0} \epsilon^{i}
R_{j}^{i} \biggr] 
+ {\mathcal O} (x^2) \, , 
\ee
where:
\bea
a^{2} R^{-2}_{-1} & = & \frac{\zeta(2)}{2} \, , \\
a^{2} R^{-1}_{-1} & = & \frac{\zeta(3)}{2} 
                        - \zeta(2) L  \, , \\
a^{2} R^{0}_{-1} & = & - \frac{\zeta^{2}(2)}{10}
		       - \zeta(3) L
                       + \zeta(2) L^{2} \, , \\
a^{2} R^{-2}_{0} & = & -1 
                       + \frac{\zeta(2)}{2} \, , \\
a^{2} R^{-1}_{0} & = & - 4
                        + 2 \zeta(2)
                        + \frac{\zeta(3)}{2}
                        + L 
                        - \zeta(2) L   \, , \\
a^{2} R^{0}_{0} & = & - 14 \! 
                       + 5 \zeta(2) 
		       - \frac{\zeta^{2}(2)}{10} \! 
                       + 2 \zeta(3)  \! 
		       + 7 L \! 
                       - 4 \zeta(2) L 
                       - \zeta(3) L 
                       - \frac{3}{2} L^{2} , \\
a^{2} R^{-2}_{1} & = & - \frac{3}{4}
                       + \frac{\zeta(2)}{2} \, , \\
a^{2} R^{-1}_{1} & = &  - 5
                        + 3 \zeta(2)
                        + \frac{\zeta(3)}{2}
                        + \frac{5}{4} L 
                        - \zeta(2) L  \, , \\
& & 
                       + \zeta(2) L^{2} \, , \\
a^{2} R^{0}_{1} & = & - \frac{187}{8} 
                       + \frac{53 \zeta(2)}{4} 
		       - \frac{\zeta^{2}(2)}{10}
                       + 3 \zeta(3)
                       + \frac{77}{8} L 
                       - 6 \zeta(2) L 
                       - \zeta(3) L \nn\\
& & 
                       - \frac{11}{8} L^{2}
                       + \zeta(2) L^{2} \, , 
\eea

\section{Conclusions}

We have presented an exact evaluation of the master integrals 
with at most one massive propagator entering the electroweak form 
factor. Due to the rather simple threshold structure, we have succeeded 
in obtaining exact representations of all the master integrals in 
terms of one-dimensional harmonic polylogarithms of $x$.
By an expansion of the harmonic polylogarithms for large and small
values of the argument, we could directly obtain the large and small 
momentum expansions of the master integrals.
We checked our results with previous ones present in the literature 
finding complete agreement 
\cite{fleischer,Gonsalves83,VanNeerven86,kodaira,smirnov}.
The remaining master integrals, containing two or three massive 
propagators, have thresholds not only in $s=0$ and $s=m^2$ but also in 
$s=4m^2$ and their exact evaluation requires probably the introduction 
of new classes of special functions.

\section{Acknowledgement}

We wish to thank E.~Remiddi for discussions and for use of the {\tt C} 
program {\tt SOLVE} to solve the linear systems generated by the ibp 
identities.

U.A. wishes to thank T. Gehrmann for discussions about two-loop 
computations and D. Comelli, M. Ciafaloni and P. Ciafaloni for 
discussions about the ew form factor.

R.B. wishes to thank the Fondazione Della Riccia for supporting his 
permanence at CERN, the Theory Division of CERN for the hospitality 
during a large portion of this work, and the Universit\`a di Roma 
``La Sapienza'' for hospitality during the final part of this work.

\appendix

\section{Propagators \label{app1}}

In this appendix we list all the denominators entering 
the two-loop integrals computed in this paper (see section 2.2).
\begin{eqnarray}
{\mathcal D}_{1} & = & k_{1}^{2} \, , \\
{\mathcal D}_{2} & = & k_{2}^{2} \, , \\
{\mathcal D}_{3} & = & (k_{1}+k_{2})^{2} \, , \\
{\mathcal D}_{4} & = & (p_{1}-k_{1})^{2} \, , \\
{\mathcal D}_{5} & = & (p_{2}+k_{1})^{2} \, , \\
{\mathcal D}_{6} & = & (p_{2}-k_{2})^{2} \, , \\
{\mathcal D}_{7} & = & (p_{1}-k_{1}+k_{2})^{2} \, , \\
{\mathcal D}_{8} & = & (p_{2}+k_{1}-k_{2})^{2} \, , \\
{\mathcal D}_{9} & = & (p_{1}+p_{2}-k_{1})^{2} \, , \\
{\mathcal D}_{10} & = & (p_{1}+p_{2}-k_{2})^{2} \, , \\
{\mathcal D}_{11} & = & (p_{1}+p_{2}-k_{1}-k_{2})^{2} \, , \\
{\mathcal D}_{12} & = & k_{1}^{2} + a \, , \\
{\mathcal D}_{13} & = & k_{2}^{2} + a \, , \\
{\mathcal D}_{14} & = & (k_{1}+k_{2})^{2} + a \, , \\
{\mathcal D}_{15} & = & (p_{1}-k_{1})^{2} + a \, , \\
{\mathcal D}_{16} & = & (p_{2}+k_{1})^{2} + a \, , \\
{\mathcal D}_{17} & = & (p_{2}-k_{2})^{2} + a \, , \\
{\mathcal D}_{18} & = & (p_{1}-k_{1}+k_{2})^{2} + a \, , \\
{\mathcal D}_{19} & = & (p_{2}+k_{1}-k_{2})^{2} + a \, , \\
{\mathcal D}_{20} & = & (p_{1}+p_{2}-k_{1})^{2} + a \, , \\
{\mathcal D}_{21} & = & (p_{1}+p_{2}-k_{2})^{2} + a \, , \\
{\mathcal D}_{22} & = & (p_{1}+p_{2}-k_{1}-k_{2})^{2} + a
\, .
\end{eqnarray}

\section{One-loop amplitudes \label{app2}}

%% Labels: \label{appb?}
In this appendix we present the results for the one-loop diagrams
which are necessary for the computation of the factorized two-loop 
amplitudes. We have recomputed them with the method of differential 
equations described in the main body of the paper in terms of HPL's. 
In some cases, we found that it was necessary to compute the one-loop 
amplitudes up to the third order in $\epsilon$.

\subsection{Tadpole}

\begin{eqnarray}
\parbox{20mm}{
\begin{fmfgraph*}(15,15)
\fmfleft{i}
\fmfright{o}
\fmfforce{0.5w,0.1h}{v1}
\fmfforce{0.25w,0.62h}{v3}
\fmfforce{0.5w,0.9h}{v7}
\fmfforce{0.74w,0.62h}{v11}
\fmf{plain,left=.1}{v1,v3}
\fmf{plain,left=.5}{v3,v7}
\fmf{plain,left=.5}{v7,v11}
\fmf{plain,left=.1}{v11,v1}
\end{fmfgraph*}} & = & \mu^{(4-D)} \int \{ d^{D}k \}
\frac{1}{k^{2} +a} \nn\\
& = & \left( \frac{\mu^{2}}{a} \right) ^{\epsilon} 
\sum_{i=-1}^{3} \epsilon^{i} A_{i} + {\mathcal O} \left( 
\epsilon^{4} \right) \, ,
\label{appb1}
\end{eqnarray}
where:
\be
\frac{A_{i}}{a} = -1 \, . 
\ee

\subsection{Bubbles}

\begin{eqnarray}
\! \! \! \! \! \! \parbox{20mm}{
\begin{fmfgraph*}(15,15)
\fmfleft{i}
\fmfright{o}
\fmf{plain}{i,v1}
\fmf{plain}{v2,o}
\fmf{photon,tension=.15,left}{v1,v2}
\fmf{photon,tension=.15,left}{v2,v1}
\end{fmfgraph*} } & = & \mu^{(4-D)}
\int \{ d^{D}k \} \frac{1}{
k^{2} \, (p-k)^{2}} \nn\\
\! \! \! \! \! \! & = & \left( \frac{\mu^{2}}{a} \right) ^{\epsilon} 
\sum_{i=-1}^{3} \epsilon^{i} E_{i} 
+ {\mathcal O} \left( \epsilon^{4}
\right) \, , 
\label{appb3}
\end{eqnarray}
where:
\begin{eqnarray}
E_{-1} & = & 1 \, , \\
E_{0} & = & 2 - H(0,x) \, , \\
E_{1} & = & 4 - \zeta(2) - 2H(0,x) + H(0,0,x) \, , \\
E_{2} & = &  8  -2 \bigl( \zeta(2) + \zeta(3) \bigr)
+ \bigl( \zeta(2) -4 \bigr) H(0,x) + 2 H(0,0,x) \nn\\
& & - H(0,0,0,x) \, , \\
E_{3} & = & 16 - 4 \zeta(2) - \frac{9}{10} \zeta^{2}(2) - 4
\zeta(3) - 8 H(0,x) + 2 \bigl( \zeta(2) \nn\\
& & + \zeta(3) \bigr) H(0,x) + \bigl( 4- \zeta(2) 
\bigr) H(0,0,x)  - 2 H(0,0,0,x) \nn\\
& & + H(0,0,0,0,x) \, ,
\end{eqnarray}

The above amplitude --- the simplest massless one --- is easily computed with
a Feynman parameter or with the general formula in the appendix of \cite{Chet}.

\begin{eqnarray}
\! \! \! \! \! \! \parbox{20mm}{
\begin{fmfgraph*}(15,15)
\fmfleft{i}
\fmfright{o}
\fmf{plain}{i,v1}
\fmf{plain}{v2,o}
\fmf{plain,tension=.15,left}{v1,v2}
\fmf{photon,tension=.15,left}{v2,v1}
\end{fmfgraph*} } & = & \mu^{(4-D)}
\int \{ d^{D}k \} \frac{1}{
k^{2} \, [(p-k)^{2}+a]} \nn\\
\! \! \! \! \! \! & = & \left( \frac{\mu^{2}}{a} \right) ^{\epsilon} 
\sum_{i=-1}^{2} \epsilon^{i} F_{i} 
+ {\mathcal O} \left( \epsilon^{3} \right) \, , 
\label{appb4}
\end{eqnarray}
where:
\begin{eqnarray}
F_{-1} & = & 1 \, , \\
F_{0} & = & 2 - \left[ 1 + \frac{1}{x} \right] H(-1,x) \, , \\
F_{1} & = & 4 - \left[ 1 + \frac{1}{x} \right] \Bigl[ 2 H(-1,x) + 
H(0,-1,x) -2 H(-1,-1,x) \Bigr] , \\
F_{2} & = & 8 - \left[ 1 + \frac{1}{x} \right]
\Bigl[ 4 H(-1,x) + 2 H(0,-1,x) \nn\\
& & - 4 H(-1,-1,x)  + 4 H(-1,-1,-1,x) - 2 H(-1,0,-1,x) \nn\\
& & - 2 H(0,-1,-1,x) + H(0,0,-1,x) \Bigr] \, ,\\
F_{3} & = & 16 + \left[ 1 + \frac{1}{x} \right]
\Bigl[ - 8 H(-1,x) + 8 H(-1,-1,x) - 4 H(0,-1,x) \nn\\
& & - 8 H(-1,-1,-1,x) + 4 H(-1,0,-1,x) + 4 H(0,-1,-1,x) \nn\\
& & - 2 H(0,0,-1,x) + 8 H(-1,-1,-1,-1,x) \nn\\
& & - 4 H(-1,-1,0,-1,x) - 4 H(-1,0,-1,-1,x) \nn\\
& & + 2 H(-1,0,0,-1,x) - 4 H(0,-1,-1,-1,x) \nn\\
& & + 2 H(0,-1,0,-1,x) + 2 H(0,0,-1,-1,x) \nn\\
& & - H(0,0,0,-1,x) \Bigr] \, ,
\end{eqnarray}

\begin{eqnarray}
\parbox{20mm}{
\begin{fmfgraph*}(15,15)
\fmfleft{i}
\fmfright{o}
\fmf{photon}{i,v1}
\fmf{photon}{v2,o}
\fmf{plain,tension=.15,left}{v1,v2}
\fmf{photon,tension=.15,left}{v2,v1}
\end{fmfgraph*} } & = & \mu^{(4-D)}
\int \{ d^{D}k \} \frac{1}{k^{2} \, [k^{2}+a]} \nn\\
& = & \left( \frac{\mu^{2}}{a} \right) ^{\epsilon} 
\sum_{i=-1}^{3} \epsilon^{i} G_{i} + {\mathcal O} \left( 
\epsilon^{4} \right) \, ,
\label{appb5}
\end{eqnarray}
where:
\begin{eqnarray}
G_{i} & = & 1 \, .
\end{eqnarray}

\subsection{Vertices}

\begin{eqnarray}
\! \! \! \! \! \! \parbox{20mm}{\begin{fmfgraph*}(15,15)
\fmfleft{i1,i2}
\fmfright{o}
\fmf{photon}{i1,v1}
\fmf{photon}{i2,v2}
\fmf{plain}{v3,o}
\fmf{plain,tension=.3}{v2,v3}
\fmf{photon,tension=.3}{v1,v3}
\fmf{photon,tension=0}{v2,v1}
\end{fmfgraph*} } & = & \mu^{4-D}
\int \{ d^{D}k \} \frac{1}{
k^{2} \, [(p_{1}-k)^{2}+a] \, (p_{2}+k)^{2}} \nn\\
\! \! \! \! \! \! & = & \frac{1}{a(1+x)} \Biggl[ \frac{1-2 \epsilon}{
\epsilon} \, \, 
\parbox{15mm}{
\begin{fmfgraph*}(15,15)
\fmfleft{i}
\fmfright{o}
\fmf{plain}{i,v1}
\fmf{plain}{v2,o}
\fmf{plain,tension=.15,left}{v1,v2}
\fmf{photon,tension=.15,left}{v2,v1}
\end{fmfgraph*} } + \frac{1- \epsilon}{\epsilon} \, \, 
\parbox{15mm}{
\begin{fmfgraph*}(15,15)
\fmfleft{i}
\fmfright{o}
\fmfforce{0.5w,0.1h}{v1}
\fmfforce{0.25w,0.62h}{v3}
\fmfforce{0.5w,0.9h}{v7}
\fmfforce{0.74w,0.62h}{v11}
\fmf{plain,left=.1}{v1,v3}
\fmf{plain,left=.5}{v3,v7}
\fmf{plain,left=.5}{v7,v11}
\fmf{plain,left=.1}{v11,v1}
\end{fmfgraph*}} \Biggr] \\
\! \! \! \! \! \! & = & \left( \frac{\mu^{2}}{a} \right) ^{\epsilon} 
\sum_{i=-1}^{2} \epsilon^{i} L_{i} + {\mathcal O} 
\left( \epsilon^{3} \right) ,
\label{appb7}
\end{eqnarray}
where:
\begin{eqnarray}
a L_{-1} & = & - \frac{1}{x} H(-1,x) \, , \\
a L_{0} & = & \frac{1}{x} \bigl[ 2 H(-1,-1,x) - H(0,-1,x) \bigr] \, , \\
a L_{1} & = & \frac{1}{x} \bigl[ 2 H(-1,0,-1,x)+ 2 H(0,-1,-1,x)
- 4 H(-1,-1,-1,x) \nn\\
& & - H(0,0,-1,x) \bigr] \, , \\
a L_{2} & = & \frac{1}{x} \bigl[ 8 H(-1,-1,-1,-1,x) - 4 H(-1,-1,0,-1,x)
\nn\\
& & - 4H(-1,0,-1,-1,x) + 2H(-1,0,0,-1,x) \nn\\
& & - 4H(0,-1,-1,-1,x) + 2H(0,-1,0,-1,x) \nn\\
& & + 2H(0,0,-1,-1,x) - H(0,0,0,-1,x) \bigr] \, .
\end{eqnarray}

The above amplitude is not a master integral.

\begin{eqnarray}
\! \! \! \! \! \! \parbox{20mm}{\begin{fmfgraph*}(15,15)
\fmfleft{i1,i2}
\fmfright{o}
\fmf{photon}{i1,v1}
\fmf{photon}{i2,v2}
\fmf{plain}{v3,o}
\fmf{photon,tension=.3}{v2,v3}
\fmf{photon,tension=.3}{v1,v3}
\fmf{plain,tension=0}{v2,v1}
\end{fmfgraph*} } & = & \mu^{(4-D)}
\int \{ d^{D}k \} \frac{1}{
[k^{2}+a] \, (p_{1}-k)^{2} \, (p_{2}+k)^{2}} \nn\\
\! \! \! \! \! \! & = & \left( \frac{\mu^{2}}{a} \right) ^{\epsilon} 
\sum_{i=0}^{2} \epsilon^{i} M_{i} + {\mathcal O} 
\left( \epsilon^{3} \right) ,
\label{appb8}
\end{eqnarray}
where:
\begin{eqnarray}
\! \! \! \! \! \! a M_{0} & = & - \frac{1}{x} H(1,0,x) \, , \\
\! \! \! \! \! \! a M_{1} & = & - \frac{1}{x} \bigl[ \zeta(2) H(1,x)
+ H(0,1,0,x) - H(1,0,0,x) + H(1,1,0,x) \bigr] , \\
\! \! \! \! \! \! a M_{2} & = & \frac{1}{x} \bigl[ - 2 \zeta(3) H(1,x)
- \zeta(2) \bigl( H(0,1,x) - H(1,0,x) + H(1,1,x) \bigr) \nn\\
\! \! \! \! \! \! & & - H(0,0,1,0,x) + 
H(0,1,0,0,x) - H(0,1,1,0,x) - H(1,0,0,0,x) \nn\\
\! \! \! \! \! \! & & - H(1,0,1,0,x) + H(1,1,0,0,x) - H(1,1,1,0,x) 
\bigr] \, ,
\end{eqnarray}

The above amplitude does not have infrared $1/\epsilon$ poles
because a non-vanishing mass in the boson line
acts as an infrared cutoff on the transverse momenta: $k_t > m$ 
\cite{ciafcom}.

\begin{eqnarray}
\! \! \! \! \! \! \parbox{20mm}{\begin{fmfgraph*}(15,15)
\fmfleft{i1,i2}
\fmfright{o}
\fmf{photon}{i1,v1}
\fmf{photon}{i2,v2}
\fmf{plain}{v3,o}
\fmf{photon,tension=.3}{v2,v3}
\fmf{photon,tension=.3}{v1,v3}
\fmf{photon,tension=0}{v2,v1}
\end{fmfgraph*} } & = & \mu^{(4-D)}
\int \{ d^{D}k \} \frac{1}{
k^{2} \, (p_{1}-k)^{2} \, (p_{2}+k)^{2}} \nn\\
\! \! \! \! \! \! & = & \frac{(1-2 \epsilon)}{\epsilon} \frac{1}{ax} \, 
\parbox{15mm}{
\begin{fmfgraph*}(15,15)
\fmfleft{i}
\fmfright{o}
\fmf{plain}{i,v1}
\fmf{plain}{v2,o}
\fmf{photon,tension=.15,left}{v1,v2}
\fmf{photon,tension=.15,left}{v2,v1}
\end{fmfgraph*} } \\
\! \! \! \! \! \! & = & \left( \frac{\mu^{2}}{a} \right) ^{\epsilon} 
\sum_{i=-2}^{2} \epsilon^{i} N_{i} + {\mathcal O} 
\left( \epsilon^{3} \right) ,
\label{appb9}
\end{eqnarray}
where:
\begin{eqnarray}
a N_{-2} & = & \frac{1}{x} \, , \\
a N_{-1} & = & - \frac{1}{x} H(0,x) \, , \\
a N_{0} & = & - \frac{1}{x} \bigl[ \zeta(2) - H(0,0,x) \bigr] \, , \\
a N_{1} & = & - \frac{1}{x} \bigl[ 2 \zeta(3) - \zeta(2) H(0,x) +
H(0,0,0,x) \bigr] \, , \\
a N_{2} & = & - \frac{1}{x} \Biggl[ \frac{9}{10} \zeta^{2}(2) - 2 \zeta(3)
H(0,x) + \zeta(2) H(0,0,x) \nn\\
& & - H(0,0,0,0,x) \Biggr]
\, .
\end{eqnarray}

\section{Independent two-loop amplitudes \label{app3}}

In this appendix we give the expressions of the reducible diagrams of
figures \ref{fig2}, \ref{fig3} and \ref{fig4}.
\bea
\parbox{20mm}{\begin{fmfgraph*}(15,15)
\fmfleft{i}
\fmfright{o}
\fmfforce{0.2w,0.5h}{v1}
\fmfforce{0.5w,0.2h}{v2}
\fmfforce{0.8w,0.5h}{v3}
\fmf{plain}{i,v1}
\fmf{plain}{v3,o}
\fmf{photon,left}{v1,v3}
\fmf{photon,right=.4}{v1,v2}
\fmf{photon,right=.4}{v2,v3}
\fmf{photon,left=.6}{v2,v3}
\end{fmfgraph*} }  & = & \mu^{2(4-D)} 
\int \{ d^{D}k_{1} \} \{ d^{D}k_{2} \}
\frac{1}{{\mathcal D}_{1} {\mathcal D}_{2} {\mathcal D}_{9} 
{\mathcal D}_{11} } \\
& = & \left( \frac{\mu^{2}}{a} \right) ^{2 \epsilon} 
\sum_{i=-2}^{2} \epsilon^{i} R^{(1)}_{i} + {\mathcal O} \left( 
\epsilon^{3} \right) , 
\eea
where:
\bea
R^{(1)}_{-2} & = & \frac{1}{2} \, , \\
R^{(1)}_{-1} & = & \frac{5}{2} - H(0,x) \, , \\
R^{(1)}_{0} & = & \frac{19}{2} - \zeta(2) - 5 H(0,x) 
+ 2 H(0,0,x)  \, , \\
R^{(1)}_{1} & = & \frac{65}{2} - 5 \zeta(2) - 5 \zeta(3) - [ 19 - 2 \zeta(2) ]
H(0,x) \nn\\
& & + 10 H(0,0,x) - 4 H(0,0,0,x)  \, , \\
R^{(1)}_{2} & = & \frac{211}{2} - 19 \zeta(2) - \frac{11 \zeta^{2}
(2)}{5} - 25 \zeta(3) - \bigl[ 65 - 10 \zeta(2) \nn\\
& & - 10 \zeta(3) \bigr] H(0,x) + \bigl[ 38 - 4 \zeta(2) 
\bigr] H(0,0,x) - 20 H(0,0,0,x) \nn\\
& & + 8 H(0,0,0,0,x) 
\, , 
\eea
\bea
\parbox{20mm}{\begin{fmfgraph*}(15,15)
\fmfleft{i}
\fmfright{o}
\fmfforce{0.2w,0.5h}{v1}
\fmfforce{0.5w,0.2h}{v2}
\fmfforce{0.8w,0.5h}{v3}
\fmf{plain}{i,v1}
\fmf{plain}{v3,o}
\fmf{photon,left}{v1,v3}
\fmf{photon,right=.4}{v1,v2}
\fmf{photon,right=.4}{v2,v3}
\fmf{plain,left=.6}{v2,v3}
\end{fmfgraph*}}  & = & \mu^{2(4-D)} 
\int \{ d^{D}k_{1} \} \{ d^{D}k_{2} \}
\frac{1}{{\mathcal D}_{1} {\mathcal D}_{9} {\mathcal D}_{11} 
{\mathcal D}_{13} } \\
& = & \left( \frac{\mu^{2}}{a} \right) ^{2 \epsilon} 
\sum_{i=-2}^{2} \epsilon^{i} R^{(2)}_{i} + {\mathcal O} \left( 
\epsilon^{3} \right) , 
\eea
where:
\bea
R^{(2)}_{-2} & = & \frac{1}{2} \, , \\
R^{(2)}_{-1} & = & \frac{5}{2} - H(0,x) \, , \\
R^{(2)}_{0} & = & \frac{19}{2} - 2 \zeta(2) - 3 H(0,x) 
- 2 \left[ 1 + \frac{1}{x} \right] H(-1,x) + H(0,0,x) \nn\\
& & + \left[ 1 - \frac{1}{x} \right] H(0,-1,x) \, , \\
R^{(2)}_{1} & = & \frac{65}{2} - 4 \zeta(2) - \zeta(3) - [ 7 - \zeta(2) ] 
H(0,x) - 12 \left[ 1 + \frac{1}{x} \right] H(-1,x) \nn\\
& & + 3 H(0,0,x) - \biggl[ 1 + \frac{5}{x} \biggr]
H(0,-1,x) + 8 \biggl[ 1 + \frac{1}{x} \biggr]
H(-1,-1,x) \nn\\
& & - H(0,0,0,x) + \biggl[ 1 - \frac{1}{x} 
\biggr] \bigl[ H(0,0,-1,x) - 4 H(0,-1,-1,x) \bigr] \, , \\
R^{(2)}_{2} & = & \frac{211}{2} - 2 \zeta(2) - \frac{27 \zeta^{2}
(2)}{10} - 5 \zeta(3) - \bigl[ 15 - 3 \zeta(2) - 2 \zeta(3) \bigr] 
H(0,x) \nn\\
& & - \bigl[ 50 + 4 \zeta(2) \bigr] \biggl[ 1 + \frac{1}{x} \biggr] 
H(-1,x) + [ 7 \! - \! \zeta(2) ] H(0,0,x) - \! \biggl[ 17 \! - \! 2 
\zeta(2) \nn\\
& & + \frac{(19 + 2 \zeta(2))}{x}
\biggr] H(0,-1,x) + 48 \biggl[ 1 + \frac{1}{x} \biggr]
H(-1,-1,x) \nn\\
& & - 3 H(0,0,0,x) - \biggl[ 1 + \frac{5}{x} \biggr] \bigl[ 
H(0,0,-1,x) - 4 H(0,-1,-1,x) \bigr]
\nn\\
& & + \biggl[ 1 + \frac{1}{x} \biggr] \bigl[ 12 H(-1,0,-1,x) - 32 
H(-1,-1,-1,x) \bigr] + \biggl[ 1 \nn\\
& & - \frac{1}{x} \biggr] \bigl[ H(0,0,0,-1,x) - 6 H(0,-1,0,-1,x) \nn\\
& & + 16 H(0,-1,-1,-1,x) - 4 H(0,0,-1,-1,x) \bigr] \nn\\
& & + H(0,0,0,0,x)
\, , 
\eea
\bea
\parbox{20mm}{\begin{fmfgraph*}(15,15)
\fmfleft{i}
\fmfright{o}
\fmfforce{0.2w,0.5h}{v1}
\fmfforce{0.5w,0.2h}{v2}
\fmfforce{0.8w,0.5h}{v3}
\fmf{plain}{i,v1}
\fmf{plain}{v3,o}
\fmf{plain,left}{v1,v3}
\fmf{photon,right=.4}{v1,v2}
\fmf{photon,right=.4}{v2,v3}
\fmf{photon,left=.6}{v2,v3}
\end{fmfgraph*} }  & = & \mu^{2(4-D)} 
\int \{ d^{D}k_{1} \} \{ d^{D}k_{2} \}
\frac{1}{{\mathcal D}_{2} {\mathcal D}_{9} {\mathcal D}_{10} 
{\mathcal D}_{12} } \\
& = & \left( \frac{\mu^{2}}{a} \right) ^{2 \epsilon} 
\sum_{i=-2}^{1} \epsilon^{i} R^{(3)}_{i} + {\mathcal O} \left( 
\epsilon^{2} 
\right) , 
\eea
where:
\bea
R^{(3)}_{-2} & = & \frac{1}{2} \, , \\
R^{(3)}_{-1} & = & \frac{5}{2} - \biggl[ 1 + \frac{1}{x} \biggr]
H(-1,x) \, , \\
R^{(3)}_{0} & = & \frac{19}{2} + \zeta(2) - \biggl[ 1 + \frac{1}{x} \biggr]
\bigl[ 5 H(-1,x) + H(0,-1,x) - 4 H(-1,-1,x) \bigr] \nn\\
& & - H(0,-1,x)
\, , \\
R^{(3)}_{1} & = & \frac{65}{2} + 5 \zeta(2) - \zeta(3) 
- \bigl[ 19 + 2 \zeta(2) \bigr] \biggl[ 1 + \frac{1}{x}
\biggr] H(-1,x) \nn\\
& &  - 10 \biggl[ 1 + \frac{1}{x} \biggr] \bigl[
H(0,-1,x) - 2 H(-1,-1,x) \bigr] + 5 H(0,-1,x) \nn\\
& & - \biggl[ 1 + \frac{1}{x} \biggr] \bigl[ H(0,0,-1,x) - 4 
H(0,-1,-1,x) - 6 H(-1,0,-1,x) \nn\\
& & + 16 H(-1,-1,-1,x) \bigr] + 4 H(0,-1,-1,x) -  H(0,0,-1,x)
\, , 
\eea
\bea
\parbox{20mm}{\begin{fmfgraph*}(15,15)
\fmfleft{i}
\fmfright{o}
\fmfforce{0.2w,0.5h}{v1}
\fmfforce{0.5w,0.2h}{v2}
\fmfforce{0.8w,0.5h}{v3}
\fmf{photon}{i,v1}
\fmf{photon}{v3,o}
\fmf{plain,left}{v1,v3}
\fmf{photon,right=.4}{v1,v2}
\fmf{photon,right=.4}{v2,v3}
\fmf{photon,left=.6}{v2,v3}
\end{fmfgraph*} }  & = & \mu^{2(4-D)} 
\int \{ d^{D}k_{1} \} \{ d^{D}k_{2} \}
\frac{1}{{\mathcal D}_{2} {\mathcal D}_{5} {\mathcal D}_{8} 
{\mathcal D}_{12} } \\
& = & \left( \frac{\mu^{2}}{a} \right) ^{2 \epsilon} 
\sum_{i=-2}^{2} \epsilon^{i} R^{(4)}_{i} + {\mathcal O} \left( 
\epsilon^{3} \right) , 
\eea
where:
\bea
R^{(4)}_{-2} & = & \frac{1}{2} \, , \\
R^{(4)}_{-1} & = & \frac{3}{2} \, , \\
R^{(4)}_{0} & = & \frac{7}{2} + \zeta(2) \, , \\
R^{(4)}_{1} & = & \frac{15}{2} + 3 \zeta(2) - \zeta(3) \, , \\
R^{(4)}_{2} & = & \frac{31}{2} + 7 \zeta(2) + \frac{9 \zeta^{2}(2)
}{5} - 3 \zeta(3) \, , 
\eea
\bea
\parbox{20mm}{\begin{fmfgraph*}(15,15)
\fmfleft{i}
\fmfright{o}
\fmfforce{0.2w,0.5h}{v1}
\fmfforce{0.5w,0.2h}{v2}
\fmfforce{0.8w,0.5h}{v3}
\fmf{photon}{i,v1}
\fmf{photon}{v3,o}
\fmf{photon,left}{v1,v3}
\fmf{photon,right=.4}{v1,v2}
\fmf{photon,right=.4}{v2,v3}
\fmf{plain,left=.6}{v2,v3}
\end{fmfgraph*} }  & = & \mu^{2(4-D)} 
\int \{ d^{D}k_{1} \} \{ d^{D}k_{2} \}
\frac{1}{{\mathcal D}_{1} {\mathcal D}_{5} {\mathcal D}_{8} 
{\mathcal D}_{13} } \\
& = & \left( \frac{\mu^{2}}{a} \right) ^{2 \epsilon} 
\sum_{i=-2}^{2} \epsilon^{i} R^{(5)}_{i} + {\mathcal O} \left( 
\epsilon^{3} \right) , 
\eea
where:
\bea
R^{(5)}_{-2} & = & - \frac{1}{2} \, , \\
R^{(5)}_{-1} & = & - \frac{1}{2} \, , \\
R^{(5)}_{0} & = & - \frac{1}{2} - \zeta(2) \, , \\
R^{(5)}_{1} & = & - \frac{1}{2} - \zeta(2) + \zeta(3) \, , \\
R^{(5)}_{2} & = & - \frac{1}{2} - \zeta(2) - \frac{9 \zeta^{2}(2)
}{5} + \zeta(3) \, , 
\eea
\bea
\parbox{20mm}{\begin{fmfgraph*}(15,15)
\fmfleft{i}
\fmfright{o}
\fmfforce{0.2w,0.5h}{v1}
\fmfforce{0.5w,0.8h}{v2}
\fmfforce{0.8w,0.5h}{v3}
\fmf{photon}{i,v1}
\fmf{photon}{v3,o}
\fmf{photon,right}{v1,v3}
\fmf{plain,left=.4}{v1,v2}
\fmf{photon,left=.4}{v2,v3}
\fmf{photon,right=.6}{v2,v3}
\end{fmfgraph*}}  & = & \mu^{2(4-D)} 
\int \{ d^{D}k_{1} \} \{ d^{D}k_{2} \}
\frac{1}{{\mathcal D}_{1} {\mathcal D}_{2} {\mathcal D}_{7} {\mathcal D}_{15} } \\
& = & \left( \frac{\mu^{2}}{a} \right) ^{2 \epsilon} 
\sum_{i=-2}^{2} \epsilon^{i} R^{(6)}_{i} + {\mathcal O} \left( 
\epsilon^{3} \right) , 
\eea
where:
\bea
R^{(6)}_{-2} & = & \frac{1}{2} \, , \\
R^{(6)}_{-1} & = & \frac{3}{2} \, , \\
R^{(6)}_{0} & = & \frac{7}{2} + \zeta(2) \, , \\
R^{(6)}_{1} & = & \frac{15}{2} + 3 \zeta(2) - \zeta(3) \, , \\
R^{(6)}_{2} & = & \frac{31}{2} + 7 \zeta(2) + \frac{9 \zeta^{2}(2)}{5} - 3 
\zeta(3) \, .
\eea
\bea
\parbox{20mm}{\begin{fmfgraph*}(15,15)
\fmfbottom{v5}
\fmftop{v4}
\fmfleft{i}
\fmfright{o}
\fmf{photon}{i,v1}
\fmf{plain}{v3,o}
\fmf{photon}{v5,v2} 
\fmf{phantom}{v2,v4} 
\fmf{photon,tension=.2,left}{v1,v2}
\fmf{plain,tension=.2,right}{v1,v2}
\fmf{photon,tension=.2,left}{v2,v3}
\fmf{photon,tension=.2,right}{v2,v3}
\end{fmfgraph*} } & = & \mu^{2(4-D)} 
\int \{ d^{D}k_{1} \} \{ d^{D}k_{2} \}
\frac{1}{{\mathcal D}_{2} {\mathcal D}_{4} {\mathcal D}_{10} 
{\mathcal D}_{12} } \\
& = & \left( \frac{\mu^{2}}{a} \right) ^{2 \epsilon} 
\sum_{i=-2}^{2} \epsilon^{i} R^{(7)}_{i} + {\mathcal O} \left( 
\epsilon^{3} \right) , 
\eea
where:
\bea
R^{(7)}_{-2} & = & 1 \, , \\
R^{(7)}_{-1} & = & 3 - H(0,x) \, , \\
R^{(7)}_{0} & = & 7 - \zeta(2) - 3 H(0,x) + H(0,0,x) \, , \\
R^{(7)}_{1} & = & 15 - 3 \zeta(2) - 2 \zeta(3) - \bigl[ 7 - \zeta(2) \bigr] 
H(0,x) \nn\\
& & + 3 H(0,0,x) - H(0,0,0,x) \, , \\
R^{(7)}_{2} & = & 31 - 7 \zeta(2) - \frac{9 \zeta^{2}(2)}{10}  - 2 \zeta(3) - 
\bigl[ 15 - 3 \zeta(2) - 2 \zeta(3) \bigr] H(0,x)
\nn\\
& & + 7 H(0,0,x) - 3 H(0,0,0,x) + H(0,0,0,0,x) \, , 
\eea
\bea
\parbox{20mm}{\begin{fmfgraph*}(15,15)
\fmfleft{i1,i2}
\fmfright{o}
\fmf{photon}{i1,v1}
\fmf{photon}{i2,v2}
\fmf{plain}{v3,o}
\fmf{photon,tension=.3}{v2,v3}
\fmf{photon,tension=.3}{v1,v3}
\fmf{photon,tension=0}{v2,v1}
\fmf{plain,tension=0,left=.5}{v1,v3}
\end{fmfgraph*} }   & = & \mu^{2(4-D)} 
\int \{ d^{D}k_{1} \} \{ d^{D}k_{2} \}
\frac{1}{{\mathcal D}_{1} {\mathcal D}_{4} {\mathcal D}_{8} 
{\mathcal D}_{13} } \\
& = & \left( \frac{\mu^{2}}{a} \right) ^{2 \epsilon} 
\sum_{i=-2}^{1} \epsilon^{i} R^{(8)}_{i} + {\mathcal O} \left( 
\epsilon^{2} \right) , 
\eea
where:
\bea
R^{(8)}_{-2} & = & - \frac{1}{2} \, , \\
R^{(8)}_{-1} & = & - \frac{5}{2} + \left[ 1 + \frac{1}{x} \right] H(-1,x) +
\frac{1}{x} H(0,-1,x) \, , \\
R^{(8)}_{0} & = & - \frac{19}{2} \! - \zeta(2) \! + \left[ 1 \! + \! 
\frac{1}{x} \right] \bigl[ 5 H(-1,x) - 4 H(-1,-1,x) + 3 H(0,-1,x) 
\bigr] \nn\\
& & - H(0,-1,x) - \frac{1}{x} \bigl[ 4 H(0,-1,-1,x) - H(0,0,-1,x) \bigr]
\, , \nn\\
R^{(8)}_{1} & = & - \frac{65}{2} - 5 \zeta(2) + \zeta(3)
+ \biggl[ (19+2
\zeta(2)) \left( 1 + \frac{1}{x} \right) \biggr] H(-1,x) \nn\\
& & - 20 \left[ 1 + \frac{1}{x} \right] H(-1,-1,x) + \biggl[ 10 + 
\bigl( 9 + 2 \zeta(2) \bigr) \frac{1}{x}  \biggr] H(0,-1,x) \nn\\
& & + \biggl[ 1 + \frac{1}{x} \biggr] \bigl[ 3 H(0,0,-1,x) \!
+ \! 16 H(-1, \! -1, \! -1,x) \! - \! 6 H(-1,0, \! -1,x) \nn\\
& & - 12 H(0,-1,-1,x) \bigr] - H(0,0,-1,x) + 4 H(0,-1,-1,x) \nn\\
& & + \frac{1}{x} \bigl[ 16 H(0,-1,-1,-1,x) - 6 
H(0,-1,0,-1,x) \nn\\
& & - 4 H(0,0,-1,-1,x) + H(0,0,0,-1,x) \bigr]
 \, .
\eea
\bea
\parbox{20mm}{\begin{fmfgraph*}(15,15)
\fmfleft{i1,i2}
\fmfright{o}
\fmf{photon}{i1,v1}
\fmf{photon}{i2,v2}
\fmf{plain}{v3,o}
\fmf{plain,tension=.3}{v2,v3}
\fmf{photon,tension=.3}{v1,v3}
\fmf{photon,tension=0}{v2,v1}
\fmf{photon,tension=0,left=.5}{v1,v3}
\end{fmfgraph*} } & = & \mu^{2(4-D)} 
\int \{ d^{D}k_{1} \} \{ d^{D}k_{2} \}
\frac{1}{{\mathcal D}_{1} {\mathcal D}_{2} {\mathcal D}_{8} 
{\mathcal D}_{15} } \\
& = & \left( \frac{\mu^{2}}{a} \right) ^{2 \epsilon} 
\sum_{i=-2}^{2} \epsilon^{i} R^{(9)}_{i} + {\mathcal O} \left( 
\epsilon^{3} \right) , 
\eea
where:
\bea
R^{(9)}_{-2} & = & \frac{1}{2} \, , \\
R^{(9)}_{-1} & = & \frac{3}{2} \, , \\
R^{(9)}_{0} & = & \frac{5}{2} + \zeta(2) + \left[ 1 + \frac{1}{x} \right] 
H(-1,x) \nn\\
& & -  H(0,-1,x) \, , \\
R^{(9)}_{1} & = & - \frac{1}{2} + 3 \zeta(2) - \zeta(3)
+ \left[ 1 + \frac{1}{x} \right]  \bigl[ 7 H(-1,x) - 4 H(-1,-1,x) \nn\\
& & + H(0,-1,x) \bigr] - 3 H(0,-1,x) + 4 H(0,-1,-1,x) \nn\\
& & - H(0,0,-1,x) \, , \\
R^{(9)}_{2} & = & - \frac{51}{2} + 5 \zeta(2) - 3 \zeta(3) + 
\frac{9 \zeta^{2}(2)}{5} + \biggl[ (33+2\zeta(2)) \biggl( 1 + 
\frac{1}{x} \biggr) \biggr] H(-1,x) \nn\\
& & - 28 \biggl( 1 + \frac{1}{x} \biggr) H(-1,-1,x) + \biggl[ 2 - 
2\zeta(2) + \frac{7}{x} \biggr] H(0,-1,x)\nn\\
& & + \biggl[ 1 + \frac{1}{x} \biggr] \bigl[ 16 H(-1,-1,-1,x)
- 6 H(-1,0,-1,x) \bigr] \nn\\
& & + \biggl[ 2 - \frac{1}{x} \biggr] \bigl[ 4 H(0,-1,-1,x) - 
H(0,0,-1,x) \bigr] \nn\\
& & - 16 H(0,-1,-1,-1,x) +
6 H(0,-1,0,-1,x) \nn\\
& & + 4 H(0,0,-1,-1,x) - H(0,0,0,-1,x) \, .
\eea
\bea
\parbox{20mm}{\begin{fmfgraph*}(15,15)
\fmfleft{i1,i2}
\fmfright{o}
\fmf{photon}{i1,v1}
\fmf{photon}{i2,v2}
\fmf{plain}{v3,o}
\fmf{photon,tension=.3}{v2,v3}
\fmf{photon,tension=.3}{v1,v3}
\fmf{photon,tension=0}{v2,v1}
\fmf{photon,tension=0,right=.5}{v2,v3}
\end{fmfgraph*} }  & = & \mu^{2(4-D)} 
\int \{ d^{D}k_{1} \} \{ d^{D}k_{2} \}
\frac{1}{{\mathcal D}_{1} {\mathcal D}_{2} {\mathcal D}_{5} 
{\mathcal D}_{7} } \\
& = & \left( \frac{\mu^{2}}{a} \right) ^{2 \epsilon} 
\sum_{i=-2}^{2} \epsilon^{i} R^{(10)}{i} + {\mathcal O} \left( 
\epsilon^{3} \right) , 
\eea
where:
\bea
R^{(10)}{-2} & = & - \frac{1}{2} \, , \\
R^{(10)}{-1} & = & - \frac{5}{2} + H(0,x) \, , \\
R^{(10)}{0} & = & - \frac{19}{2} + \zeta(2) + 5 H(0,x) 
- 2 H(0,0,x) \, , \\
R^{(10)}{1} & = & - \frac{65}{2} + 5 \bigl[ \zeta(2) + \zeta(3)
\bigr]  + \biggl[ 19 - 2 \zeta(2) \biggr] H(0,x) -
10 H(0,0,x) \nn\\
& & + 4 H(0,0,0,x) \, , \\
R^{(10)}{2} & = & - \frac{211}{2} + 19 \zeta(2) + 
\frac{11 \zeta^{2}(2)}{5} + 25 \zeta(3) + \biggl[ 65 - 10 \bigl( 
\zeta(2) \nn\\
& & + \zeta(3) \bigr) \biggr] H(0,x) - \biggl[ 38 - 4 \zeta(2) \biggr] 
H(0,0,x) + 20 H(0,0,0,x) \nn\\
& & - 8 H(0,0,0,0,x) \, .
\eea
\bea
\parbox{20mm}{\begin{fmfgraph*}(15,15)
\fmfleft{i1,i2}
\fmfright{o}
\fmf{photon}{i1,v1}
\fmf{photon}{i2,v2}
\fmf{plain}{v3,o}
\fmf{photon,tension=.3}{v2,v3}
\fmf{photon,tension=.3}{v1,v3}
\fmf{photon,tension=0}{v2,v1}
\fmf{plain,right=45}{v3,v3}
\end{fmfgraph*} }   & = & \mu^{2(4-D)} 
\int \{ d^{D}k_{1} \} \{ d^{D}k_{2} \}
\frac{1}{{\mathcal D}_{1} {\mathcal D}_{4} {\mathcal D}_{5} 
{\mathcal D}_{13} } \\
& = & \left( \frac{\mu^{2}}{a} \right) ^{2 \epsilon} 
\sum_{i=-3}^{1} \epsilon^{i} R^{(11)}_{i} + {\mathcal O} \left( 
\epsilon^{2} \right) , 
\eea
where:
\bea
R^{(11)}_{-3} & = & - \frac{1}{x} \, , \\
R^{(11)}_{-2} & = & - \frac{1}{x} \bigl[ 1 - H(0,x) \bigr] \, , \\
R^{(11)}_{-1} & = & - \frac{1}{x} \bigl[ 1 - \zeta(2) - H(0,x) + H(0,0,x) 
\bigr] \, , \\
R^{(11)}_{0} & = & - \frac{1}{x} \bigl[ 1 - \zeta(2) - 2 \zeta(3) - (1-
\zeta(2)) H(0,x) + H(0,0,x) \nn\\
& & - H(0,0,0,x) \bigr] \, , \\
R^{(11)}_{1} & = & - \frac{1}{x} \biggl[ 1 - \zeta(2) - 2 \zeta(3) - \frac{9
\zeta^{2}(2)}{10} - (1 - \zeta(2) - 2 \zeta(3)) H(0,x) \nn\\
& & + (1-\zeta(2)) H(0,0,x) - H(0,0,0,x) + H(0,0,0,0,x) \biggr] \, .
\eea
\begin{eqnarray}
\parbox{20mm}{\begin{fmfgraph*}(15,15)
\fmfforce{0.2w,0.5h}{v1}
\fmfforce{0.5w,0.8h}{v2}
\fmfforce{0.5w,0.2h}{v3}
\fmfforce{0.8w,0.5h}{v4}
\fmfleft{i}
\fmfright{o}
\fmf{plain}{i,v1}
\fmf{plain}{v4,o}
\fmf{photon,tension=.2,left=.4}{v1,v2}
\fmf{photon,tension=.2,right=.4}{v1,v3}
\fmf{photon,tension=.2,left=.4}{v2,v4}
\fmf{photon,tension=.2,right=.4}{v3,v4}
\fmf{photon,tension=0}{v2,v3}
\end{fmfgraph*} } & = & \mu^{2(4-D)} 
\int \{ d^{D}k_{1} \} \{ d^{D}k_{2} \}
\frac{1}{{\mathcal D}_{1} {\mathcal D}_{2} {\mathcal D}_{3} 
{\mathcal D}_{9} {\mathcal D}_{11}} \\
& = & \left( \frac{\mu^{2}}{a} \right) ^{2 \epsilon} 
\sum_{i=0}^{1} \epsilon^{i} R^{(12)}_{i} + {\mathcal O} \left( 
\epsilon^{2} \right) , 
\eea
where:
\bea
aR^{(12)}_{0} & = & \frac{6 \zeta(3)}{x} \, , \\
aR^{(12)}_{1} & = & \frac{1}{x} \biggl[ \frac{18}{5} \zeta^{2}(2) + 12 \zeta(3)
- 12 \zeta(3) H(0,x) \biggr] \, . 
\eea
\begin{eqnarray}
\parbox{20mm}{\begin{fmfgraph*}(15,15)
\fmfforce{0.2w,0.5h}{v1}
\fmfforce{0.5w,0.8h}{v2}
\fmfforce{0.5w,0.2h}{v3}
\fmfforce{0.8w,0.5h}{v4}
\fmfleft{i}
\fmfright{o}
\fmf{plain}{i,v1}
\fmf{plain}{v4,o}
\fmf{plain,tension=.2,left=.4}{v1,v2}
\fmf{photon,tension=.2,right=.4}{v1,v3}
\fmf{photon,tension=.2,left=.4}{v2,v4}
\fmf{photon,tension=.2,right=.4}{v3,v4}
\fmf{photon,tension=0}{v2,v3}
\end{fmfgraph*} } & = & \mu^{2(4-D)} 
\int \{ d^{D}k_{1} \} \{ d^{D}k_{2} \}
\frac{1}{{\mathcal D}_{2} {\mathcal D}_{3} {\mathcal D}_{9} 
{\mathcal D}_{11} {\mathcal D}_{12}} \\
& = & \left( \frac{\mu^{2}}{a} \right) ^{2 \epsilon} 
\sum_{i=0}^{1} \epsilon^{i} R^{(13)}_{i} + {\mathcal O} \left( 
\epsilon^{2} \right) , 
\eea
where:
\bea
a R^{(13)}_{0} & = & \frac{1}{x} \bigl[ \zeta(2) H(-1,x)  \nn\\
& & + 2 H(-1,0,-1,x) + H(-1,0,0,x) + 2 H(0,-1,-1,x) \nn\\
& & - H(0,-1,0,x) \bigr] \, , \\
a R^{(13)}_{1} & = & \frac{1}{x} \bigl[ 2( \zeta(2) \! - \! 2 \zeta(3) )H(-1,x)
\! - \! 2 \zeta(2) H(0, \! -1,x) \! + \! \zeta(2) H( \! -1,0,x)) \nn\\
& & - 2 \zeta(2) H(-1,-1,x) - 4 H(-1,0,-1,x) + 2 H(-1,0,0,x) \nn\\
& & + 4 H(0,-1,-1,x) - 2 H(0,-1,0,x) + 4 H(-1,-1,0,-1,x) \nn\\
& & - 2 H( \! -1, \! -1,0,0,x) \! + \! 4 H( \! -1,0, \! -1, \! -1,x) \! 
+ \! 2 H( \! -1,0, \! -1,0,x) \nn\\
& & - 2 H(-1,0,0,-1,x) \! - \! 3 H(-1,0,0,0,x) \! - \! 12 
H(0,-1,-1,-1,x) \nn\\
& & + 2 H(0,-1,-1,0,x) + 4 H(0,-1,0,-1,x) + 2 H(0,-1,0,0,x) \nn\\
& & + 2 H(0,0,-1,-1,x) - H(0,0,-1,0,x) \bigr]
\, ,
\eea
\begin{eqnarray}
\parbox{20mm}{\begin{fmfgraph*}(15,15)
\fmfforce{0.2w,0.5h}{v1}
\fmfforce{0.5w,0.8h}{v2}
\fmfforce{0.5w,0.2h}{v3}
\fmfforce{0.8w,0.5h}{v4}
\fmfleft{i}
\fmfright{o}
\fmf{photon}{i,v1}
\fmf{photon}{v4,o}
\fmf{photon,tension=.2,left=.4}{v1,v2}
\fmf{plain,tension=.2,right=.4}{v1,v3}
\fmf{photon,tension=.2,left=.4}{v2,v4}
\fmf{photon,tension=.2,right=.4}{v3,v4}
\fmf{photon,tension=0}{v2,v3}
\end{fmfgraph*} } & = & \mu^{2(4-D)} 
\int \{ d^{D}k_{1} \} \{ d^{D}k_{2} \}
\frac{1}{{\mathcal D}_{1} {\mathcal D}_{2} {\mathcal D}_{3} 
{\mathcal D}_{5} {\mathcal D}_{17}} \\
& = & \left( \frac{\mu^{2}}{a} \right) ^{2 \epsilon} 
\sum_{i=-2}^{1} \epsilon^{i} R^{(14)}_{i} + {\mathcal O} \left( 
\epsilon^{2} \right) , 
\eea
where:
\bea
a R^{(14)}_{-2} & = & \frac{1}{2} \, , \\
a R^{(14)}_{-1} & = & \frac{1}{2}  \, , \\
a R^{(14)}_{0} & = & \frac{1}{2} + \zeta(2) \, , \\
a R^{(14)}_{1} & = & \frac{1}{2} + \zeta(2) - \zeta(3) \, ,
\eea
\begin{eqnarray}
\parbox{20mm}{\begin{fmfgraph*}(15,15)
\fmfleft{i1,i2}
\fmfright{o}
\fmfforce{0.8w,0.5h}{v4}
\fmf{photon}{i1,v1}
\fmf{photon}{i2,v2}
\fmf{plain}{v4,o}
\fmf{photon,tension=.15}{v2,v4}
\fmf{photon,tension=.4}{v1,v3}
\fmf{photon,tension=.2}{v3,v4}
\fmf{photon,tension=0}{v2,v1}
\fmf{photon,tension=0}{v2,v3}
\end{fmfgraph*} } & = & \mu^{2(4-D)} 
\int \{ d^{D}k_{1} \} \{ d^{D}k_{2} \}
\frac{1}{{\mathcal D}_{1} {\mathcal D}_{2} {\mathcal D}_{5} 
{\mathcal D}_{7} {\mathcal D}_{8}} \\
& = & \left( \frac{\mu^{2}}{a} \right) ^{2 \epsilon} 
\sum_{i=-2}^{0} \epsilon^{i} R^{(15)}_{i} + {\mathcal O} \left( 
\epsilon \right) , 
\eea
where:
\bea
a R^{(15)}_{-2} & = & \frac{\zeta(2)}{2 x} \, , \\
a R^{(15)}_{-1} & = & \frac{1}{2 x} \bigl[ \zeta(3) - 2 \zeta(2) H(0,x) 
\bigr] \, , \\
a R^{(15)}_{0} & = & - \frac{1}{x} \biggl[ \frac{\zeta^{2}(2)}{10} + \zeta(3)
H(0,x) - 2 \zeta(2) H(0,0,x) \biggr] \, ,
\eea
\begin{eqnarray}
\parbox{20mm}{\begin{fmfgraph*}(15,15)
\fmfleft{i1,i2}
\fmfright{o}
\fmfforce{0.8w,0.5h}{v4}
\fmf{photon}{i1,v1}
\fmf{photon}{i2,v2}
\fmf{plain}{v4,o}
\fmf{photon,tension=.4}{v1,v3}
\fmf{photon,tension=.2}{v3,v4}
\fmf{photon,tension=.15}{v2,v4}
\fmf{photon,tension=0}{v2,v1}
\fmf{plain,tension=0,left=.5}{v3,v4}
\end{fmfgraph*} } & = & \mu^{2(4-D)} 
\int \{ d^{D}k_{1} \} \{ d^{D}k_{2} \}
\frac{1}{{\mathcal D}_{1} {\mathcal D}_{4} {\mathcal D}_{5} 
{\mathcal D}_{8} {\mathcal D}_{13}} \\
& = & \left( \frac{\mu^{2}}{a} \right) ^{2 \epsilon} 
\sum_{i=-3}^{1} \epsilon^{i} R^{(16)}_{i} + {\mathcal O} \left( 
\epsilon^{2} \right) , 
\eea
where:
\bea
a R^{(16)}_{-3} & = & \frac{1}{x} \, , \\
a R^{(16)}_{-2} & = & \frac{1}{x} \bigl[ 1 - H(0,x) \bigr] \, , \\
a R^{(16)}_{-1} & = & \frac{1}{x} \bigl[ 2 - \zeta(2) - H(0,x) 
- \biggl( 1 + \frac{1}{x} \biggr) H(-1,x) + H(0,0,x) \nn\\
& & + H(0,-1,x)\bigr] \, , \\
a R^{(16)}_{0} & = & \frac{1}{x} \biggl[ 4 - \zeta(2) - 2 \zeta(3) - ( 1 -
\zeta(2) ) H(0,x) - 3 \biggl( 1 + \frac{1}{x} \biggr) H(-1,x) \nn\\
& &  + H(0,0,x) + 4 \biggl( 1 + \frac{1}{x} \biggr) H(-1,-1,x) - 
H(0,-1,x) \nn\\
& & - H(0,0,0,x) + H(0,0,-1,x) - 4 H(0,-1,-1,x) \biggr] \, , \\
a R^{(16)}_{1} & = & \frac{1}{x} \biggl[ 8 + \zeta(2) - \frac{9 
\zeta^{2}(2)}{10} - 2 \zeta(3)- ( 1 - \zeta(2) - 2 \zeta(3) ) H(0,x)
\nn\\
& & - ( 7 + 2 \zeta(2) ) \biggl( 1 + \frac{1}{x} \biggr) H(-1,x) + 
( 1 - \zeta(2) ) H(0,0,x) - ( 5 \nn\\
& & - 2 \zeta(2) ) H(0,-1,x) + 12 \biggl( 1 + \frac{1}{x} \biggr) 
H(-1,-1,x)  - H(0,0,0,x) \nn\\
& & - H(0,0,-1,x) + 4 H(0,-1,-1,x) + \biggl( 1 + \frac{1}{x} \biggr) 
\bigr[ 6 H(-1,0,-1,x) \nn\\
& & - 16 H(-1,-1,-1,x) \bigr] + H(0,0,0,0,x) + H(0,0,0,-1,x) \nn\\
& & + 16 H(0,-1,-1,-1,x) - 6 H(0,-1,0,-1,x) \nn\\
& & - 4 H(0,0,-1,-1,x) \biggr]
\, ,
\eea
\begin{eqnarray}
\parbox{20mm}{\begin{fmfgraph*}(15,15)
\fmfleft{i1,i2}
\fmfright{o}
\fmfforce{0.8w,0.5h}{v4}
\fmf{photon}{i1,v1}
\fmf{photon}{i2,v2}
\fmf{plain}{v4,o}
\fmf{photon,tension=.4}{v1,v3}
\fmf{photon,tension=.2}{v3,v4}
\fmf{photon,tension=.15}{v2,v4}
\fmf{plain,tension=0}{v2,v1}
\fmf{photon,tension=0,left=.5}{v3,v4}
\end{fmfgraph*} } & = & \mu^{2(4-D)} 
\int \{ d^{D}k_{1} \} \{ d^{D}k_{2} \}
\frac{1}{{\mathcal D}_{2} {\mathcal D}_{4} {\mathcal D}_{5} 
{\mathcal D}_{8} {\mathcal D}_{12}} \\
& = & \left( \frac{\mu^{2}}{a} \right) ^{2 \epsilon} 
\sum_{i=-1}^{1} \epsilon^{i} R^{(17)}_{i} + {\mathcal O} \left( 
\epsilon^{2} \right) , 
\eea
where:
\bea
a R^{(17)}_{-1} & = & - \frac{1}{x} H(1,0,x) \, , \\
a R^{(17)}_{0} & = & - \frac{1}{x} \bigl[ 2 \zeta(2) H(1,x) + 2 H(1,0,x) +
H(0,1,0,x) \nn\\
& & - 2 H(1,0,0,x) + 2 H(1,1,0,x) \bigr] \, , \\
a R^{(17)}_{1} & = & - \frac{1}{x} \bigl[ 4 ( \zeta(2) \! + \! \zeta(3) ) 
H(1,x) \! + \! 2 \zeta(2) H(0,1,x) \! + \! 2 ( 2 \! - \! \zeta(2)) 
H(1,0,x) \nn\\
& & + 4 \zeta(2) H(1,1,x) + 2 H(0,1,0,x) - 4 H(1,0,0,x) \nn\\
& & + 4 H(1,1,0,x) + H(0,0,1,0,x) - 2 H(0,1,0,0,x) \nn\\
& & + 2 H(0,1,1,0,x) + 4 H(1,0,0,0,x) + 2 H(1,0,1,0,x) \nn\\
& & - 4 H(1,1,0,0,x) + 4 H(1,1,1,0,x) \bigr]
\, ,
\eea
\begin{eqnarray}
\parbox{20mm}{\begin{fmfgraph*}(15,15)
\fmfleft{i1,i2}
\fmfright{o}
\fmfforce{0.8w,0.5h}{v4}
\fmf{photon}{i1,v1}
\fmf{photon}{i2,v2}
\fmf{plain}{v4,o}
\fmf{photon,tension=.4}{v1,v3}
\fmf{photon,tension=.2}{v3,v4}
\fmf{plain,tension=.15}{v2,v4}
\fmf{photon,tension=0}{v2,v1}
\fmf{photon,tension=0,left=.5}{v3,v4}
\end{fmfgraph*} } & = & \mu^{2(4-D)} 
\int \{ d^{D}k_{1} \} \{ d^{D}k_{2} \}
\frac{1}{{\mathcal D}_{1} {\mathcal D}_{2} {\mathcal D}_{4} 
{\mathcal D}_{8} {\mathcal D}_{15}} \\
& = & \left( \frac{\mu^{2}}{a} \right) ^{2 \epsilon} 
\sum_{i=-2}^{0} \epsilon^{i} R^{(18)}_{i} + {\mathcal O} \left( 
\epsilon \right) , 
\eea
where:
\bea
a R^{(18)}_{-2} & = & - \frac{1}{2x} H(-1,x) \, , \\
a R^{(18)}_{-1} & = & - \frac{1}{2x} \bigl[ 2 H(-1,x) - 4 H(-1,-1,x) + 
H(0,-1,x) \bigr] \, , \\
a R^{(18)}_{0} & = & - \frac{1}{2x} \bigl[ 2 ( 2 + \zeta(2)) H(-1,x) + 2 
H(0,-1,x) - 8 H(-1,-1,x) \nn\\
& & + 16 H(-1,-1,-1,x) - 6 H(-1,0,-1,x) - 4 H(0,-1,-1,x) \nn\\
& & + H(0,0,-1,x) \bigr] \, ,
\eea
\begin{eqnarray}
\parbox{20mm}{\begin{fmfgraph*}(15,15)
\fmfleft{i1,i2}
\fmfright{o}
\fmfforce{0.8w,0.5h}{v4}
\fmf{photon}{i1,v1}
\fmf{photon}{i2,v2}
\fmf{plain}{v4,o}
\fmf{photon,tension=.4}{v1,v3}
\fmf{photon,tension=.2}{v3,v4}
\fmf{photon,tension=.15}{v2,v4}
\fmf{photon,tension=0}{v2,v1}
\fmf{photon,tension=0,left=.5}{v3,v4}
\end{fmfgraph*} } & = & \mu^{2(4-D)} 
\int \{ d^{D}k_{1} \} \{ d^{D}k_{2} \}
\frac{1}{{\mathcal D}_{1} {\mathcal D}_{2} {\mathcal D}_{4} 
{\mathcal D}_{5} {\mathcal D}_{8}} \\
& = & \left( \frac{\mu^{2}}{a} \right) ^{2 \epsilon} 
\sum_{i=-3}^{1} \epsilon^{i} R^{(19)}_{i} + {\mathcal O} \left( 
\epsilon^{2} \right) , 
\eea
where:
\bea
a R^{(19)}_{-3} & = & \frac{1}{2x}  \, , \\
a R^{(19)}_{-2} & = & \frac{1}{x} \bigl[ 1 - H(0,x) \bigr]  \, , \\
a R^{(19)}_{-1} & = & \frac{1}{x} \bigl[ 2 - \zeta(2) - 2 H(0,x) + 2 H(0,0,x)
\bigr]   \, , \\
a R^{(19)}_{0} & = & \frac{1}{x} \bigl[ 4 - 2 \zeta(2) - 5 \zeta(3) - 2 ( 2 -
\zeta(2) ) H(0,x) + 4 H(0,0,x) \nn\\
& & - 4 H(0,0,0,x) \bigr] \, , \\
a R^{(19)}_{1} & = & \frac{1}{x} \biggl[ 8 - 4 \zeta(2) - 
\frac{11 \zeta^{2}(2)}{5} - 10 \zeta(3) - 2 ( 4 - 2 \zeta(2) - 
5 \zeta(3) ) H(0,x) \nn\\
& & + 4 ( 2 \! - \! \zeta(2) ) H(0,0,x) \! - \! 8 H(0,0,0,x) \! + \! 8 
H(0,0,0,0,x) \bigr] ,
\eea
\begin{eqnarray}
\parbox{20mm}{\begin{fmfgraph*}(15,15)
\fmfleft{i1,i2}
\fmfright{o}
\fmfforce{0.8w,0.5h}{v4}
\fmf{photon}{i1,v1}
\fmf{photon}{i2,v2}
\fmf{plain}{v4,o}
\fmf{plain,tension=.4}{v1,v3}
\fmf{photon,tension=.2}{v3,v4}
\fmf{photon,tension=.15}{v2,v4}
\fmf{photon,tension=0}{v2,v1}
\fmf{photon,tension=0,left=.5}{v3,v4}
\end{fmfgraph*} } & = & \mu^{2(4-D)} 
\int \{ d^{D}k_{1} \} \{ d^{D}k_{2} \}
\frac{1}{{\mathcal D}_{1} {\mathcal D}_{2} {\mathcal D}_{5} 
{\mathcal D}_{7} {\mathcal D}_{15}} \\
& = & \left( \frac{\mu^{2}}{a} \right) ^{2 \epsilon} 
\sum_{i=-2}^{1} \epsilon^{i} R^{(20)}_{i} + {\mathcal O} \left( 
\epsilon^{2} \right) , 
\eea
where:
\bea
a R^{(20)}_{-2} & = & - \frac{1}{x} H(1,x) \, , \\
a R^{(20)}_{-1} & = & - \frac{1}{x} \bigl[ 2 H(-1,x) - 2 H(-1,-1,x) - 
H(-1,0,x) \nn\\
& & + H(0,-1,x) \bigr]  \, , \\
a R^{(20)}_{0} & = & \frac{1}{x} \bigl[ - 4 H(-1,x) + 4 H(-1,-1,x) + 2 
H(-1,0,x) - 2 H(0,-1,x) \nn\\
& & - 4 H(-1,-1,-1,x) - 2 H(-1,-1,0,x) + 2 H(-1,0,-1,x) \nn\\
& & - 2 H(-1,0,0,x) + 2 H(0,-1,-1,x) + H(0,-1,0,x) \nn\\
& & - H(0,0,-1,x) \bigr] \, , \\
a R^{(20)}_{1} & = & \frac{1}{32 x} \bigl[ ( 8 - 6 \zeta(3) ) H(-1,x)
+ 4 H(0,-1,x) + 2 \zeta(2) H(-1,0,x) \nn\\
& & - 4 H(-1,0,x) - 8 H(-1,-1,x) + 4 H(-1,0,0,x) \nn\\
& & - 4  H(-1,0,-1,x) - 4 H(0,-1,-1,x) + 2 H(0,0,-1,x)  \nn\\
& & - 2 H(0,-1,0,x) + 4 H(-1,-1,0,x) + 8 H(-1,-1,-1,x) \nn\\
& & - 8 H(-1,-1,-1,-1,x) - 4 H(-1,-1,-1,0,x) \nn\\
& & + 4 H(-1,-1,0,-1,x) - 4 H(-1,-1,0,0,x)  \nn\\
& & + 4 H(-1,0,-1,-1,x)  + 2 H(-1,0,-1,0,x) \nn\\
& & - 2 H(-1,0,0,-1,x) - 4 H(-1,0,0,0,x) \nn\\
& & + 4 H(0,-1,-1,-1,x) + 2 H(0,-1,-1,0,x) \nn\\
& & - 2 H(0,-1,0,-1,x) + 2 H(0,-1,0,0,x) \nn\\
& & - 2 H(0,0,-1,-1,x) - H(0,0,-1,0,x) \nn\\
& & + H(0,0,0,-1,x) \bigr]
\, ,
\eea
\begin{eqnarray}
\parbox{20mm}{\begin{fmfgraph*}(15,15)
\fmfleft{i1,i2}
\fmfright{o}
\fmfforce{0.2w,0.9h}{v2}
\fmfforce{0.2w,0.1h}{v1}
\fmfforce{0.2w,0.5h}{v3}
\fmfforce{0.8w,0.5h}{v4}
\fmf{photon}{i1,v1}
\fmf{photon}{i2,v2}
\fmf{plain}{v4,o}
\fmf{photon,tension=0}{v2,v3}
\fmf{photon,tension=0}{v3,v4}
\fmf{photon,tension=0}{v1,v4}
\fmf{plain,tension=0}{v2,v4}
\fmf{photon,tension=0}{v1,v3}
\end{fmfgraph*} } & = & \mu^{2(4-D)} 
\int \{ d^{D}R^{(21)}_{1} \} \{ d^{D}R^{(21)}_{2} \}
\frac{1}{{\mathcal D}_{1} {\mathcal D}_{2} {\mathcal D}_{3} 
{\mathcal D}_{6} {\mathcal D}_{15}} \\
& = & \left( \frac{\mu^{2}}{a} \right) ^{2 \epsilon} 
\sum_{i=-2}^{0} \epsilon^{i} R^{(21)}_{i} + {\mathcal O} \left( 
\epsilon \right) , 
\eea
where:
\bea
a R^{(21)}_{-2} & = & \frac{1}{2x} H(0,-1,x) \, , \\
a R^{(21)}_{-1} & = & \frac{1}{2x} \bigl[ H(0,0,-1,x) - 4 H(0,-1,-1,x)
\bigr]  \, , \\
a R^{(21)}_{0} & = & \frac{1}{2x} \bigl[ 2 \zeta(2) H(0,-1,x) + H(0,0,0,-1,x) -
4 H(0,0,-1,-1,x) \nn\\
& & - 6 H(0,-1,0,-1,x) + 16 H(0,-1,-1,-1,x) \bigr]  \, ,
\eea 
\begin{eqnarray}
\parbox{20mm}{\begin{fmfgraph*}(15,15)
\fmfleft{i1,i2}
\fmfright{o}
\fmfforce{0.2w,0.9h}{v2}
\fmfforce{0.2w,0.1h}{v1}
\fmfforce{0.2w,0.5h}{v3}
\fmfforce{0.8w,0.5h}{v4}
\fmf{photon}{i1,v1}
\fmf{photon}{i2,v2}
\fmf{plain}{v4,o}
\fmf{photon,tension=0}{v1,v3}
\fmf{photon,tension=0}{v3,v4}
\fmf{photon,tension=0}{v2,v4}
\fmf{photon,tension=0}{v2,v3}
\fmf{photon,tension=0}{v1,v4}
\end{fmfgraph*} } & = & \mu^{2(4-D)} 
\int \{ d^{D}k_{1} \} \{ d^{D}k_{2} \}
\frac{1}{{\mathcal D}_{1} {\mathcal D}_{2} {\mathcal D}_{3} 
{\mathcal D}_{4} {\mathcal D}_{6}} \\
& = & \left( \frac{\mu^{2}}{a} \right) ^{2 \epsilon} 
\sum_{i=-4}^{0} \epsilon^{i} R^{(22)}_{i} + {\mathcal O} \left( 
\epsilon \right) , 
\eea
where:
\bea
a R^{(22)}_{-4} & = & \frac{1}{4x}  \, , \\
a R^{(22)}_{-3} & = & - \frac{1}{2x} H(0,x) \, , \\
a R^{(22)}_{-2} & = & - \frac{1}{2x} \bigl[ \zeta(2) - 2 H(0,0,x) \bigr] 
\, , \\
a R^{(22)}_{-1} & = & - \frac{1}{2x} \bigl[ 5 \zeta(3) - 2 \zeta(2)  H(0,x) 
+ 4 H(0,0,0,x) \bigr] \, , \\
a R^{(22)}_{0} & = & - \frac{1}{x} \biggl[ \frac{11 \zeta^{2}(2)}{10} 
- 5 \zeta(3) H(0,x) + 2 \zeta(2) H(0,0,x) \nn\\
& & - 4 H(0,0,0,0,x) \biggr] \, ,
\eea
\begin{eqnarray}
\parbox{20mm}{\begin{fmfgraph*}(15,15)
\fmfleft{i1,i2}
\fmfright{o}
\fmfforce{0.2w,0.9h}{v2}
\fmfforce{0.2w,0.1h}{v1}
\fmfforce{0.2w,0.55h}{v3}
\fmfforce{0.2w,0.15h}{v5}
\fmfforce{0.8w,0.5h}{v4}
\fmf{photon}{i1,v1}
\fmf{photon}{i2,v2}
\fmf{plain}{v4,o}
\fmf{photon}{v2,v3}
\fmf{photon,left}{v3,v5}
\fmf{photon,right}{v3,v5}
\fmf{photon}{v1,v4}
\fmf{photon}{v2,v4}
\end{fmfgraph*} } & = & \mu^{2(4-D)} 
\int \{ d^{D}k_{1} \} \{ d^{D}k_{2} \}
\frac{1}{{\mathcal D}_{1} {\mathcal D}_{2} {\mathcal D}_{3} 
{\mathcal D}_{4} {\mathcal D}_{5}} \\
& = & \left( \frac{\mu^{2}}{a} \right) ^{2 \epsilon} 
\sum_{i=-3}^{1} \epsilon^{i} R^{(23)}_{i} + {\mathcal O} \left( 
\epsilon^{2} \right) , 
\eea
where:
\bea
a R^{(23)}_{-3} & = & \frac{1}{4x} \, , \\
a R^{(23)}_{-2} & = & \frac{1}{2x} \bigl[ 1 - H(0,x) \bigr]  \, , \\
a R^{(23)}_{-1} & = & \frac{1}{x} \bigl[ 1 - H(0,x) + H(0,0,x)\bigr]  \, , \\
a R^{(23)}_{0} & = & \frac{2}{x} \bigl[ 1 - \zeta(3) - H(0,x) + H(0,0,x)
- H(0,0,0,x) \bigr]  \, , \\
a R^{(23)}_{1} & = & \frac{4}{x} \biggl[ 1 - \zeta(3) - 
\frac{3 \zeta^{2}(2)}{10} - ( 1 - \zeta(3) ) H(0,x) + H(0,0,x) \nn\\
& & - H(0,0,0,x) + H(0,0,0,0,x) \bigr]  \, ,
\eea
\begin{eqnarray}
\parbox{20mm}{\begin{fmfgraph*}(15,15)
\fmfleft{i1,i2}
\fmfright{o}
\fmf{photon}{i1,v1}
\fmf{photon}{i2,v2}
\fmf{plain}{v4,o}
\fmf{photon,tension=.3}{v2,v3}
\fmf{photon,tension=.3}{v1,v3}
\fmf{photon,tension=0}{v2,v1}
\fmf{photon,tension=.2,left}{v3,v4}
\fmf{photon,tension=.2,right}{v3,v4}
\end{fmfgraph*} } & = & \mu^{2(4-D)} 
\int \{ d^{D}k_{1} \} \{ d^{D}k_{2} \}
\frac{1}{{\mathcal D}_{1} {\mathcal D}_{2} {\mathcal D}_{4} 
{\mathcal D}_{5} {\mathcal D}_{10}} \\
& = & \left( \frac{\mu^{2}}{a} \right) ^{2 \epsilon} 
\sum_{i=-3}^{1} \epsilon^{i} R^{(24)}_{i} + {\mathcal O} \left( 
\epsilon^{2} \right) , 
\eea
where:
\bea
a R^{(24)}_{-3} & = & \frac{1}{x}  \, , \\
a R^{(24)}_{-2} & = & \frac{2}{x} \bigl[ 1 - H(0,x) \bigr] \, , \\
a R^{(24)}_{-1} & = & \frac{2}{x} \bigl[ 2 - \zeta(2) - 2 H(0,x) + 2 H(0,0,x) 
\bigr] 
\, , \\
a R^{(24)}_{0} & = & \frac{4}{x} \bigl[ 2 - \zeta(2) - \zeta(3) - ( 2 -
\zeta(2) ) H(0,x) + 2 H(0,0,x) \nn\\
& & - 2  H(0,0,0,x) \bigr] \, , \\
a R^{(24)}_{1} & = & \frac{4}{x} \biggl[  4 - 2 \zeta(2) -
\frac{\zeta^{2}(2)}{5}  - 2 \zeta(3)  - 2 ( 2 - \zeta(2) -  \zeta(3) )
H(0,x) \nn\\
& & + 2 ( 2 \! - \! \zeta(2) ) H(0, \! 0,x) - 4 H(0, \! 0, \! 0,x) + 4 
H(0, \! 0, \! 0, \! 0,x) \biggr] ,
\eea
\begin{eqnarray}
\parbox{20mm}{\begin{fmfgraph*}(15,15)
\fmfleft{i1,i2}
\fmfright{o}
\fmf{photon}{i1,v1}
\fmf{photon}{i2,v2}
\fmf{plain}{v4,o}
\fmf{plain,tension=.3}{v2,v3}
\fmf{photon,tension=.3}{v1,v3}
\fmf{photon,tension=0}{v2,v1}
\fmf{photon,tension=.2,left}{v3,v4}
\fmf{photon,tension=.2,right}{v3,v4}
\end{fmfgraph*} }  & = & \mu^{2(4-D)} 
\int \{ d^{D}k_{1} \} \{ d^{D}k_{2} \}
\frac{1}{{\mathcal D}_{1} {\mathcal D}_{2} {\mathcal D}_{4} 
{\mathcal D}_{5} {\mathcal D}_{15}} \\
& = & \left( \frac{\mu^{2}}{a} \right) ^{2 \epsilon} 
\sum_{i=-2}^{1} \epsilon^{i} R^{(25)}_{i} + {\mathcal O} \left( 
\epsilon^{2} \right) , 
\end{eqnarray}
where:
\bea
a R^{(25)}_{-2} & = &  - \frac{1}{x} H(-1,x) \, , \\
a R^{(25)}_{-1} & = & - \frac{1}{x} \bigl[ 2 H(-1,x) + H(-1,0,x) - 2 
H(-1,-1,x) \bigr] \, , \\
a R^{(25)}_{0} & = & - \frac{1}{x} \bigl[ ( 4 - \zeta(2)) H(-1,x) - 2 
H(-1,0,x) - 4 H(-1,-1,x) \nn\\
& & + 4 H(-1, \! -1, \! -1,x) + \! 2 H(-1, \! -1,0,x) \! + \! 
H(-1,0,0,x) \bigr] , \\
a R^{(25)}_{1} & = & - \frac{1}{x} \bigl\{ 2 ( 4 - \zeta(2) - \zeta(3)) H(-1,x) 
- ( 4 - \zeta(2)) \bigl[ H(-1,0,x)  \nn\\
& & - 2 (4-\zeta(2)) H(-1,-1,x) \bigr] + 2 H(-1,0,0,x) \nn\\
& & + 4 H(-1,-1,0,x) + 8 H(-1,-1,-1,x) - H(-1,0,0,0,x) \nn\\
& & - 2 H(-1,-1,0,0,x) - 4 H(-1,-1,-1,0,x) \nn\\
& & - 8 H(-1,-1,-1,-1,x) \bigr\}
\, ,
\eea
\begin{eqnarray}
\parbox{20mm}{\begin{fmfgraph*}(15,15)
\fmfleft{i1,i2}
\fmfright{o}
\fmfforce{0.1w,0.1h}{v1}
\fmfforce{0.4w,0.3h}{v2}
\fmfforce{0.1w,0.9h}{v3}
\fmfforce{0.4w,0.7h}{v4}
\fmfforce{0.9w,0.5h}{v5}
\fmf{photon}{i1,v1}
\fmf{phantom}{i2,v3}
\fmf{plain}{v5,o}
\fmf{photon,left}{v1,v2}
\fmf{plain,right}{v1,v2}
\fmf{photon}{v3,v4}
\fmf{photon}{v2,v4}
\fmf{photon}{v4,v5}
\fmf{photon}{v2,v5}
\end{fmfgraph*} } & = & \mu^{2(4-D)} 
\int \{ d^{D}k_{1} \} \{ d^{D}k_{2} \}
\frac{1}{{\mathcal D}_{1} {\mathcal D}_{2} {\mathcal D}_{4} 
{\mathcal D}_{5} {\mathcal D}_{17}} \\
& = & \left( \frac{\mu^{2}}{a} \right) ^{2 \epsilon} 
\sum_{i=-3}^{1} \epsilon^{i} R^{(26)}_{i} + {\mathcal O} \left( 
\epsilon^{2} \right) , 
\eea
where:
\bea
a R^{(26)}_{-3} & = & \frac{1}{x} \, , \\
a R^{(26)}_{-2} & = & \frac{1}{x} \bigl[ 1 - H(0,x) \bigr]  \, , \\
a R^{(26)}_{-1} & = & \frac{1}{x} \bigl[ 1 - \zeta(2) - H(0,x) + H(0,0,x) 
\bigr]  \, , \\
a R^{(26)}_{0} & = & \frac{1}{x} \bigl[ 1 - \zeta(2) - 2 \zeta(3) - ( 1 -
\zeta(2) ) H(0,x) + H(0,0,x) \nn\\
& & - H(0,0,0,x) \bigr]  \, , \\
a R^{(26)}_{1} & = & \frac{1}{x} \biggl[ 1 - \zeta(2) - 2 \zeta(3) - 
\frac{9 \zeta^{2}(2)}{10} - ( 1 - \zeta(2) - 2 \zeta(3) ) H(0,x) \nn\\
& & + ( 1 - \zeta(2) ) H(0,0,x) - H(0,0,0,x) + H(0,0,0,0,x) \bigr]  \, .
\eea

%
%
%                BIBLIOGRAFIA
%
%

\end{fmffile}


\begin{thebibliography}{99} 
\def    \np     #1#2#3{{\it Nucl. Phys.} {\bf #1} (19#2) #3}
\def    \nptwoth     #1#2#3{{\it Nucl. Phys.} {\bf #1} (20#2) #3}
\def    \prep   #1#2#3{{\it Phys. Rep.} {\bf #1}  (19#2) #3}
\def    \pl     #1#2#3{{\it Phys. Lett.} {\bf #1} (19#2) #3}
\def    \pltwoth     #1#2#3{{\it Phys. Lett.} {\bf #1} (20#2) #3}
\def    \plold  #1#2#3{{\it Phys. Lett.} {\bf #1B} (19#2) #3}
\def    \prl    #1#2#3{{\it Phys. Rev. Lett.} {\bf #1}  (19#2) #3} 
\def    \pr     #1#2#3{{\it Phys. Rev.} {\bf #1}  (19#2) #3}
\def    \prd    #1#2#3{{\it Phys. Rev.} {\bf D#1}  (19#2) #3} 
\def    \zeit   #1#2#3{{\it Z. Phys.} {\bf C#1}  (19#2) #3}
\def    \cmp    #1#2#3{{\it Comm. Math. Phys.} {\bf #1}  (19#2) #3}
\def    \ibid   #1#2#3{{\it ibid.} {\bf #1} (19#2) #3}
\def    \nc     #1#2#3{{\it Nuovo Cim.} {\bf #1} (19#2) #3}
\def    \acta   #1#2#3{{\it Acta Phys. Polon.} {\bf #1} (19#2) #3}
\def    \tmp    #1#2#3{{\it Theor. Math. Phys.} {\bf #1} (19#2) #3}
\def    \comp    #1#2#3{{\it Comput. Phys. Commun.} {\bf #1} (20#2) #3}
\def    \hepph  #1 {{\tt hep-ph/#1}}
\def    \hepex  #1 {{\tt hep-ex/#1}}
\parskip 0pt
\itemsep=0pt


%
%%%%%%%%%%%%%% 2-loop logs %%%%%%%%%%%%%%%%%%%
%

\bibitem{ciafcom}
  P. Ciafaloni and D. Comelli, \pl{B446}{99}{278} ({\tt hep-ph/9809321}).
%%CITATION = HEP-PH 9809321;%%

%%%%%%%%%%%%%%% Harm-Polylog %%%%%%%%%%%%%%%%%%%%%%%%%%%%%%%%%%%%

\bibitem{Polylog}
  E.~Remiddi and J.~A.~M.~Vermaseren, {\it Int. J. Mod. Phys.} 
  {\bf A15} (2000) 725 ({\tt hep-ph/9905237}). 
%%CITATION = HEP-PH 9905237;%%

\bibitem{Polylog3}
  T. Gehrmann and E. Remiddi, \comp{141}{01}{296} 
  ({\tt hep-ph/0107173}).
%%CITATION = HEP-PH 0107173;%%

%%%%%%%%%%%%%%% Polylog %%%%%%%%%%%%%%%%%%%%%%%%%%%%%%%%%%%%

\bibitem{Nielsen}
  N. Nielsen, {\it Nova Acta Leopoldiana} (Halle) {\bf 90} (1909) 123.

\bibitem{Kolbig}
  K. S. K\"olbig, J. A. Mignaco and E. Remiddi, {\it BIT} {\bf 10} 
  (1970) 38.

%
%%%%%%%%%%%%%%%%%% Dim Reg %%%%%%%%%%%%%%%%%%%%%%%%%%%%%
%

\bibitem{DimReg}
  G. 't Hooft and M. Veltman, \np{B44}{72}{189}.\\
%%CITATION = NUPHA,B44,189;%%
  C. G. Bollini and J. J. Giambiagi, \plold{40}{72}{566}; 
  \nc{12B}{72}{20}. \\
%%CITATION = NUCIA,B12,20;%%
  J. Ashmore, {\it Lett. Nuovo Cimento} {\bf 4} (1972) 289.\\
%%CITATION = NCLTA,4,289;%%
  G. M. Cicuta and E. Montaldi, {\it Lett. Nuovo Cimento} {\bf 4} 
  (1972) 289. \\
%%CITATION = NCLTA,4,329;%%
  R. Gastmans and R. Meuldermans, \np{B63}{73}{277}.
%%CITATION = NUPHA,B63,277;%%

%%%%%%%%%%%%%%%%%% IBP's %%%%%%%%%%%%%%%%%%%%%%%%%%%%%%%%% 

\bibitem{Chet}
  F.V. Tkachov, \pl{B100}{81}{65}.\\
%%CITATION = PHLTA,B100,65;%%
  K.G. Chetyrkin and F.V. Tkachov, \np{B192}{81}{159}.
%%CITATION = NUPHA,B192,159;%%

%%%%%%%%%%%%%%%%%% Diff. Eqs. Method %%%%%%%%%%%%%%

\bibitem{Kotikov1} 
  A. V. Kotikov, \pl{B254}{91}{158}.
%%CITATION = PHLTA,B254,158;%%

\bibitem{Kotikov2} 
  A. V. Kotikov, \pl{B259}{91}{314}.
%%CITATION = PHLTA,B259,314;%%

\bibitem{Kotikov3} 
  A. V. Kotikov, \pl{B267}{91}{123}.
%%CITATION = PHLTA,B267,123;%%

\bibitem{Rem1} 
  E. Remiddi, \nc{110A}{97}{1435} ({\tt hep-th/9711188}).
%%CITATION = HEP-TH 9711188;%%

\bibitem{Rem2}
  M. Caffo, H. Czy\.{z}, S. Laporta and E. Remiddi, 
  \acta{B29}{98}{2627} ({\tt hep-ph/9807119}).\\
%%CITATION = HEP-TH 9807119;%%
  M. Caffo, H. Czy\.{z}, S. Laporta and E. Remiddi, 
  \nc{A111}{98}{365} ({\tt hep-ph/9805118}).
%%CITATION = HEP-TH 9805118;%%


%%%%%%%%%%%%%%%%%% LI %%%%%%%%%%%%%%%%%%%%%%%%%%%%%%%%%

\bibitem{Rem3}
  T. Gehrmann and E. Remiddi, \nptwoth{B580}{00}{485} 
  ({\tt hep-ph/9912329}). 
%%CITATION = HEP-PH 9912329;%%
  
%%%%%%%%%%%%%%%%%% Sym %%%%%%%%%%%%%%%%%%%%%%%%%%%%%%%%%%%%%

\bibitem{Bon1}
  R.~Bonciani, P.~Mastrolia and E.~Remiddi,
{\tt hep-ph/0301170}.
%%CITATION = HEP-PH 0301170;%%


%%%%%%%%%%%%% proiezione su fattori di forma invarianti %%%%

\bibitem{remiddivecchio}
  R.~Barbieri, J.~A.~Mignaco and E.~Remddi, \nc{11A}{72}{824}; 
  \nc{11A}{72}{865}. 
%%CITATION = NUCIA,A11,824;%%

%%%%%%%%%%%%% unica referenza diagramma ausiliario %%%%%%%%%

\bibitem{glover}
  E.~W.~N.~Glover, M.~E.~Tejeda-Yeomans, 
  {\it Nucl. Phys. Proc. Suppl.}  {\bf 89} (2000) 196
  ({\tt hep-ph/0010031}).
%%CITATION = HEP-PH 0010031;%%

\bibitem{Lap}
  S. Laporta and E. Remiddi, \pl{B379}{96}{283} ({\tt hep-ph/9602417}).
%%CITATION = APPOA,B28,959;%%

\bibitem{timovanritbergen}
  T.~van~Ritbergen and R.~G.~Stuart, 
  \nptwoth{B564}{00}{343} ({\tt hep-ph/9904240}).
%%CITATION = HEP-PH 9904240;%%

%%%%%%%%%%%%%%%%%%% Programs %%%%%%%%%%%%%%%%%%%%%%%%%%%%

\bibitem{FORM} J.A.M.\ Vermaseren, Symbolic Manipulation with
               {\tt FORM}, Version 2, CAN, Amsterdam, 1991; \\
               New features of {\tt FORM}, [math-ph/0010025].
%%CITATION = MATH-TH 0010025;%%


\bibitem{Mathe}
  {\it Mathematica 4.2}, Copyright 1988-2002 Wolfram Research, Inc.
  
\bibitem{SOLVE}
  {\tt SOLVE}, by E. Remiddi.

\bibitem{Maple}
  {\it MAPLE 7}, Copyright 2001 by Waterloo Maple 
  Software and the University of Waterloo.

\bibitem{fleischer}
  J. Fleischer, A. V. Kotikov and O. L. Veretin, 
   \np{B547}{99}{343}  ({\tt hep-ph/9808242}).
%%CITATION = HEP-PH 9808242;%%

\bibitem{fleischer0}
  J. Fleischer, A. V. Kotikov and O. L. Veretin, \pl{B417}{98}{163} 
  ({\tt hep-ph/9707492}).
%%CITATION = HEP-PH 9707492;%%

\bibitem{kodaira}
  M. Hori, H. Kawamura and J. Kodaira 
\pltwoth{B491}{00}{275} {(\tt hep-ph/0007329)}.
%%CITATION = HEP-PH 0007329;%%

%%%%%%%%%  vertici massless a 2 loops

\bibitem{Gonsalves83}
  R. J. Gonsalves, \prd{28}{83}{1542}.
%%CITATION = PHRVA,D28,1542;%%

\bibitem{VanNeerven86}
  W. L. van Neerven, \np{B268}{86}{453}.
%%CITATION = NUPHA,B268,453;%%

%%%%%% confronto risultati precedenti

\bibitem{smirnov}
  V. A. Smirnov, \pl{B404}{97}{101} ({\tt hep-ph/9703357}).
%%CITATION = HEP-PH 9703357;%%


\end{thebibliography}
\end{document}